\documentclass[a4paper]{article}

\pdfoutput=1

\RequirePackage{cmap}   % Для того, чтобы работал поиск русского текста в PDF файле

\usepackage{ifthen}

\newboolean{eng}
\newboolean{rus}
\newlength{\lengrus}

\setboolean{eng}{true}
\setboolean{rus}{true}
\setboolean{rus}{false}     %%% Clean/insert  "%" in the beginning of the line for exclusion/inclusion of the Russian

\usepackage{engrus}

\usepackage{ifpdf}

\usepackage{float}

\makeatletter
\renewcommand{\@oddhead }{\vbox{\hbox to\textwidth{%
\href {http://iopscience.iop.org/article/10.1088/2399-6528/aadd73/meta}{A.~A.~Chernitskii {\it J. Phys. Commun.} {\bf 2} (2018) 105013}
\hfill\thepage\strut}\hrule
\vspace{0.6ex}
\noindent
\mifengrus{\hfill}{}\mifrus{Авторский текст на русском языке}{}
}}
\makeatother

\ifpdf
\usepackage{graphicx}
  \usepackage{xcolor}
  \definecolor{darkgreen}{rgb}{0,.35,0}
   \definecolor{col1}{rgb}{0.125,0.0625,0.125}

\RequirePackage[colorlinks=true
,urlcolor=col1
,anchorcolor=col1
,citecolor=col1
,filecolor=col1
,linkcolor=col1
,menucolor=col1
,pagecolor=col1
,pdfproducer=medialab
,pdfa=true
]{hyperref}

\else
  \usepackage[dvips]{graphicx}
\usepackage[bookmarksnumbered,plainpages,hypertex,unicode,hyperindex]{hyperref}
\fi

\usepackage{latexsym}
\usepackage{amsfonts}
\usepackage{amssymb}
\usepackage{amsbsy}
\usepackage{amsmath}   %
\usepackage{supertabular}

\usepackage{color}

\usepackage{xcolor}

\usepackage{ulem}

\definecolor{lgreen}{rgb}{0.9,1,0.8}
\definecolor{mcolora}{rgb}{0.75,0.25,0}
\definecolor{mcolorr}{rgb}{0,0.75,1}
\definecolor{mcolorm}{rgb}{1,0,0}

\definecolor{mcolorab}{rgb}{1,0,0}
\definecolor{mcolorrb}{rgb}{0,0,1}
\definecolor{mcolormb}{rgb}{1,0,0}

\definecolor{mcolorab}{rgb}{0.5,0,0.5}
\definecolor{mcolorab}{rgb}{1,0.5,0}
\definecolor{mcolorab}{rgb}{1,0.3,0}
\definecolor{mcolorab}{rgb}{0,.35,0}
\definecolor{mcolorab}{rgb}{0,.25,0}
\definecolor{mcolorrb}{rgb}{0,0,0}
\definecolor{mcolormb}{rgb}{1,0,0}

\definecolor{mcolorac}{rgb}{1,0.5,0}
\definecolor{mcolorrc}{rgb}{0,0.5,1}

\def\textaddb#1{\textcolor{mcolorab}{#1}}
\def\textremb#1{\textcolor{mcolorrb}{\sout{#1}}}
\def\mmparb#1{\marginpar{\textcolor{mcolormb}{#1}}}
\def\textaddb#1{#1}
\def\textremb#1{}
\def\mmparb#1{}

\def\textaddc#1{\textcolor{mcolorac}{#1}}
\def\textremc#1{\textcolor{mcolorrc}{\sout{#1}}}
\def\textaddc#1{#1}
\def\textremc#1{}

\def\textadd1#1{\textcolor{mcolora}{#1}}
\def\textrem1#1{\textcolor{mcolorr}{\sout{#1}}}
\def\mmpar1#1{\marginpar{\textcolor{mcolorm}{#1}}}
\def\textadd1#1{#1}
\def\textrem1#1{}
\def\mmpar1#1{}

\def\textaddd#1{\textcolor{mcolorab}{#1}}
\def\textremd#1{\textcolor{mcolorrb}{\sout{#1} }}

\def\textaddd#1{#1}
\def\textremd#1{}

\makeatletter
\begingroup\expandafter\expandafter\expandafter\endgroup
\expandafter\ifx\csname subequations\endcsname\relax
\else
\let\HyOrg@subequations\subequations
\def\subequations{%
\stepcounter{equation}%
\protected@edef\theHparentequation{%
\@ifundefined{theHequation}\theequation\theHequation
}%
\addtocounter{equation}{-1}%
\HyOrg@subequations
\def\theHequation{\theHparentequation\alph{equation}}%
\ignorespaces
}%
\fi
\makeatother

\usepackage[usenames,dvipsnames]{pstricks}
\ifpdf
 \else
  \usepackage{epsfig}
  \usepackage{pst-solides3d}
\fi

\makeatletter
    \@addtoreset{equation}{section}
    \renewcommand{\theequation}{\arabic{section}.\arabic{equation}}
    \@addtoreset{figure}{section}
    
\makeatother

\def\p{\partial}
\def\dfrac#1#2{{\displaystyle\frac{#1}{#2}}}
\def\stTD#1#2{\hbox to 0em{\mathsurround=0em $\stackrel{#1}{\makebox[0pt]{} #2}$\hss} \phantom{#2}}\def\stscript#1#2{\hbox to 0em{\mathsurround=0em ${\scriptstyle\stackrel{#1}{\makebox[0pt]{} #2}}$\hss} \phantom{#2}}\def\stscriptscript#1#2{\hbox to 0em{\mathsurround=0em ${\scriptscriptstyle\stackrel{#1}{\makebox[0pt]{} #2}}$\hss} \phantom{#2}}
\def\comb#1#2#3{{\mathsurround 0pt\hbox to 0pt {\hspace*{#3}\raisebox{#2}{${#1}$}\hss}}}
\def\combs#1#2#3{{\mathsurround 0pt\hbox to 0pt {\hspace*{#3}\raisebox{#2}{${\scriptstyle #1}$}\hss}}}
\def\combss#1#2#3{{\mathsurround 0pt\hbox to 0pt {\hspace*{#3}\raisebox{#2}{${\scriptscriptstyle #1}$}\hss}}}
\def\e#1{\mathrm{e}^{#1}}

\def\df{\mathrm{d}}

\def\metr{\mathfrak{m}}

\def\metrp{\mathchoice{\comb{-}{-0.9ex}{0ex}\mathfrak{m}}{\comb{-}{-0.9ex}{0ex}\mathfrak{m}}{\combs{-}{-0.75ex}{-0.1ex}\mathfrak{m}}{}{}}

\def\bje{{\mathsurround 0pt\lower.0ex\hbox{${\scriptscriptstyle \mathbf{e}}$}\mspace{-3.4mu}\mathbf{j}}}
\def\je{{\mathsurround 0pt\lower.0ex\hbox{${\scriptscriptstyle e}$}\mspace{-4.5mu}j}}
\def\bjm{{\mathsurround 0pt\lower.0ex\hbox{${\scriptscriptstyle \mathbf{m}}$}\mspace{-5.6mu}\mathbf{j}}}
\def\jm{{\mathsurround 0pt\lower.0ex\hbox{${\scriptscriptstyle m}$}\mspace{-7.0mu}j}}

\def\p{\partial}

\def\him{\imath}

\def\hconj#1{\mathchoice{{{}^{\boldsymbol{*}}\mspace{-4mu}#1}}{{{}^{\boldsymbol{*}}\mspace{-4mu}#1}}{{{}^{\boldsymbol{*}}\mspace{-4mu}#1}}{}{}}

\def\eqdef{\doteqdot}
\def\EMT{\mathchoice{\combs{\to}{0.3ex}{-0.2ex}{T}}{\combs{\to}{0.3ex}{-0.2ex}{T}}{\combss{\to}{0.2ex}{-0.2ex}{T}}{\combss{\to}{0.2ex}{-0.2ex}{T}}}
\def\EMTc{\mathchoice{\combs{-}{1.5ex}{0.2ex}{\EMT}}{\combs{-}{1.5ex}{0.2ex}{\EMT}}{\combss{-}{1.1ex}{0.05ex}{\EMT}}{\combss{-}{1.1ex}{0.05ex}{\EMT}}}
\def\EMTi{\mathchoice{\combss{\infty}{1.8ex}{0.15ex}\EMT}{\combss{\infty}{1.85ex}{0.15ex}\EMT}{\combss{\infty}{1.25ex}{-0.12ex}\EMT}{\combss{\infty}{1.2ex}{-0.12ex}\EMT}}
\def\AMT{\mathchoice{\combs{\circ}{0.9ex}{0.9ex}{M}}{\combs{\circ}{0.9ex}{0.9ex}{M}}{\combss{\circ}{0.7ex}{0.65ex}{M}}{\combss{\circ}{0.7ex}{0.65ex}{M}}}

\def\Vols{V}
\def\vdlina{l_{\!v}}
\def\dVols{\df\mspace{-2mu}\Vols}
\def\Phase{\theta}
\def\bPhase{\overline{\Phase}}

\def\cE{\mathcal{E}}

\def\bcP{\boldsymbol{\mathcal{P}}}
\def\cP{\mathcal{P}}
\def\bcM{\boldsymbol{\mathcal{J}}}
\def\cM{\mathcal{J}}
\def\OOO#1#2{\mathcal{O}\!\left(#1\right)_{#2}}

\def\Cron{\delta}
\def\LCh{\epsilon}
\def\Energy{\mathbb{E}}
\def\bEMV{\pmb{\mathbb{P}}}
\def\EMV{\mathbb{P}}
\def\bAMV{\pmb{\mathbb{J}}}
\def\AMV{\mathbb{J}}
\def\Act{\mathcal{A}}

\def\Vol{\overline{V}}
\def\dVol{\df\mspace{-2mu}\Vol}

\def\xxx{\chi}

\def\ffun{\Phi}

\def\dffun{\ffun}

\def\ffind{\Upsilon}
\def\q{\bar{q}}
\def\br{\bar{r}}
\def\qL{\bar{\bar{q}}} %\breve{q}}
\def\brrho{\bar{\bar{\rho}}} %\breve{\rho}}
\def\LF{\mathchoice{\combs{-}{0.3ex}{-0.1ex}\mathcal{L}}{\combs{-}{0.3ex}{-0.1ex}\mathcal{L}}{\combss{-}{0.25ex}{-0.12ex}\mathcal{L}}{}{}}
\def\LFo{\mathchoice{\combs{=}{0.3ex}{-0.1ex}\mathcal{L}}{\combs{=}{0.3ex}{-0.1ex}\mathcal{L}}{\combss{=}{0.25ex}{-0.12ex}\mathcal{L}}{}{}}

\def\const{{\rm const}}
\def\fcE{f^{\cE}}
\def\bta{\overline{o}}
\def\btam{\overline{\rho}}
\def\btap{\overline{\varphi}}
\def\pint{\mathcal{I}}
\def\pcoef{\mathcal{C}}
\def\mxi{\xi}
\def\mxic{\hconj{\xi}}

\def\mtxi{\mathchoice{{\comb{\tilde{}}{0.45ex}{0.6ex}{\xi}}}{{\comb{\tilde{}}{0.45ex}{0.6ex}{\xi}}}{{\combs{\tilde{}}{0.35ex}{0.25ex}{\xi}}}{}{}}
\def\mtcxi{\hconj{\mtxi}}
\def\mtrho{\mathchoice{{\comb{\tilde{}}{0ex}{0.7ex}{\rho}}}{{\comb{\tilde{}}{0ex}{0.7ex}{\rho}}}{{\combs{\tilde{}}{0.01ex}{0.6ex}{\rho}}}{}{}}
\def\mtphi{\mathchoice{{\comb{\tilde{}}{0ex}{0.7ex}{\varphi}}}{{\comb{\tilde{}}{0ex}{0.7ex}{\varphi}}}{{\combs{\tilde{}}{0.01ex}{0.6ex}{\varphi}}}{}{}}
\def\romega{\tilde{\omega}}
\def\cbrho{\mathchoice{\combs{\boldsymbol{-}}{0.1ex}{-0.1ex}{\rho}}{\combs{\boldsymbol{-}}{0.1ex}{-0.1ex}{\rho}}{\combss{\boldsymbol{-}}{0.1ex}{-0.15ex}{\rho}}{}{}}
\def\cbphi{\mathchoice{\combs{\boldsymbol{-}}{0.1ex}{-0.1ex}{\varphi}}{\combs{\boldsymbol{-}}{0.1ex}{-0.1ex}{\varphi}}{\combss{\boldsymbol{-}}{0.1ex}{-0.15ex}{\varphi}}{}{}}
\def\cbxi{\mathchoice{\combs{\boldsymbol{-}}{-0.1ex}{-0.25ex}{\xi}}{\combs{\boldsymbol{-}}{-0.1ex}{-0.25ex}{\xi}}{\combss{\boldsymbol{-}}{-0.06ex}{-0.30ex}{\xi}}{}{}}
\def\mxis{\overline{\xi}}
\def\trhoinf{\tilde{\rho}_{\!\!\infty}}
\def\zdlina{l_{\!s}}
\def\Wspos{W}
\def\Sentr{S}
\def\Ugas{U}
\def\NVols{N}
\def\gVols{E}
\def\uVols{u}
\def\Tenerg{\mathcal{T}}
\def\kBol{k_{\!B}}
\def\Tabs{\mathchoice{\combs{\circ}{0.4ex}{0.9ex}{T}}{\combs{\circ}{0.4ex}{0.9ex}{T}}{\combss{\circ}{0.3ex}{0.75ex}{T}}{\combss{\circ}{0.3ex}{1.6ex}{T}}}
\def\pollog#1#2{\mathchoice{\mathrm{L\mspace{-4mu}i\mspace{0.5mu}}_{#1}\mspace{-5mu}\left(#2\right)}{\mathrm{L\mspace{-4mu}i\mspace{0.5mu}}_{#1}\mspace{-5mu}\left(#2\right)}{\mathrm{L\mspace{-4mu}i\mspace{0mu}}_{#1}\mspace{-3mu}\left(#2\right)}{\mathrm{L\mspace{-4mu}i\mspace{0mu}}_{#1}\mspace{-3mu}\left(#2\right)}}

\begin{document}

\begin{titlepage}
\phantom{.}

\phantom{.}
\vspace{-3.2pc}
\vbox{\hbox to\textwidth{%
%A.~A.~Chernitskii {\it J. Phys. Commun.} {\bf 2} (2018) 105013
\href {http://iopscience.iop.org/article/10.1088/2399-6528/aadd73/meta}{A.~A.~Chernitskii {\it J. Phys. Commun.} {\bf 2} (2018) 105013}
\hfill \href {https://doi.org/10.1088/2399-6528/aadd73}{\path{doi:10.1088/2399-6528/aadd73}}\strut}\hrule
\vspace{0.6ex}
\noindent
\mifengrus{\hfill}{}\mifrus{Авторский текст на русском языке}{}
}

\mifeng{%
\begin{center}
{\bf \LARGE LIGHTLIKE SHELL SOLITONS\\[1.25ex]OF EXTREMAL SPACE-TIME FILM}\\[4ex]
{\bf\large Alexander A. Chernitskii}\\[2ex]
\small $^1$ \it Department  of Mathematics,\\
\small\it St. Petersburg State Chemical Pharmaceutical University,\\
\small\it Prof. Popov str. 14, St. Petersburg, 197022, Russia\\[1ex]
\small $^2$\it A.~Friedmann Laboratory for Theoretical Physics,\\
\small St.-Petersburg, Russia\\[1ex]
\small\rm AAChernitskii@mail.ru
\end{center}
\vspace{1ex}
\hrule
\vspace{2ex}
}{}

\mifrus{%
\selectlanguage{russian}
\begin{center}
{\bf\LARGE СВЕТОПОДОБНЫЕ\\[0.5ex] ОБОЛОЧЕЧНЫЕ СОЛИТОНЫ\\[0.5ex] ЭКСТРЕМАЛЬНОЙ\\[0.5ex] ПРОСТРАНСТВЕННО-ВРЕМЕННОЙ\\[0.5ex] ПЛЁНКИ}\\[4ex]
{\bf\large Александр А. Черницкий}\\[2ex]
\small $^1$ \it Кафедра высшей математики,\\
\small\it Санкт-Петербургский Химико-Фармацевтический Университет,\\
\small\it ул. Проф. Попова 14, Санкт-Петербург, 197022, Россия\\[1ex]
\small $^2$\it Фридмановская Лаборатория Теоретической Физики,\\
\small Санкт-Петербург, Россия\\[1ex]
\small\rm AAChernitskii@mail.ru
\end{center}
}{}

\vspace{1ex}

\mifrus{\hrule}{}

\vspace{5ex}

\mifengrus{\newpage}{}
\mifeng{%
\begin{center}
{\bf Abstract}
\end{center}
New exact solution class of Born -- Infeld type nonlinear scalar field model is obtained.
The variational principle of this model has a specific form which is
characteristic for extremal four-dimensional hypersurface or
%\textremc{hyper film}
\textaddc{hyper-film} in five-dimensional space-time.
Obtained solutions are singular solitons propagating with speed of light and
having energy, momentum, and angular momentum which can be calculated
for explicit conditions. Such solitons will be called the lightlike ones.
The soliton singularity has a form of moving two-dimensional surface
or shell.
The lightlike soliton can have a set of tubelike singular shells with the appropriate cavities.
A twisted lightlike soliton is considered.
 It is notable that its energy is proportional to its angular momentum in high-frequency approximation.
  A case with one tubelike cavity is considered.
In this case the soliton shell is diffeomorphic to a cylindrical surface with
threads by  multifilar helix.
The shell transverse size of the appropriate finite energy soliton can be converging to zero at infinity.
 The ideal gas of such lightlike solitons with minimal twist parameter is considered in a finite volume.
Explicit conditions provide that the angular momentum of each soliton in the volume equals Planck constant.
 The equilibrium energy spectral density for the solitons is obtained.
It has the form of Planck distribution in some approximation.
 A beam of \textaddc{the} twisted lightlike solitons is considered.
 The representation of arbitrary polariza\-tion for \textaddc{the} beam with \textaddc{the} twisted lightlike solitons is discussed.
 It is shown that
% \textremc{this beam \textadd1{can} provides}
 the effect of mechanical angular momentum transfer to absorbent
by \textaddc{the} circularly polarized beam \textaddc{can be provided}. This effect \textadd1{is} well known for photon beam.
 Thus the soliton solution which have determinate likeness with photon is obtained in particular.
\vspace{2ex}
}{}

\mifrus{%
\selectlanguage{russian}
\begin{center}
{\bf Резюме}
\end{center}
Получен новый класс точных решений нелинейной скалярной полевой модели типа Борна -- Инфельда.
Вариационный принцип этой модели имеет специфический вид, характерный для экстремальной четырёхмерной гиперповерхности
или гиперплёнки в пятимерном пространстве-времени. Полученные решения представляют собой сингулярные солитоны,
распространяющиеся со скоростью света и имеющие энергию, импульс и момент импульса, которые могут быть вычислены
для определённых условий. Такие солитоны будут называться светоподобными. Солитонная сингулярность имеет вид двумерной движущейся поверхности или оболочки. Светоподобный солитон может иметь множество трубчатых сингулярных оболочек с соответствующими полостями. Рассмотрен закрученный светоподобный солитон. Примечательно, что
 в высокочастотном приближении его энергия пропорциональна его моменту импульса. Рассмотрен случай с одной трубчатой полостью. В этом случае солитонная оболочка диффеоморфна  цилиндрической поверхности с нарезами многозаходной спиралью. Поперечный размер оболочки соответствующего солитона конечной энергии может сходиться к нулю на бесконечности.
Рассмотрен  идеальный газ таких светоподобных солитонов с минимальным параметром закрученности в конечном объёме.
Определённые условия обеспечивают равенство момента импульса каждого солитона постоянной Планка.
Получена равновесная спектральная плотность энергии солитонов. Она имеет вид распределения Планка в некотором приближении. Рассмотрен луч закрученных светоподобных солитонов. Обсуждается представление произвольной поляризации для луча
с закрученными светоподобными солитонами. Показано, что может быть обеспечен эффект передачи механического момента импульса поглотителю  циркулярно поляризованным лучём.
Этот эффект хорошо известен для фотонного луча.
Таким образом, в частности, получено солитонное решение, имеющее
определённое сходство с фотоном.
}{}

\mifengrus{\vspace{10ex}}{\mifeng{\newpage}{\vspace{15ex}}}

\hrule

\vspace{1ex}

\tableofcontents

\vspace{4ex}

\hrule

\vspace{4ex}

\end{titlepage}

%\newpage

\engrus{1ex}{3ex}{%
\section{Introduction}
\label{introd}
}{%
\mifeng{\addtocounter{section}{-1}}{}
\section{Введение}
\mifeng{}{\label{introd}}
}
\engrus{0.5ex}{0.5ex}{%
A nonlinear space-time scalar field model considered here is known for a long time sufficiently.
This model is related to well known Born -- Infeld nonlinear electrodynamics \textadd1{\cite{BornInfeld1934a,Chernitskii2004a}}.
\textremb{, and it}\textaddb{It} is sometimes called \textaddb{also} Born -- Infeld type scalar field model
\textadd1{\cite{BarbChern1967-1e}}.
}{%
Рассматриваемая здесь нелинейная пространственно-временная скалярная полевая модель известна достаточно давно.
Эта модель имеет отношение к хорошо известной нелинейной электродинамике Борна -- Инфельда \cite{BornInfeld1934a,Chernitskii2004a}.
Её также иногда называют скалярной полевой моделью типа Борна -- Инфельда
\cite{BarbChern1967-1e}.
}

\engrus{0.5ex}{0.5ex}{%
This model is attractive because it has \textremb{relatively}\textaddb{comparatively} simple and geometrically clear form.
It can be considered as a relativistic generalization of the minimal surface or
minimal thin film model in three-dimensional space. \textremd{(see, for example,\cite{DubrNovFom1992e})}
}{%
Эта модель привлекательна поскольку имеет относительно простой и геометрически ясный вид.
Она может рассматриваться как релятивистское обобщение модели минимальной поверхности или минимальной тонкой плёнки в трёхмерном пространстве.
}

\engrus{0.5ex}{0.5ex}{%
In this generalization we have an extremal four-dimensional film in five-dimensional space-time.
But the model equation appears as differential one for \textaddd{a} scalar field in four-dimensional space-time.
}{%
В этом обобщении мы имеем экстремальную четырёхмерную плёнку в пятимерном про\-странстве-времени. Однако модельное уравнение
представляет собой дифференциальное уравнение для скалярного поля в четырёхмерном пространстве-времени.
}

\engrus{0.5ex}{0.5ex}{%
On the other hand, this model can
\textremb{provide}\textaddb{describe}
the \textremb{necessary}\textaddb{observable physical} effects%
\textaddb{,}
which
\textaddb{is a necessary requirement} \textremb{are required}%
for \textremb{a}\textaddb{the} realistic filed model.
}{%
С другой стороны, эта модель может
описывать
наблюдаемые физические эффекты, что является необходимым требованием для реалистической полевой модели.
}

\engrus{0.5ex}{0.5ex}{%
In particular, the model under consideration has a static spherically symmetric solution,
which is identical \textaddb{in form} to \textaddb{a} zero \textremb{four-vector}component of \textaddb{the} electromagnetic \textaddb{four-vector} potential for dyon solution
of Born -- Infeld electrodynamics \cite{Chernitskii1999,Chernitskii2012be}.
This static solution of the scalar model gives
 the appropriate moving soliton solution
with the aid of Lorentz transform.
}{%
В частности, рассматриваемая модель имеет статическое сферически симметричное решение, которое совпадает по виду с нулевой компонентой электромагнитного
четырёхвекторного потенциала для дионного решения электродинамики Борна -- Инфельда \cite{Chernitskii1999,Chernitskii2012be}.
Это статическое решение скалярной модели даёт соответствующее движущееся решение посредством преобразования Лоренца.
}

\engrus{0.5ex}{0.5ex}{%
As it was shown in the \textrem1{cited}work \cite{Chernitskii1999},
in the case of nonlinear electrodynamics there are the conformity between \textadd1{the} long-range interaction
of solitons and two known long-range interactions of physical particles, \textadd1{those are} \textrem1{that is}electromagnetic
and gravitational ones. But the methods which was used for the  investigation of soliton long-range interaction
are independent of the field model. The appropriate instruments are \textadd1{the} integral conservation laws and \textadd1{the} characteristic equation.
}{%
Как было показано в работе \cite{Chernitskii1999}, в случае нелинейной электродинамики имеется соответствие между дальнем взаимодействием солитонов и двумя известными дальними взаимодействиями
физических частиц, а именно электромагнитным и гравитационным. Однако методы, которые были использованы для исследования солитонного дальнего взаимодействия являются независимыми от
полевой модели. Соответствующие инструменты -- это интегральные законы сохранения и характеристическое уравнение.
}

\engrus{0.5ex}{0.5ex}{%
These methods applying to the scalar model under consideration give the results, which
are similar to ones for \textadd1{the} nonlinear electrodynamics \cite{Chernitskii2016b,Chernitskii2017a}.
\textremd{These results in detail must be \textadd1{a} matter for another article.}%
Here we briefly discuss the \textrem1{obtaining of Lorentz force for interacting}\textadd1{interaction of such} scalar solitons in the next section.
}{%
Эти методы, применённые к рассматриваемой скалярной полевой модели дают результаты, которые сходны с результатами, полученными для нелинейной электродинамики \cite{Chernitskii2016b,Chernitskii2017a}.
%Эти результаты в деталях должны быть предметом другой статьи.
Здесь мы кратко обсудим взаимодействие подобных скалярных солитонов в следующей секции.
}

\engrus{0.5ex}{0.5ex}{%
An essential difference of the scalar field model from the nonlinear electrodynamics
is obviously caused by the different tensor character of the fields.
In particular, a weak single-component scalar wave can not
\textremb{provide}\textaddb{describe}
the transverse polarization of \textadd1{an} electromagnetic wave.
}{%
Существенное отличие скалярной полевой модели от нелинейной электродинамики, очевидно, обусловлено различным тензорным характером полей.
В частности, слабая однокомпонентная скалярная волна не может описать поперечную поляризацию электромагнитной волны.
}

\engrus{0.5ex}{0.5ex}{%
But at the present time we
\textaddb{can}
consider the light as \textadd1{a} photon
\textremb{beam}\textaddb{flux}
but not a weak electromagnetic wave
with \textremd{the}\textaddd{a} constant amplitude.
The photon beam could be represented by an appropriate scalar soliton beam.
In this case an essential space-time nonhomogeneous of soliton solution may
provide the necessary symmetry properties for the beam.
}{%
Однако, в настоящее время мы можем рассматривать свет как поток фотонов, а не слабую электромагнитную волну постоянной амплитуды.
Фотонный луч мог бы быть представлен соответствующим лучом скалярных солитонов.
В этом случае существенная неоднородность солитонного решения может обеспечить необходимые свойства симметрии фотонного луча.
}

\engrus{0.5ex}{0.5ex}{%
Thus at the present work we consider the model of extremal space-time film.
We obtain its singular exact soliton solutions propagating with the speed of light and
having energy, momentum, and angular momentum which can be calculated
for explicit conditions. Such solitons will be called the lightlike ones.
}{%
Таким образом в настоящей работе мы рассматриваем модель экстремальной пространст\-венно-временной плёнки.
Мы получаем её сингулярное точное солитонное решение, распространяющееся со скоростью света и имеющее энергию, импульс и момент импульса,
которые могут быть вычислены для определённых условий. Такие солитоны будут называться светоподобными.
}

\engrus{0.5ex}{0.5ex}{%
Then we investigate in detail a lightlike soliton solution having a
rotation about the direction of propagation that is twisted lightlike soliton.
}{%
Затем мы детально исследуем светоподобный солитон обладающий вращением относительно направления распространения, то есть вращающийся светоподобный солитон.
}

\engrus{0.5ex}{0.5ex}{%
We consider the ideal gas of such twisted lightlike solitons.
Using explicit assumptions we obtain Planck distribution formula
in some approximation.
}{%
Мы рассматриваем идеальный газ таких вращающихся светоподобных солитонов.
Используя определённые предположения, мы получаем формулу распределения Планка в некотором приближении.
}

\engrus{0.5ex}{0.5ex}{%
At last we consider a beam with the twisted lightlike solitons.
We show that this beam can represent \textadd1{the} photon one. In this case
\textremb{we}\textaddb{the beam can}
have \textremb{, in particular,}the polarization property and
\textaddb{can provide} the effect of \textadd1{the} mechanical angular momentum transfer
to absorbent \textremb{by \textadd1{the} circularly polarized beam.}%
\textaddb{for the circularly polarization.}
}{%
Наконец мы рассматриваем луч вращающихся светоподобных солитонов. Мы показываем, что этот луч может представлять поток фотонов.
В этом случае луч может иметь свойство поляризации и может обеспечивать эффект передачи механического момента импульса поглотителю при циркулярной поляризации.
}

\engrus{3ex}{2ex}{%
\section{Extremal space-time film}
\label{estf}
}{%
\mifeng{\addtocounter{section}{-1}}{}
\section[Экстремальная пространственно-временная плёнка]{Экстремальная пространственно-временная плёнка}
\mifeng{}{\label{estf}}
}

\begin{subequations}\label{349227101}
\engrus{0ex}{0ex}{%
Let us consider the following action
\textaddb{integral}
which has the world volume form:
}{%
Рассмотрим следующий интеграл действия, имеющий форму мирового объёма
}
 \begin{equation}
 \label{38184713}
\Act   = \int\limits_{\Vol}\sqrt{|\mathfrak{M}|}\;\left(\df x\right)^{4}
 \;,
 \end{equation}
\engrus{0ex}{0ex}{%
\noindent
where
\mbox{$\mathfrak{M} \eqdef \det(\mathfrak{M}_{\mu\nu})$},
\mbox{$\left(\df x\right)^{4} \eqdef \df x^{0}\df x^{1}\df x^{2}\df x^{3}$},
$\Vol$ is space-time volume,
}{%
\noindent
где
\mbox{$\mathfrak{M} \eqdef \det(\mathfrak{M}_{\mu\nu})$},
\mbox{$\left(\df x\right)^{4} \eqdef \df x^{0}\df x^{1}\df x^{2}\df x^{3}$},
\mbox{$\Vol$ --} про\-странст\-венно-временной объём,
}
  \begin{equation}
 \label{381936901}
 \mathfrak{M}_{\mu\nu} = \metr_{\mu\nu} + \xxx^2\,\frac{\p \ffun}{\p x^{\mu}}\,\frac{\p \ffun}{\p x^{\nu}}
 \;,
 \end{equation}
\engrus{0.5ex}{0.5ex}{%
\mifrus{\noindent}{}%
$\metr_{\mu\nu}$ are \textadd1{the} components of \textadd1{the} metric tensor for \textadd1{the} flat four-dimensional space-time,
$\ffun$ is \textadd1{a} scalar real field function, $\xxx$ is \textadd1{a} dimensional constant.
The Greek indices take \textadd1{the} values $\{0,1,2,3\}$.
}{%
\mifeng{\noindent}{}%
$\metr_{\mu\nu}$ -- компоненты метрического тензора для плоского четырёхмерного про\-странства-времени,
$\ffun$ -- скалярная действительная функция, $\xxx$ -- размерная константа.
Греческие индексы принимают значения $\{0,1,2,3\}$.
}
 \end{subequations}

\engrus{0.5ex}{0.5ex}{%\mifrus{}{}%\noindent%
\textaddd{%
As we have mentoned in introduction, the action (\ref{349227101}) is a generalization
of the appropriate expression for the mathematical model of two-dimensional minimal thin film
in the tree-dimensional space of our everyday experience.
Such generalized model is considered, for example, in the monograph \cite{DubrNovFom1992e}.
}
}{%\mifeng{}{}%\noindent%
Как было упомянуто во введении, действие (\ref{349227101}) является обобщением
соответствующего выражения для математической модели двумерной минимальной тонкой плёнки
%, например мыльной,
в трёхмерном пространстве нашего повседневного опыта.
%Надо отметить, что подинтегральная функция действия (\ref{38184713}) не обращается в $0$ на бесконечности.
Подобная обобщённая модель рассматривается, например, в монографии \cite{DubrNovFom1992e}.
%(см., например, ).
}

\engrus{0.5ex}{0.5ex}{%\mifrus{}{}%\noindent%
\textaddd{%
It is worth emphasizing that the choice of such model is consistent with the trend of geometrization
of physics, which has already proven its fruitfulness in description of processes and phenomena of  the material world. Here we have in mind the conception of general relativity.
From the standpoint of discussion for the questions related to geometrization of physical world picture,
the monographes \cite{Weyl1922E,Eddington1924,MiznerTornUillerI1973E} should be noted.
}
}{%\mifeng{}{}%\noindent%
Надо сказать, что выбор такой математической модели соответствует тенденции геометризации физики, которая уже доказала свою плодотворность в описании процессов и явлений материального мира. Здесь мы имеем ввиду концепцию общей относительности. С точки зрения обсуждения вопросов, связанных с геометризацией физической картины мира, следует отметить монографии
\cite{Weyl1922E,Eddington1924,MiznerTornUillerI1973E}.
}

\engrus{0.5ex}{0.5ex}{%\mifrus{}{}%\noindent%
\textaddd{%
A natural feature of the world volume action (\ref{349227101})
is that his density does not
%disappear
vanish
for zero derivatives of the
field function $\p\,\Phi/\p\,x^{\mu} = 0$ and, consequently, for zero hypersurface curvature.
%Likewise,
%In a similar spirit
In just the same way
an area element of a two-dimensional thin film in the three-dimensional space does
not
%disappear
vanish
where
the curvature of the minimal surface becomes zero.
}
}{%\mifeng{}{}%\noindent%
Естественной характерной чертой действия мирового объёма (\ref{349227101}) является то, что его плотность не исчезает при обращении в ноль производных полевой функции $\p\,\Phi/\p\,x^{\mu} = 0$
 и, следовательно, кривизны гиперповерхности.
Точно так же не исчезает элемент площади двумерной тонкой плёнки в трёхмерном пространстве, там где
кривизна минимальной поверхности обращается в ноль.
}

\engrus{0.5ex}{0.5ex}{%
\textaddd{Thus} the variational principle
\mbox{$\delta \Act = 0$}
with action (\ref{349227101}) corresponds to extremal four-dimensional film $\ffun (\{x^{\mu}\})$ in
\textaddd{a} five-dimensional space-time
%\mifeng{\mifrus{}{\break}}{}
$\{\ffun,x^{0}, x^{1}, x^{2}, x^{3} \}$.
}{%
Таким образом вариационный принцип \mbox{$\delta \Act = 0$}
с действием (\ref{349227101}) соответствует экстремальной четырёхмерной плёнке $\ffun (\{x^{\mu}\})$ в
пятимерном пространстве-времени $\{\ffun, x^{0}, x^{1}, x^{2}, x^{3} \}$.
}

\engrus{0.5ex}{0.5ex}{%
Determinant $\mathfrak{M}$ in (\ref{349227101}) can be represented in the form
}{%
Детерминант $\mathfrak{M}$ в (\ref{349227101}) может быть представлен в виде
}
\begin{equation}
\label{37937480}
\mathfrak{M}  = \metr \left(1 + \xxx^{2}\,\metr^{\mu\nu}\,\frac{\p \ffun}{\p x^{\mu}}\,\frac{\p \ffun}{\p x^{\nu}} \right)
\,,
\end{equation}
\engrus{0.5ex}{0.5ex}{%
\noindent
where $\metr \eqdef \det(\metr_{\mu\nu})$.
}{%
\noindent
где $\metr \eqdef \det(\metr_{\mu\nu})$.
}

\engrus{1ex}{1ex}{%
Taking into account (\ref{37937480}) we can write \textadd1{the} model action (\ref{349227101}) in the form
}{%
Учитывая (\ref{37937480}), мы можем записать модельное действие (\ref{349227101}) в виде
}
\begin{subequations}\label{37974058}
\begin{equation}
\label{382010071}
\Act  = \int\limits_{\Vol}\LF\;\dVol
\;,
\end{equation}
\engrus{1ex}{1ex}{%
\noindent
where $\dVol \eqdef \sqrt{|\metr|}\;\left(\df x\right)^{4}$ is \textadd1{a} four-dimensional volume element,
}{%
\noindent
где $\dVol \eqdef \sqrt{|\metr|}\;\left(\df x\right)^{4}$ -- четырёхмерный элемент объёма,
}
\begin{equation}
\label{383322881}
\LF  \eqdef
\sqrt{\left|1 + \xxx^{2}\,\metr^{\mu\nu}\,\frac{\p \ffun}{\p x^{\mu}}\,\frac{\p \ffun}{\p x^{\nu}}\right| }
\;.
\end{equation}
\end{subequations}

\engrus{1ex}{1ex}{%
\textadd1{The} variational principle with action (\ref{37974058}) gives the following model equation:
}{%
Вариационный принцип с действием (\ref{37974058}) даёт следующее модельное уравнение:
}
\begin{subequations}\label{408139091}
\begin{equation}
\label{407970191}
  \frac{1}{\sqrt{|\metr|}}\frac{\p }{\p x^{\mu}} \sqrt{|\metr|}\,\ffind^{\mu} = 0
\;,
\end{equation}
\engrus{0.5ex}{0.5ex}{%
\noindent%
where
}{%
\noindent%
где
}
\begin{align}
\label{421005381}
\ffind^{\mu}  &\eqdef \frac{\dffun^{\mu}}{\LF}
\;,\quad
\dffun^{\mu} = \metr^{\mu\nu}\,\dffun_{\nu}
\;,
\\
\label{35751937}
\dffun_{\nu}  &\eqdef \frac{\p\ffun}{\p x^{\nu}}
\;.
\end{align}
\end{subequations}

\engrus{0.5ex}{0.5ex}{%\mifrus{}{}%\noindent%
\textaddd{%
Note that the action (\ref{349227101}), (\ref{37974058}) and the equation (\ref{408139091})
are valid for the field outside of singularities. The
presence of field singularities requires an addition
of the appropriate singular densities to the action (\ref{349227101}), (\ref{37974058})
and in the right side of the equation (\ref{408139091}). This aspect was considered for the case
of nonlinear electrodynamics in the work \cite{Chernitskii1999} and it is
%discussed in detail
expanded
in the monograph \cite{Chernitskii2012be}. But here we do not introduce
the singular densities as inessential for the subsequent analysis.%
}
}{%\mifeng{}{}%\noindent%
Отметим, что действие (\ref{349227101}), (\ref{37974058}) и уравнение (\ref{408139091})
справедливы для поля вне сингулярностей. Наличие сингулярностей поля требует добавления
соответствующих сингулярных плотностей к действию (\ref{349227101}), (\ref{37974058}) и в правую
часть уравнения (\ref{408139091}). Для случая нелинейной электродинамики этот аспект рассматривался
в работе \cite{Chernitskii1999} и подробно обсуждается в монографии \cite{Chernitskii2012be}.
Здесь однако мы не вводим сингулярные плотности как несущественные для последующего анализа.
}

\engrus{1ex}{1ex}{%
We have the following  evident relations from (\ref{35751937}):
}{%
Из (\ref{35751937}) имеем следующие очевидные соотношения:
}
\begin{equation}
\label{300145771}
 \frac{\p \dffun_{\mu}}{\p x^{\nu}} - \frac{\p \dffun_{\nu}}{\p x^{\mu}} = 0
\;.
\end{equation}

\begin{subequations}\label{450528581}
\engrus{1ex}{1ex}{%
Inversion for relations (\ref{421005381}) gives
}{%
Инверсия соотношений (\ref{421005381}) даёт
}
\begin{equation}
\label{450419781}
 \dffun^{\mu} =  \frac{\ffind^{\mu}}{\LFo}
\;,
\end{equation}
\engrus{1ex}{1ex}{%
\noindent
where \textaddb{we put}
}{%
\noindent
где мы полагаем
}
\begin{equation}
\label{450716341}
\LFo  \eqdef
\sqrt{\left|1 - \xxx^{2}\,\metr_{\mu\nu}\,\ffind^{\mu}\,\ffind^{\nu} \right|}
\;.
\end{equation}
\end{subequations}

\engrus{0.5ex}{0.5ex}{%
\textaddb{%
It must be noted here that the presence of modulus in the both expressions (\ref{383322881}) and (\ref{450716341})
stipulates the following feature of the model under consideration.
The signature of space-time metric tensor can be changed on singular sets of solutions.
We shall consider this feature in more detail below.
}
}{%
Здесь надо отметить, что наличие модуля в обоих выражениях (\ref{383322881}) и (\ref{450716341}) обуславливает
следующую особенность рассматриваемой модели. На сингулярных множествах решений сигнатура пространственно-временной метрики может меняться.
Мы рассмотрим эту особенность более подробно ниже.
}

\engrus{1ex}{1ex}{%
Let us write also the following useful relation, which is obtained from (\ref{383322881}) and (\ref{450528581}):
}{%
Запишем также следующее полезное соотношение, получающееся из\mifeng{}{\break} (\ref{383322881}) и (\ref{450528581}):
}
\begin{equation}
\label{683761031}
\LF\,\LFo  = 1
\;.
\end{equation}

\engrus{1ex}{1ex}{%
For the case when the field invariant $\dffun^{\rho}\,\dffun_{\rho}$ \textadd1{is} relatively small ($\xxx^{2}\,|\dffun^{\rho}\,\dffun_{\rho}|\ll 1$) we can represent the action density $\LF$
\textremd{with}\textaddd{by} two first terms \textremd{in}\textaddd{of the} formal power series
\textremd{of}\textaddd{in} $\xxx^2$:
}{%
Для случая, когда полевой инвариант $\dffun^{\rho}\,\dffun_{\rho}$  достаточно мал\mifeng{}{\break} ($\xxx^{2}\,|\dffun^{\rho}\,\dffun_{\rho}|\ll 1$) мы можем представить плотность действия $\LF$
двумя первыми членами формального степенного ряда по $\xxx$:
}
\begin{equation}
\label{429878711}
\LF  = 1 + \frac{\xxx^{2}}{2}\,\metr^{\mu\nu}\,\frac{\p \ffun}{\p x^{\mu}}\,\frac{\p \ffun}{\p x^{\nu}}
+ \OOO{\xxx^4}{\xxx\to 0}
\;.
\end{equation}
\engrus{1ex}{1ex}{%
\noindent
The appropriate linearized equation has the form
}{%
\noindent
Соответствующее линеаризованное уравнение имеет вид
}
\begin{equation}
\label{432760781}
\frac{1}{\sqrt{|\metr|}}\, \frac{\p}{\p x^{\nu}}\left(\sqrt{|\metr|}\,\metr^{\mu\nu}\,\frac{\p \ffun}{\p x^{\mu}}\right)
+ \OOO{\xxx^2}{\xxx\to 0} = 0
\;.
\end{equation}
\engrus{0.5ex}{0.5ex}{%
\noindent
Also let us write the linearized relation \textadd1{for} (\ref{421005381}) \textadd1{as}:
}{%
\noindent
Также запишем линеаризованные соотношения для (\ref{421005381}) как
}
\begin{equation}
\label{418354551}
 \ffind^{\mu}  = \dffun^{\mu}
  + \OOO{\xxx^2}{\xxx\to 0}
\;.
\end{equation}

\engrus{0.5ex}{0.5ex}{%
\textadd1{The} nonlinear differential equation of second order (\ref{408139091}) for \textadd1{the} function $\ffun$
can be represented in the form of
the first order differential equation system
for four-vectors $\dffun_{\mu}$ or $\ffind^{\mu}$.
In this case we have differential field equations (\ref{407970191}) and (\ref{300145771})
\textaddd{accordingly}.
In addition \textadd1{to this} we must consider algebraical relations (\ref{421005381}) or (\ref{450528581}).
}{%
Нелинейное дифференциальное уравнение второго порядка (\ref{408139091}) для функции $\ffun$
может быть представлено в виде системы дифференциальных уравнений первого порядка для четырёхвекторов
$\dffun_{\mu}$ или $\ffind^{\mu}$.
В этом случае мы имеем дифференциальные уравнения (\ref{407970191}) и (\ref{300145771})
соответственно.
В дополнение к ним мы должны использовать алгебраические соотношения (\ref{421005381}) или (\ref{450528581}).
}

\begin{subequations}\label{43842964}
\engrus{0.5ex}{0.5ex}{%
As \textadd1{it} can be seen, the model action (\ref{37974058}) is susceptible to the choice of
\textadd1{a} metric signature.
Here we will consider both the signatures $\{+,-,-,-\}$ and $\{-,+,+,+\}$.
Thus we use the following designations for Minkowski metric:
}{%
Как можно видеть, модельное действие чувствительно к изменению сигнатуры метрики. Здесь мы будем рассматривать
обе сигнатуры $\{+,-,-,-\}$ и $\{-,+,+,+\}$. Таким образом мы используем следующие обозначения для метрики Минковского:                                                                                                                      }
\begin{align}
\label{43842964a}
&\metrp^{00} = 1
\;,
\quad
\metrp^{0i} = 0
\;,
\quad
\metrp^{ij}=-\Cron^{ij}
\\
\label{43842964b}
&\metrp^{00} = -1
\;,
\quad
\metrp^{0i} = 0
\;,
\quad
\metrp^{ij}=\Cron^{ij}
\end{align}
\engrus{0.5ex}{0.5ex}{%
\noindent
where $\Cron^{ij}$ is Kronecker symbol. The Latin indices take values $\{1,2,3\}$.
}{%
\noindent
где $\Cron^{ij}$ -- символ Кронекера. Латинские индексы принимают значения $\{1,2,3\}$.
}
\end{subequations}

\engrus{0.5ex}{0.5ex}{%
\textaddb{%
The choice between the two metric signatures (\ref{43842964}) can be made if a comparison of theoretical results and experiments
provides this opportunity.
We can be also guided by \textremd{a}\textaddd{some} formal \textaddd{mathematical} reasonings. \textremd{such that the requirement of continuity of the field function.}
}
}{%
Выбор одной из двух сигнатур метрики (\ref{43842964}) может быть сделан, если сопоставление теоретических результатов с экспериментами
даёт эту возможность. Мы можем также руководствоваться некоторыми формальным математическими соображениями.
}

\engrus{0.5ex}{0.5ex}{%
As mentioned above, we assume that the signature of metric can be changed on singular sets
of solutions for the model under consideration.
As will be shown (see sec. \ref{genlls} below), this assumption allows to obtain more physically acceptable
solutions in the case of the presence of singularities.
}{%
Как отмечалось выше, мы допускаем, что сигнатура метрики может меняться на сингулярных множествах
решений рассматриваемой модели.
Как будет показано (см. далее секцию \ref{genlls}), это допущение позволяет в случае наличия сингулярностей получить более физически приемлемые решения.
}

\begin{subequations}\label{670504501}
\engrus{0.5ex}{0.5ex}{%
The signature of \textadd1{the} metric (\ref{43842964a}) allows the same spherically
symmetric solution of the model that was obtained for \textadd1{the} nonlinear electrodynamics
by M.~Born and L.~Infeld in their classical work \textaddd{\cite{BornInfeld1934a}}:
}{%
Сигнатура метрики (\ref{43842964a}) обеспечивает то же сферически-симметричное решение модели, которое было получено
для нелинейной электродинамики М.~Борном и Л.~Инфельдом в их классической работе
\cite{BornInfeld1934a}.
}
\begin{equation}
\label{670504501a}
 \ffind_r = \frac{\q}{r^{2}}
\;,\quad
\frac{\p \ffun}{\p r}  = \frac{\q}{\sqrt{r^{4}+\br^{4}}}
\;,
\end{equation}
\engrus{0.5ex}{0.5ex}{%
\noindent
where $\q$ is \textaddb{a} constant, $r$ is \textaddb{the} radial spherical coordinate, \mbox{$\br  \eqdef \sqrt{|\q\,\xxx|}$}.
}{%
\noindent
где $\q$ -- константа, $r$ -- радиальная сферическая координата, \mbox{$\br  \eqdef \sqrt{|\q\,\xxx|}$}.
}

\engrus{0.5ex}{0.5ex}{%
It is evident that \textadd1{the} solution (\ref{670504501}) give birth to the class of soliton solutions \textremd{with}\textaddd{through}
Lorentz transformations.
Such solutions in this model also can be considered as point charged particles
because their long-range \textrem1{interactions have electromagnetic character }%
\textadd1{interaction has features of electromagnetic one in particular}.
}{%
Очевидно, что решение (\ref{670504501}) порождает класс солитонных решений посредством преобразований
Лоренца.
Подобные решения в этой модели также могут рассматриваться как точечные заряженные частицы, поскольку их дальнее взаимодействие, в частности,
имеет характерные черты электромагнитного.
}

\engrus{0.5ex}{0.5ex}{%
Indeed to investigate the \textaddb{soliton} interactions we can use the method based on
\textaddb{the}
integral conservation law of momentum (for Born -- Infeld nonlinear electrodynamics see \cite{Chernitskii1999,Chernitskii2012be}). Let us
consider the long-range interaction
\textremb{of an appropriate to (\ref{670504501}) moving }\textaddb{for the rest}
soliton-particles
\textremb{with the rest one }\textaddb{of the type}
 (\ref{670504501}).
In this case the method gives \textaddb{the} pure \textaddb{electrostatic}
\textremb{electrical }interaction
between the particles. Then we can transform the obtained
law of
\textremb{particle movement with electrical force }\textaddb{motion for a particle}
to
\textremb{another }\textaddb{a} moving reference frame.
In this case Lorentz transform of the
\textaddb{obtained electrical}
force gives its magnetic component.
In this connection it should be noted that the Lorentz transform of the force was presented by A. Einstein in last section of his classical work on special relativity \cite{Einstein1905aE}.
}{%
Действительно, чтобы исследовать солитонные взаимодействия мы можем использовать метод, основанный на интегральном законе сохранения импульса
(для нелинейной электродинамики Борна -- Инфельда см.
\cite{Chernitskii1999,Chernitskii2012be}).
Рассмотрим дальнее взаимодействие покоящихся солитонов-частиц вида\mifeng{}{\break} (\ref{670504501}).
 В этом случае метод даёт чисто электростатическое взаимодействие между частицами.
Затем мы можем преобразовать полученный закон движения для частицы
к движущейся системе отсчёта.
В этом случае преобразование Лоренца для полученной электрической силы даёт её магнитную компоненту.
В этой связи надо отметить, что преобразование Лоренца для силы было представлено А.~Эйнштейном в последней секции его классической работы по частной относительности  \cite{Einstein1905aE}.
}

\engrus{0.5ex}{0.5ex}{%
It should be mentioned  that using the another metric signature  (\ref{43842964b}) for action (\ref{37974058})
leads to the following
spherically symmetric solution instead of (\ref{670504501a}):
}{%
Надо заметить, что использование другой сигнатуры метрики (\ref{43842964b}) для действия (\ref{37974058})
приводит к следующему сферически симметричному решению вместо (\ref{670504501a}):
}
\begin{equation}
\label{670504501b}
 \ffind_r = \frac{\q}{r^{2}}
\;,\quad
\frac{\p \ffun}{\p r}  = \frac{\q}{\sqrt{\left|r^{4}-\br^{4}\right|}}
\;.
\end{equation}
\engrus{0.5ex}{0.5ex}{%
%\textrem1{Here we must consider the area $r \geqslant \br$.}
In this case we have infinity values for $\ffun_{r}$ and $\LF$ on the sphere $r = \br$.
But, as it can be shown, the field function $\ffun$
\textaddb{of the solution}
is finite on this sphere.
}{%
В этом случае мы имеем бесконечные значения для $\ffun_{r}$ и $\LF$ на сфере $r = \br$.
Однако, как можно показать, полевая функция $\ffun$ решения конечна на этой сфере.
}
\end{subequations}

\engrus{0.5ex}{0.5ex}{%
\textaddb{%
More detail investigation of the long-range interaction for the solitons of type (\ref{670504501}) is presented in the work \cite{Chernitskii2017a}.
}
}{%
Более детальное исследование дальнего взаимодействия солитонов вида (\ref{670504501}) представлено в работе \cite{Chernitskii2017a}.
}

\engrus{0.5ex}{0.5ex}{%
Instead relations (\ref{421005381}) and (\ref{450528581}) between four-vectors $\{\ffun_{\mu}\}$ and $\{\ffind^{\nu}\}$
we can consider
relations between quadruples of components $ \{\ffun_{0},\ffind_{1},\ffind_{2},\ffind_{3}\}$
and $ \{\ffind_{0},\ffun_{1},\ffun_{2},\ffun_{3}\}$ in Minkowski metric (\ref{43842964}).
This representation can be preferable for some problems.
}{%
Вместо соотношений (\ref{421005381}) и (\ref{450528581}) между четырёхвекторами $\{\ffun_{\mu}\}$ и $\{\ffind^{\nu}\}$
мы можем рассматривать соотношения между четвёрками компонент
 $ \{\ffun_{0},\ffind_{1},\ffind_{2},\ffind_{3}\}$
и $ \{\ffind_{0},\ffun_{1},\ffun_{2},\ffun_{3}\}$ в метрике Минковского (\ref{43842964}).
Это представление может быть предпочтительным для некоторых задач.
}

\begin{subequations}\label{375035851}
\engrus{0.5ex}{0.5ex}{%
The appropriate solution of equations (\ref{421005381}) gives the following relations:
}{%
Соответствующее решение уравнений (\ref{421005381}) даёт следующие соотношения:
}
\begin{align}
\label{375035851a}
\ffind_{0} =
\sqrt{\left|\frac{1 \pm \xxx^{2}\,\ffind_{j}\,\ffind_{j}}{1 \pm \xxx^{2}\,\ffun_{0}^{2}}\right|}\;\ffun_{0}
\;,\quad
\ffun_{i} = \sqrt{\left|\frac{1 \pm \xxx^{2}\,\ffun_{0}^{2}}{1 \pm \xxx^{2}\,\ffind_{j}\,\ffind_{j}}\right|}\;\ffind_{i}
\;,
\\
\label{375035851b}
\ffun_{0} = \sqrt{\left|\frac{1 \mp \xxx^{2}\,\ffun_{j}\,\ffun_{j}}{1 \mp \xxx^{2}\,\ffind_{0}^{2}}\right|}\;\ffind_{0}
\;,\quad
\ffind_{i} = \sqrt{\left|\frac{1 \mp \xxx^{2}\,\ffind_{0}^{2}}{1 \mp \xxx^{2}\,\ffun_{j}\,\ffun_{j}}\right|}\;\ffun_{i}
\;.
\end{align}
\engrus{0.5ex}{0.5ex}{%
Here and below \textremb{in this section }%
top and bottom signs are appropriate to metrics (\ref{43842964a}) and  (\ref{43842964b}) accordingly,
\textaddb{unless otherwise specified}.
}{%
Здесь и далее верхний и нижний знаки отвечают метрикам (\ref{43842964a}) и (\ref{43842964b})
соответственно, если не указано иначе.
}
\end{subequations}

\engrus{0.5ex}{0.5ex}{%
\textadd1{The}
comparison \textadd1{of the} relations (\ref{375035851}) and (\ref{421005381}) gives the following expressions
for the action density $\LF$ in Cartesian coordinates:
}{%
Сравнение соотношений (\ref{375035851}) и (\ref{421005381}) даёт следующие выражения для плотности
действия $\LF$ в декартовых координатах:
}
\begin{equation}
\label{488528741}
\LF  = \sqrt{\left|\frac{1 \pm \xxx^{2}\,\ffun_{0}^{2}}{1 \pm \xxx^{2}\,\ffind_{j}\,\ffind_{j}}\right|}
= \sqrt{\left|\frac{1 \mp \xxx^{2}\,\ffun_{j}\,\ffun_{j}}{1 \mp \xxx^{2}\,\ffind_{0}^{2}}\right|}
\;.
\end{equation}

\engrus{0.5ex}{0.5ex}{%
Now let us write the field equation (\ref{408139091}) in Cartesian coordinates with \textadd1{the} metric
(\ref{43842964}).
After \textadd1{the} differentiation $\ffind^{\mu}$ (\ref{421005381}) in (\ref{407970191}) and \textadd1{the} multiplication
the equation by $\LF^{3}$ we obtain
}{%
Теперь запишем полевое уравнение (\ref{408139091}) в декартовых координатах с метрикой
(\ref{43842964}).
После дифференцирования $\ffind^{\mu}$ (\ref{421005381}) в (\ref{407970191}) и умножения
 уравнения на $\LF^{3}$ получаем
}
 \begin{equation}
 \label{371394071}
 \left(\metrp^{\mu\nu}\,\left(1 + \xxx^{2}\,\metr^{\alpha\beta}\,
 \ffun_{\alpha}\,\ffun_{\beta}
 %\frac{\p \ffun}{\p x^{\alpha}}\,\frac{\p \ffun}{\p x^{\beta}}
 \right)
 - \xxx^{2}\,\dffun^{\mu}\,\dffun^{\nu}\right)
 \frac{\p^{2}\,\ffun}{\p x^{\mu}\,\p x^{\nu}}  = 0
 \;.
 \end{equation}
\engrus{0.5ex}{0.5ex}{%
As we see, \textadd1{the} obtained equation does not include radicals.
\textaddb{It is significant that the general form of this homogeneous equation does not depend on a sign of the expression \mbox{$1 + \xxx^{2}\,\metr^{\alpha\beta}\,\ffun_\alpha\,\ffun_\beta$}
in the action density (\ref{383322881}).}
}{%
Как мы видим, полученное решение не содержит радикалов. Важно отметить, что общий вид этого однородного уравнения не зависит от знака выражения \mbox{$1 + \xxx^{2}\,\metr^{\alpha\beta}\,\ffun_\alpha\,\ffun_\beta$}
 в плотности действия (\ref{383322881}).
}

\engrus{0.5ex}{0.5ex}{%
It is evident that the model under consideration keeps \textadd1{the} invariance for
\textadd1{the} space-time rotation and \textadd1{the} scale transformation. Thus any solution gives birth to the
appropriate class of solutions with the following transform:
}{%
Очевидно, что рассматриваемая модель инвариантна относительно про\-странственно-вре\-мен\-ных поворотов
и масштабного преобразования. Таким образом любое решение порождает соответствующий класс решений
посредством следующего преобразования:
}
 \begin{equation}
 \label{485617551}
\ffun (\{x^{\mu}\})\;   \to\; a\,\ffun (\{L^{\mu}_{.\nu}\,x^{\nu}/a\})
\;,
 \end{equation}
\engrus{0.5ex}{0.5ex}{%
\noindent
where $L^{\mu}_{.\nu}$ are \textadd1{the} components of \textadd1{the} space-time rotation matrix,
$a$ is \textadd1{a} scale parameter.
}{%
\noindent
где $L^{\mu}_{.\nu}$ -- компоненты матрицы пространственно-временного поворота,
$a$ -- масштабный параметр.
}

\engrus{3ex}{2ex}{%
\section{Energy-momentum and angular momentum}
}{%
\mifeng{\addtocounter{section}{-1}}{}
\section{Энергия-импульс и момент импульса}
\mifeng{}{\label{enmomam}}
}

\engrus{0.5ex}{0.5ex}{%
\textadd1{The} customary method gives the following \textadd1{expression for} canonical energy-momentum density tensor of the model  \textaddd{outside of singularities} in Cartesian coordinates
}{%
Обычный метод даёт следующее выражение для канонического тензора энергии-импульса модели вне сингулярностей в декартовых прямоугольных координатах:
}
\begin{equation}
\label{442613621}
\EMTc^{\mu\nu}   =
%\frac{\varkappa}{4\pi\,\xxx^2\,\LF}
\frac{1}{4\pi\,\xxx^2\,\LF}
\biggl(\xxx^2\,\dffun^{\mu}\,\dffun^{\nu} - \metrp^{\mu\nu}\,\Bigr(1 + \xxx^2\,\metrp^{\alpha\beta}\,\dffun_{\alpha}\,\dffun_{\beta}\Bigl)\biggr)
\;.
\end{equation}

\engrus{0.5ex}{0.5ex}{%\mifrus{}{}%\noindent%
\textaddd{%
The possible field singularities can require a modification of the energy-mo\-men\-tum tensor (\ref{442613621}) or the
appropriate differential conservation law by including singular densities.
But there are solutions with singularities for which this modification is inessential because
the appropriate integral conservation law is unchanged. Here we consider such solutions only.%
}
}{%\mifeng{}{}%\noindent%
Возможные полевые сингулярности могут требовать модификации тензора энергии-им\-пуль\-са (\ref{442613621}) или соответствующего дифференциального закона сохранения включением сингулярных плотностей.
Однако существуют решения с сингулярностями, для которых эта модификация не существенна поскольку
не изменяет соответствующий интегральный закон сохранения. Здесь мы рассматриваем только такие решения.
}

\engrus{0.5ex}{0.5ex}{%
Notice \textaddd{also} that the canonical energy-momentum density tensor is defined here uniformly for the both cases of the representation for the expression under modulus
\mbox{$\pm(1 + \xxx^2\,\metrp^{\alpha\beta}\,\dffun_{\alpha}\,\dffun_{\beta})$}
 in the action density (\ref{383322881}). This
 \textaddd{can}
 provide the positive definiteness of the energy density
 \textremd{ for the certain cases when a sign of}\textaddd{in space-time regions where}
the expression under modulus
\textremd{is changed.}\textaddd{has negative values.}
}{%
Обратим внимание также на то, что канонический тензор плотности энергии-им\-пуль\-са определён здесь единообразно для обоих случаев раскрытия выражения под знаком модуля
\mbox{$\pm(1 + \xxx^2\,\metrp^{\alpha\beta}\,\dffun_{\alpha}\,\dffun_{\beta})$}
в плотности действия (\ref{383322881}).
Это \textaddd{может} обеспечить положительную определённость плотности энергии
\textremd{в определённых случаях, когда}\textaddd{в пространственно-временных областях, где}
 выражение под знаком модуля \textremd{ меняет знак.}\textaddd{имеет отрицательные значения.}
}

\engrus{0.5ex}{0.5ex}{%\mifrus{}{}%\noindent%
\textaddd{Here we use an allowed mathematical arbitrariness in the definition of the energy-momentum tensor for a field model,
which does not violate its differential conservation law.
But, of course, the cases for negative values of the expression under modulus in the action
density (\ref{383322881}) must be considered individually from a viewpoint of their physical relevance.
}
}{%\mifeng{}{}%\noindent%
\textaddd{
Здесь мы используем допустимый математический произвол в определении тензора энер\-гии-импульса полевой модели, не нарушающий дифференциальный закон его сохранения.
Однако, конечно, случаи отрицательных значений выражения под знаком модуля
в плотности действия (\ref{383322881})
следует рассматривать отдельно с точки зрения их физической релевантности.
}
}

\engrus{0.5ex}{0.5ex}{%
As we see, the canonical tensor (\ref{442613621}) is symmetrical.
}{%
Как мы видим, канонический тензор (\ref{442613621}) симметричен.
}

\engrus{0.5ex}{0.5ex}{%
To \textremb{use }\textaddb{obtain the} finite integral characteristics of solutions in infinite space-time we introduce \textaddb{the} regularized energy-momentum density tensor
\textremb{with }\textaddb{in} the following \textremb{formula: }\textaddb{way:}
}{%
Чтобы получить конечные интегральные характеристики решений в бесконечном пространстве мы вводим регуляризованный тензор энергии-им\-пуль\-са следующим способом:
}
\begin{equation}
 \label{809745461}
\EMT^{\mu\nu}   = \EMTc^{\mu\nu} - \EMTi^{\mu\nu}
 \;,
 \end{equation}
\engrus{0ex}{0.5ex}{%
\noindent
where $\EMTi^{\mu\nu}$ is \textadd1{a} regularizing symmetrical energy-momentum density tensor which can
be defined depending on \textadd1{the} class of solutions under consideration. Here we will use
\textaddb{the} constant regularizing tensor
}{%
\noindent
где $\EMTi^{\mu\nu}$ -- регуляризующий симметричный тензор энергии-импульса, который может быть определён в зависимости от класса рассматриваемых решений.
Здесь мы будем использовать постоянный регуляризующий тензор
}
\begin{equation}
\label{431999421}
 \EMTi^{\mu\nu} = -\frac{1}{4\pi\,\xxx^2}\,\metrp^{\mu\nu}
\;.
\end{equation}

\engrus{0.5ex}{0.5ex}{%
We have \textadd1{the} conservation law for \textaddb{the} regularized energy-momentum density tensor in Cartesian coordinates
}{%
Имеем закон сохранения для регуляризованного тензора плотности энергии-импульса
}
\begin{equation}
 \label{80647273}
\frac{\p \EMT^{\mu\nu}}{\p x^{\nu}}   = 0
 \;.
 \end{equation}

\engrus{0.5ex}{0.5ex}{%
Let us define \textadd1{the} angular momentum density tensor by customary way.
We have the following appropriate conservation law:
}{%
Определим тензор плотности момента энергии-импульса обычным образом. Имеем следующий соответствующий закон сохранения:
}
\begin{equation}
\frac{\partial
\AMT^{\mu\nu\rho}}{\partial x^\rho} = 0\;,
\label{ConsLaw:An}
\end{equation}
\engrus{0.5ex}{0.5ex}{%
\noindent
where
}{%
\noindent
где
}
\begin{equation}
\label{69808677}
\AMT^{\mu\nu\rho} \doteqdot x^{\mu}\,\EMT^{\nu\rho} - x^{\nu}\,\EMT^{\mu\rho}
\;.
\end{equation}

\begin{subequations}\label{481319051}
\engrus{0.5ex}{0.5ex}{%
We introduce the following special designations for energy, momentum vector, and
angular momentum vector densities: $\cE$, $\bcP$, $\bcM$. Let us write
the appropriate expressions taking into account relations
(\ref{383322881}), (\ref{421005381}),
and (\ref{488528741}):
}{%
Вводим следующие специальные обозначения для плотностей энергии, вектора импульса и вектора момента импульса: $\cE$, $\bcP$, $\bcM$.
Запишем соответствующие выражения, учитывая соотношения (\ref{383322881}), (\ref{421005381})
и (\ref{488528741}):
}
\begin{align}
\cE &\eqdef \EMT^{00}
\label{46450798}
=\frac{1}{4\pi\,\xxx^2\,\LF}\,\Bigl(\xxx^{2}\,\bigl(\dffun_{i}\,\dffun_{i}\bigr) + \metrp^{00}\,\bigl(\LF - 1\bigr)\Bigr)
\;,\\
\label{481422932}
\cP^{i} &\eqdef
\EMT^{0i} =
\EMT^{i0}
=
\frac{1}{4\pi}\,\frac{\dffun^{0}\,\dffun^{i}}{\LF}
=
\frac{1}{4\pi}\,\dffun^{0}\,\ffind^{i}
=
\frac{1}{4\pi}\,\dffun^{i}\,\ffind^{0}
\;,\\
\label{48502203}
\cM_{i} &\eqdef \LCh_{ijk}\,x^{j}\,\cP^{k}
\;,
\end{align}
\engrus{0.5ex}{0.5ex}{%
\noindent
where $\LCh_{ijk}$ is Levi-Civita symbol ($\LCh_{123} = 1$).
}{%
\noindent
где $\LCh_{ijk}$ -- символ Леви-Чивита ($\LCh_{123} = 1$).
}
\end{subequations}

\engrus{0.5ex}{0.5ex}{%
Let us define energy, momentum, and angular momentum of \textadd1{the} field in a three-dimensional volume $\Vols$:
}{%
Определим энергию, импульс и момент импульса поля в трёхмерном объёме $\Vols$:
}
\begin{equation}
\label{576969441}
 \Energy_{\Vols} \doteqdot
\int\limits_{\Vols}\cE\,\dVols
\;,\quad
\bEMV_{\Vols}\doteqdot \int\limits_{\Vols}\bcP\,\dVols
\;,\quad
\bAMV_{\Vols} \doteqdot \int\limits_{\Vols}\bcM\,\dVols
\;.
\end{equation}

\mifengrus{\newpage}{}

\engrus{3ex}{2ex}{%
\section{General lightlike soliton}
\label{genlls}
}{%
\mifeng{\addtocounter{section}{-1}}{}
\section{Общий светоподобный солитон}
\mifeng{}{\label{genlls}}
}

 \begin{subequations}\label{426265581}
\engrus{0.5ex}{0.5ex}{%
Let us consider \textaddb{the} solution in a form of wave propagating along $x^{3}$ axis of Cartesian coordinate system with the speed of light.
Let this solution be have some transverse and longitudinal field distributions. Thus we can write
}{%
Рассмотрим решение в виде волны, распространяющейся вдоль оси $x^{3}$ декартовой системы координат со скоростью света.
Пусть это решение имеет некоторое поперечное и продольное распределение поля. Таким образом можем записать
}
\begin{align}
 \label{426404171}
 \ffun &=\ffun \left(\Phase,x^{1},x^{2}\right)
 \;,\\
 \label{426404172}
 \Phase &= \omega\,x^{0} - k_3\,x^{3}
 \;,\quad
 k_3^2 = \omega^2
 \;,\quad
 \omega > 0
 \;.
 \end{align}
  \end{subequations}

\engrus{0.5ex}{0.5ex}{%
Substitution (\ref{426265581}) to field equation (\ref{371394071}) gives the following equation:
}{%
Подстановка (\ref{426265581}) в полевое уравнение (\ref{371394071}) даёт следующее уравнение:
}
%\begin{multline}
\begin{equation}
\label{476118041}
\Biggl(
1 \mp \xxx^2\biggl(\frac{\p\ffun}{\p x^{2}}\biggr)^{\!\!2}
\Biggr)
\frac{\p^2\ffun}{(\p x^{1})^2}
\pm 2\,\xxx^2\,\frac{\p\ffun}{\p x^{1}}\,\frac{\p\ffun}{\p x^{2}}\,\frac{\p^2\ffun}{\p x^{1}\p x^{2}}
%\\
+
\Biggl(
1 \mp \xxx^2\biggl(\frac{\p\ffun}{\p x^{1}}\biggr)^{\!\!2}
\Biggr)
\frac{\p^2\ffun}{(\p x^{2})^2}
 = 0
\;,
%\end{multline}
\end{equation}
\engrus{0.5ex}{0.5ex}{%
\noindent
where top and bottom signs are appropriate to Minkowski metrics (\ref{43842964a}) and (\ref{43842964b}) accordingly.
}{%
\noindent
где верхний и нижний знаки отвечают метрикам Минковского (\ref{43842964a}) и (\ref{43842964b}) соответственно.
}

\engrus{0.5ex}{0.5ex}{%
As we see the obtained equation (\ref{476118041}) does not include derivatives on phase of wave $\Phase$ (\ref{426404172}).
}{%
Как мы видим, полученное уравнение (\ref{476118041}) не содержит производных по фазе волны $\Phase$ (\ref{426404172}).
}

\engrus{0.5ex}{0.5ex}{%
\textadd1{The} equation (\ref{476118041}) is elliptical with the following condition:
}{%
Уравнение (\ref{476118041}) -- эллиптическое при следующем условии:
}
\begin{equation}
\label{765372831}
1 \mp \xxx^2\left(\ffun_{1}^2 + \ffun_{2}^2 \right)  > 0
\;.
\end{equation}

\engrus{0.5ex}{0.5ex}{%
The similar in form (\ref{476118041}) equations were considered. About this topic see the paper by R.~Ferraro \cite{FerraroRafael1304p5506}
and references therein.
}{%
Сходные по виду (\ref{476118041}) уравнения рассматривались. Относительно этой темы см. статью Р.~Ферраро \cite{FerraroRafael1304p5506}
и содержащуюся в ней библиографию.
}

\engrus{0.5ex}{0.5ex}{%
In particular, the similar in form but different in type equation was considered by B.M.~Barbashov and N.A.~Chernikov \cite{BarbChern1966-1e}.
A Lax representation (see, for example, \cite{Whitham1974}) for this equation was presented
in \textadd1{an} article by J.C.~Brunelli and A.~Das
\cite{Brunelli1998a}.
}{%
В частности, сходное по виду, но отличающееся по типу уравнение рассматривалось\break
Б.М.~Барбашовым и Н.А.~Черниковым \cite{BarbChern1966-1e}.
Представление Лакса (см., например, \cite{Whitham1974}) для этого уравнения было представлено в статье J.C.~Brunelli и A.~Das
\cite{Brunelli1998a}.
}

\engrus{0.5ex}{0.5ex}{%
The monograph by \mbox{G.B.~Whitham} \cite{Whitham1974} contains relatively simple way for obtaining the Barbashov -- Chernikov solution
with the help of hodograph transformation (see, for example, \cite{KurantGilbertI1989e2}).
}{%
Монография Уизема \cite{Whitham1974} содержит относительно простой путь получения решения Барбашова -- Черникова при помощи преобразования годографа
 (см., например, \cite{KurantGilbertI1989e2}).
}

\engrus{0.5ex}{0.5ex}{%
Here we use in outline the Whitham method but for the elliptic (for condition (\ref{765372831})) equation (\ref{476118041}).
The qualitative difference between hyperbolic and elliptic equations causes the appropriate difference in the solution way.
}{%
Здесь мы используем в общих чертах метод Уизема, но для эллиптического (при условии (\ref{765372831})) уравнения (\ref{476118041}).
Качественное различие между гиперболическим и эллиптическим уравнениями вызывает соответствующее отличие в методе решения.
}

\engrus{0.5ex}{0.5ex}{%
Let us introduce new independent variables
}{%
Введём новые независимые переменные
}
\begin{equation}
\label{430348001}
  \xi = x^{1} + \him\,x^{2}
 \;,\quad
 \hconj{\xi} = x^{1} - \him\,x^{2}
\;,
\end{equation}
\engrus{0.5ex}{0.5ex}{%
\noindent
where $\him^{2} = -1$.
}{%
\noindent
где $\him^{2} = -1$.
}

\begin{subequations}\label{552514381}
\engrus{0.5ex}{0.5ex}{%
Also we will use cylindrical coordinates $\{\rho,\varphi,x^{3}\}$. We have the following evident relations:
}{%
Также мы будем использовать цилиндрические координаты $\{\rho,\varphi,x^{3}\}$. Имеем следующие очевидные соотношения:
}
\begin{alignat}{2}
\label{552706741}
\xi &= \rho\,\e{\him\,\varphi}\;,\quad   &
\hconj{\xi} &= \rho\,\e{-\him\,\varphi}
\;,\\
\label{552706742}
\rho &= \sqrt{\xi\,\hconj{\xi}}\;,\quad   &
\varphi &=
-\him\,\ln\!\left(\xi\!\left/\!\sqrt{\xi\,\hconj{\xi}}\right.\right)
\,.
\end{alignat}
\end{subequations}

\engrus{0.5ex}{0.5ex}{%
Using new variables (\ref{430348001}) we obtain from (\ref{476118041}) the following equation:
}{%
Используя новые переменные (\ref{430348001}), мы получаем из (\ref{476118041}) следующее уравнение:
}
%\begin{multline}
\begin{equation}
\label{436283681}
\left(
1 \mp 2\,\xxx^2\,\frac{\p \ffun}{\p \xi}\,\frac{\p \ffun}{\p \hconj{\xi}}
\right)
\frac{\p^2 \ffun}{\p\xi\,\p\hconj{\xi}}
%\\
\pm \xxx^2\left(\frac{\p \ffun}{\p \xi}\right)^{\!2}
\frac{\p^2 \ffun}{(\p\hconj{\xi})^{2}}
\pm \xxx^2\left(\frac{\p \ffun}{\p \hconj{\xi}}\right)^{\!2}
\frac{\p^2 \ffun}{(\p\xi)^2}
  = 0
\;.
%\end{multline}
\end{equation}
\engrus{0.5ex}{0.5ex}{%
Equation (\ref{436283681}) is hyperbolic with the following condition:
}{%
Уравнение (\ref{436283681}) является гиперболическим при следующем условии:
}
\begin{equation}
\label{78423193}
1 \mp 4\,\xxx^2\left(\ffun_{\xi}^2 + \ffun_{\hconj{\xi}}^2 \right)  > 0
\;.
\end{equation}

\begin{subequations}\label{592114861}
\engrus{0.5ex}{0.5ex}{%
As \textadd1{it was} noted in section \ref{estf} the field model under consideration is invariant by space-time rotation\textaddb{s} and scale transformation\textaddb{s}.
But \textadd1{the} equation (\ref{476118041}) does not contain derivatives with respect to coordinates $\{x^{0},x^{3}\}$.
Because this here we have space-time rotation and scale invariance in the planes $\{x^{1},x^{2}\}$ and $\{x^{0},x^{3}\}$
with mutually independent parameters.
Thus equation
(\ref{476118041}) is invariant with respect to rotation about $x^{3}$ axis and scale transformation
in $\{x^{1},x^{2}\}$ plane.
}{%
Как было показано в секции \ref{estf}, рассматриваемая полевая модель инвариантна относительно пространственно-временных поворотов и масштабных преобразований.
Однако уравнение (\ref{476118041}) не содержит производных по координатам $\{x^{0},x^{3}\}$.
Поэтому здесь мы имеем пространственно-временную поворотную и масштабную инвариантность в плоскостях $\{x^{1},x^{2}\}$ и $\{x^{0},x^{3}\}$
с взаимно независимыми параметрами.
Таким образом уравнение
(\ref{476118041}) инвариантно относительно вращения вокруг оси $x^{3}$ и масштабного преобразования в плоскости
$\{x^{1},x^{2}\}$.
}

\engrus{0.5ex}{0.5ex}{%
%\textremb{As \textadd1{it is} applied to \textadd1{the} equation (\ref{436283681}),}
Taking into account relations (\ref{552514381}) and (\ref{485617551}), these
two types of invariance \textaddb{applied to equation (\ref{436283681})} are provided by the following general substitution:
}{%
Учитывая соотношения (\ref{552514381}) и (\ref{485617551}), эти два типа инвариантности применительно к уравнению
(\ref{436283681}) обеспечиваются следующей общей подстановкой:
}
\begin{equation}
 \label{495228201}
\ffun (\xi,\hconj{\xi})\;  \to\; \sqrt{\bta\,\hconj{\bta}}\;\ffun \left(\xi/\bta,\hconj{\xi}/\hconj{\bta}\right)
 \;,
 \end{equation}
\engrus{0.5ex}{0.5ex}{%
\noindent
where $\bta$ is arbitrary complex constant with respect to coordinates $\{\xi,\hconj{\xi}\}$,
$\hconj{\bta}$ is complex conjugate to $\bta$ quantity. The constant $\bta$ will be called the scale-rotation parameter of solution in the plane $\{x^{1},x^{2}\}$.
}{%
\noindent
где $\bta$ -- произвольная комплексная константа по отношению к координатам $\{\xi,\hconj{\xi}\}$,
$\hconj{\bta}$ -- ком\-плекс\-но-сопряжённая к $\bta$ величина. Константа $\bta$ будет называться масштабно-поворотным параметром решения в плоскости $\{x^{1},x^{2}\}$.
}

\engrus{0.5ex}{0.5ex}{%
The complex constant $\bta$ can be written in the form
}{%
Комплексная константа $\bta$ может быть записана в виде
}
\begin{equation}
\label{586498021}
\bta  = \btam\,\e{\him\,\btap}
\;,
\end{equation}
\engrus{0.5ex}{0.5ex}{%
\noindent
where $\btam$ and $\btap$ are real constants with respect to coordinates $\{x^{1},x^{2}\}$.
But in general case these constants can be depend on phase of soliton $\Phase$ (\ref{426404172}):
}{%
\noindent
где $\btam$ и $\btap$ -- действительные константы по отношению к координатам $\{x^{1},x^{2}\}$.
Однако в общем случае эти константы  могут зависеть от фазы солитона $\Phase$ (\ref{426404172}):
}\begin{equation}
\label{591083061}
\btam=\btam(\Phase)\;,\quad
\btap=\btap(\Phase)
\;.
\end{equation}
\engrus{0.5ex}{0.5ex}{%
Because this we will call $\bta(\Phase)$ the scale-rotation function of the soliton.
It is evident that the function $\btam(\Phase)$ defines the phase dependence of transversal scale and the  function $\btap(\Phase)$
defines the phase dependence of rotation about $x^{3}$ axis.
}{%
Поэтому мы будем называть $\bta(\Phase)$ масштабно-поворотной функцией решения.
Очевидно, что функция $\btam(\Phase)$ определяет фазовую зависимость поперечного масштаба и функция $\btap(\Phase)$
определяет фазовую зависимость поворота относительно оси $x^{3}$.
}
\end{subequations}

\engrus{0.5ex}{0.5ex}{%
Thus if we have a solution $\ffun (\xi,\hconj{\xi})$ to equation (\ref{436283681}) then by means of
invariant substitution (\ref{592114861}) we obtain wave propagating along $x^{3}$ axis and
preserving its transversal form. Longitudinal form of the wave defined by scale-rotation phase function $\bta(\Phase)$ is also preserved.
}{%
Таким образом, если мы имеем решение $\ffun (\xi,\hconj{\xi})$ уравнения (\ref{436283681}), то посредством инвариантной подстановки
 (\ref{592114861}) мы получаем волну, распространяющуюся вдоль оси $x^{3}$ и сохраняющую свою поперечную форму.
 Продольная форма волны, определённая масштабно-поворотной фазовой функцией $\bta(\Phase)$ также сохраняется.
}

\engrus{0.5ex}{0.5ex}{%
As \textadd1{a} result, we have in (\ref{592114861}) a wave packet propagating with speed of light and preserving its shape.
It will be called the lightlike soliton.
}{%
В результате мы имеем в (\ref{592114861}) волновой пакет, распространяющийся со скоростью света и сохраняющий свою форму.
Он будет называться светоподобным солитоном.
}

\engrus{0.5ex}{0.5ex}{%
Hereafter up to formulas (\ref{406914401}) we consider the equation (\ref{436283681}) for the case of top signs that is \textaddb{for}  the metric (\ref{43842964a}).
But the obtained solution will be represented in general form which is appropriate to both metrics  (\ref{43842964}).
}{%
Далее до формул (\ref{406914401}) мы рассматриваем уравнение (\ref{436283681}) для случая верхних знаков, то есть для метрики (\ref{43842964a}).
Но полученное решение будет представлено в общем виде, который отвечает обоим метрикам  (\ref{43842964}).
}

\engrus{0.5ex}{0.5ex}{%
Thus \textadd1{the} equation (\ref{436283681}) with top signs is equivalent to the following first order system:
}{%
Таким образом уравнение (\ref{436283681}) с верхними знаками эквивалентно следующей системе первого порядка:
}
\begin{subequations}\label{492564371}
\begin{align}
\label{492624601}
\frac{\p \ffun_{\xi}}{\p \hconj{\xi}} - \frac{\p \ffun_{\hconj{\xi}}}{\p \xi} &= 0
\;,\\
\label{492624602}
\left(1 - 2\,\xxx^2\,\ffun_{\xi}\,\ffun_{\hconj{\xi}}\right)\frac{\p \ffun_{\xi}}{\p\hconj{\xi}}
+\xxx^2\,\ffun_{\xi}^2\,\frac{\p \ffun_{\hconj{\xi}}}{\p\hconj{\xi}}
+\xxx^2\,\ffun_{\hconj{\xi}}^2\,\frac{\p \ffun_{\xi}}{\p\xi}
&= 0
\;,
\end{align}
\end{subequations}
\engrus{0.5ex}{0.5ex}{%
\noindent
where
}{%
\noindent
где
}
\begin{equation}
\label{482705972}
 \ffun_{\xi}  \eqdef \frac{\p \ffun}{\p \xi}
 \;,\quad
 \ffun_{\hconj{\xi}} \eqdef \frac{\p \ffun}{\p \hconj{\xi}}
\;.
\end{equation}

\begin{subequations}\label{554143861}
\engrus{0.5ex}{0.5ex}{%
By interchanging the roles of the dependent and independent variables in (\ref{492564371}) we obtain the linear system
}{%
Изменяя роли зависимых и независимых переменных в (\ref{492564371}), получаем линейную систему
}
\begin{align}
\label{55451527}
\frac{\p \xi}{\p \ffun_{\hconj{\xi}}} - \frac{\p \hconj{\xi}}{\p \ffun_{\xi}} &= 0
\;,\\
\label{55458093}
\left(1 - 2\,\xxx^2\,\ffun_{\xi}\,\ffun_{\hconj{\xi}}\right)\frac{\p \xi}{\p\ffun_{\hconj{\xi}}}
-\xxx^2\,\ffun_{\xi}^2\,\frac{\p \xi}{\p\ffun_{\xi}}
-\xxx^2\,\ffun_{\hconj{\xi}}^2\,\frac{\p \hconj{\xi}}{\p \ffun_{\hconj{\xi}}}
&= 0
\;,
\end{align}
\end{subequations}
\engrus{0.5ex}{0.5ex}{%
\noindent
which is equivalent to \textadd1{the} single equation
}{%
\noindent
которая эквивалентна одному уравнению
}
\begin{multline}
\label{739169291}
\left(1 - 2\,\xxx^2\,\ffun_{\xi}\,\ffun_{\hconj{\xi}}\right)\frac{\p^2 \xi}{\p\ffun_{\xi}\p\ffun_{\hconj{\xi}}}
 - \xxx^2\,\ffun_{\xi}^2\,\frac{\p^2 \xi}{(\p\ffun_{\xi})^2}
  - \xxx^2\,\ffun_{\hconj{\xi}}^2\,\frac{\p^2 \xi}{(\p\ffun_{\hconj{\xi}})^2}
\\
- 2\,\xxx^2\,\ffun_{\xi}\,\frac{\p\xi}{\p\ffun_{\xi}}
- 2\,\xxx^2\,\ffun_{\hconj{\xi}}\,\frac{\p\xi}{\p\ffun_{\hconj{\xi}}}
 = 0
\;.
\end{multline}

\engrus{0.5ex}{0.5ex}{%
Let us introduce new independent variables  \textadd1{in (\ref{739169291})}
}{%
Введём новые независимые переменные в (\ref{739169291})
}
\begin{subequations}\label{743719371}
\begin{align}
\label{743982531}
\eta_1
&=
\phantom{-}
\him\,\frac{\sqrt{1-4\,\xxx^2\,\ffun_{\xi}\,\ffun_{\hconj{\xi}}}-1}{2\,\xxx\,\ffun_{\hconj{\xi}}}
\;,\\
\label{743982532}
\eta_2
&= -\him\,\frac{\sqrt{1-4\,\xxx^2\,\ffun_{\xi}\,\ffun_{\hconj{\xi}}}-1}{2\,\xxx\,\ffun_{\xi}}
\;.
\end{align}
\end{subequations}
\engrus{0.5ex}{0.5ex}{%
\noindent
\textadd1{The} inversion for relations (\ref{743719371}) gives
}{%
\noindent
Инверсия соотношений (\ref{743719371}) даёт
}
\begin{subequations}\label{749508201}
\begin{align}
\label{749535151}
\ffun_{\xi} &=
\phantom{-}
\frac{\him\,\eta_1}{\xxx\left(1 + \eta_1\,\eta_2\right)}
\;,\\
\label{749535152}
\ffun_{\hconj{\xi}} &=
-
\frac{\him\,\eta_2}{\xxx\left(1 + \eta_1\,\eta_2\right)}
\;.
\end{align}
\end{subequations}

\begin{subequations}\label{438309761}
\engrus{0.5ex}{0.5ex}{%
Substituting (\ref{743719371}) and (\ref{749508201}) into (\ref{554143861}), we obtain
}{%
Подставляя (\ref{743719371}) и (\ref{749508201}) в (\ref{554143861}), получаем
}
\begin{align}
\label{438384571}
\eta_1^2\,\frac{\p \xi}{\p\eta_1} + \frac{\p\hconj{\xi}}{\p\eta_1} &= 0
\;,\\
\label{438384572}
\frac{\p \xi}{\p\eta_2} + \eta_2^2\,\frac{\p\hconj{\xi}}{\p\eta_2} &= 0
\;.
\end{align}
\end{subequations}

\engrus{0.5ex}{0.5ex}{%
\textadd1{The} sequential elimination each of the dependent variables $\hconj{\xi}(\eta_1,\eta_2)$ and $\xi(\eta_1,\eta_2)$
from system (\ref{438309761})
gives two simple equations
}{%
Последовательное исключение каждой из зависимых переменных $\hconj{\xi}(\eta_1,\eta_2)$ и $\xi(\eta_1,\eta_2)$
из системы (\ref{438309761}) даёт два простых уравнения
}
\begin{equation}
\label{445015541}
\frac{\p^2\xi}{\p\eta_1\,\p\eta_2}  = 0
\;,\quad
\frac{\p^2\hconj{\xi}}{\p\eta_1\,\p\eta_2}  = 0
\;,
\end{equation}
\engrus{0.5ex}{0.5ex}{%
\noindent
solutions of which have the form
}{%
\noindent
решения которых имеют вид
}
\begin{equation}
\label{446726991}
\xi  = \xi_{1}(\eta_1) + \xi_{3}(\eta_2)
\;,\quad
\hconj{\xi}  =
{\xi}_{2}(\eta_2) + {\xi}_{4}(\eta_1)
\;,
\end{equation}
\engrus{0.5ex}{0.5ex}{%
\noindent
where $\xi_{1}(\eta_1)$, $\xi_{2}(\eta_2)$, ${\xi}_{3}(\eta_2)$, ${\xi}_{4}(\eta_1)$ are arbitrary functions.
}{%
\noindent
где $\xi_{1}(\eta_1)$, $\xi_{2}(\eta_2)$, ${\xi}_{3}(\eta_2)$, ${\xi}_{4}(\eta_1)$ -- произвольные функции.
}

\engrus{0.5ex}{0.5ex}{%
Substitution (\ref{446726991}) to (\ref{438309761}) gives
}{%
Подстановка (\ref{446726991}) в (\ref{438309761}) даёт
}
\begin{equation}
\label{477781131}
\eta^2\,\xi_{1}^{\prime} + {\xi}_{4}^{\prime}  = 0
\;,\quad
\xi_{3}^{\prime} + \eta_2^2\,{\xi}_{2}^{\prime}  = 0
\;.
\end{equation}

\begin{subequations}\label{481079141}
\engrus{0.5ex}{0.5ex}{%
Taking into consideration (\ref{446726991}) and \ref{477781131}), we can write the following relations for general solution of the system (\ref{438309761}):
}{%
Учитывая (\ref{446726991}) и \ref{477781131}), мы можем записать следующие соотношения для общего решения системы (\ref{438309761}):
}
\begin{alignat}{2}
\label{481597181}
\df\xi &= \df\mtxi_{1} -\eta_2^2\,\mtxi_{2}^{\prime}\,\df\eta_2
{\;}&=&{\;} \df\mtxi_{1} -\eta_2^2\,\df\mtxi_{2}
\;,\\
\label{481597182}
\df\hconj{\xi} &= \df\mtxi_{2} -\eta^2\,\mtxi_{1}^{\prime}\,\df\eta_1
{\;}&=&{\;} \df\mtxi_{2} -\eta^2\,\df\mtxi_{1}
\;.
\end{alignat}
\engrus{0.5ex}{0.5ex}{%
Thus the general solution of the system (\ref{438309761}) contains only two arbitrary functions
$\mtxi_{1}(\eta_1)$ and $\mtxi_{2}(\eta_2)$.
}{%
Таким образом общее решение системы (\ref{438309761}) содержит только две произвольные функции
$\mtxi_{1}(\eta_1)$ и $\mtxi_{2}(\eta_2)$.
}
\end{subequations}

\begin{subequations}\label{528644171}
\engrus{0.5ex}{0.5ex}{%
Using (\ref{749508201}) and (\ref{481079141}), we obtain
}{%
Используя (\ref{749508201}) и (\ref{481079141}), получаем
}
\begin{alignat}{2}
\label{528668511}
\frac{\p\ffun}{\p \eta_1}
&=
\ffun_{\xi}\,\frac{\p\xi}{\p\eta_1} + \ffun_{\hconj{\xi}}\,\frac{\p\hconj{\xi}}{\p\eta_1}
&&=
\phantom{-}
\frac{\him}{\xxx}\,\eta_1\,\mtxi^{\prime}_{1}
\;,\\
\label{528668512}
\frac{\p\ffun}{\p \eta_2}
&=
\ffun_{\xi}\,\frac{\p\xi}{\p\eta_2} + \ffun_{\hconj{\xi}}\,\frac{\p\hconj{\xi}}{\p\eta_2}
&&=
-\frac{\him}{\xxx}\,\eta_2\,{\mtxi_{2}}^{\prime}
\;.
\end{alignat}
\end{subequations}

\engrus{0.5ex}{0.5ex}{%
From (\ref{528644171}) we have
}{%
Из (\ref{528644171}) имеем
}
\begin{equation}
\label{770391131}
\df\ffun  = \frac{\him}{\xxx}\left(\eta_1\,{\mtxi_{1}}^{\prime}\,\df\eta_1 -
 \eta_2\,{\mtxi_{2}}^{\prime}\,\df\eta_2 \right)
= \frac{\him}{\xxx}\left(\eta_1\,\df\mtxi_{1} -
 \eta_2\,\df{\mtxi_{2}}\right)
\;.
\end{equation}
\engrus{0.5ex}{0.5ex}{%
\noindent
Here the variables $\eta_1$ and $\eta_2$ in last expression must be considered as inverse functions for
$\mtxi_{1}(\eta_1)$ and ${\mtxi_{2}}(\eta_2)$ that are $\eta_1 = \eta_1 (\mtxi_{1})$ and
$\eta_2 = \eta_2 ({\mtxi_{2}})$.
}{%
\noindent
Здесь переменные $\eta_1$ и $\eta_2$  в последнем выражении должны рассматриваться как обратные функции для
$\mtxi_{1}(\eta_1)$ и ${\mtxi_{2}}(\eta_2)$, то есть $\eta_1 = \eta_1 (\mtxi_{1})$ и
$\eta_2 = \eta_2 ({\mtxi_{2}})$.
}

\engrus{0.5ex}{0.5ex}{%
Let us introduce the designations
}{%
Введём обозначения
}
\begin{equation}
\label{779618711}
 \frac{\him}{\xxx}\,\eta_1 \bigl(\mtxi_{1}\bigr) \eqdef \frac{\df \Xi_{1}}{\df \mtxi_{1}}
 \;,\quad
 -\frac{\him}{\xxx}\,\eta_2 \bigl(\mtxi_{2}\bigr) \eqdef \frac{\df \Xi_{2}}{\df \mtxi_{2}}
\;,
\end{equation}
\engrus{0.5ex}{0.5ex}{%
\noindent
where functions $\Xi_{1} (\mtxi_{1})$ and ${\Xi_{2}}({\mtxi_{2}})$ are arbitrary because of arbitrariness of the functions $\mtxi_{1}(\eta_1)$ and ${\mtxi_{2}}(\eta_2)$.
}{%
\noindent
где функции $\Xi_{1} (\mtxi_{1})$ и ${\Xi_{2}}({\mtxi_{2}})$ -- произвольные, ввиду произвольности функций $\mtxi_{1}(\eta_1)$ и ${\mtxi_{2}}(\eta_2)$.
}

\begin{subequations}\label{406914401}
\engrus{0.5ex}{0.5ex}{%
Then, using (\ref{770391131}) and (\ref{779618711}), we have the general solution of equation (\ref{436283681}) in the form
}{%
Тогда, используя (\ref{770391131}) и (\ref{779618711}), имеем общее решение уравнения (\ref{436283681}) в виде
}
\begin{equation}
\label{755070621}
\ffun  = \Xi_{1} (\mtxi_{1}) + \Xi_{2}(\mtxi_{2})
\;.
\end{equation}
\engrus{0.5ex}{0.5ex}{%
Here the arbitrariness of the functions $\Xi_{1} (\mtxi_{1})$ and $\Xi_{2}(\mtxi_{2})$ is restricted by reality of the field function $\ffun$.
The connection between variables $\{\mtxi_{1},\mtxi_{2}\}$ and
$\{{\xi},\hconj{{\xi}}\}$ is defined (for both metrics  (\ref{43842964})) by relations
}{%
Здесь произвольность функций $\Xi_{1} (\mtxi_{1})$ и $\Xi_{2}(\mtxi_{2})$ ограничивается действительностью полевой функции $\ffun$.
Связь между переменными $\{\mtxi_{1},\mtxi_{2}\}$ и
$\{{\xi},\hconj{{\xi}}\}$ определена (для обоих метрик (\ref{43842964})) соотношениями
}
\begin{equation}
\label{375438001}
\begin{pmatrix}
 \df\xi  \\
\df\hconj{\xi}
\end{pmatrix} =
\begin{pmatrix}
 1 &  \pm\xxx^2\,\bigl(\Xi_{2}^{\prime}\bigr)^2  \\
 \pm\xxx^2\,\bigl(\Xi_{1}^{\prime}\bigr)^2 &  1
\end{pmatrix}
\begin{pmatrix}
  \df\mtxi_{1} \\
  \df\mtxi_{2}
\end{pmatrix}
\;,
\end{equation}
\engrus{0.5ex}{0.5ex}{%
\noindent
which (for top signs) are obtained  from (\ref{481079141}) with (\ref{779618711}).
}{%
\noindent
которые (для верхних знаков) получены из (\ref{481079141}) с (\ref{779618711}).
}
\end{subequations}

\engrus{0.5ex}{0.5ex}{%
Expressions (\ref{406914401}) with scale-rotation transformation (\ref{592114861}) represent the general lightlike soliton of the model under consideration.
}{%
Выражения (\ref{406914401}) с масштабно-поворотным преобразованием (\ref{592114861}) представляет общий светоподобный солитон рассматриваемой модели.
}

\begin{subequations}\label{405109561}
\engrus{0.5ex}{0.5ex}{%
Relations (\ref{375438001}) can be inverted on the assumption of nonsingularity of the transformation matrix:
}{%
Соотношения (\ref{375438001}) могут быть инвертированы в предположении несингулярности матрицы перехода:
}
\begin{align}
\label{405569401}
&\qquad\qquad\qquad\qquad
 1 - \xxx^{4}\,\bigl(\Xi_{1}^{\prime}\bigr)^2\,\bigl(\Xi_{2}^{\prime}\bigr)^2 \neq 0
\;,
\\[1ex]
\label{303442791}
 &
 \begin{pmatrix}
  \df\mtxi_{1}  \\
  \df\mtxi_{2}
\end{pmatrix} =
\frac{1}{1 - \xxx^{4}\,\bigl(\Xi_{1}^{\prime}\bigr)^2\,\bigl(\Xi_{2}^{\prime}\bigr)^2}
\begin{pmatrix}
 1 &  \mp\xxx^2\,\bigl(\Xi_{2}^{\prime}\bigr)^2  \\
 \mp\xxx^2\,\bigl(\Xi_{1}^{\prime}\bigr)^2 &  1
\end{pmatrix}
\begin{pmatrix}
 \df\xi  \\
\df\hconj{\xi}
\end{pmatrix}
\;.
\end{align}
\end{subequations}

\engrus{0.5ex}{0.5ex}{%
\textadd1{The} obtained solution (\ref{406914401}) can be checked directly.
Substitution (\ref{755070621}) to equation (\ref{436283681}) and using (\ref{405109561}) reduce to identity.
}{%
Полученное решение (\ref{406914401}) может быть проверено непосредственно.
Подстановка (\ref{755070621}) в уравнение (\ref{436283681}) и использование (\ref{405109561}) приводит к тождеству.
}

%\engrus{0.5ex}{0.5ex}{%
%\textremb{
%It can be checked also that equation (\ref{436283681}) is  hyperbolic  \textremb{with the solution} \textaddb{for the set of functions}  (\ref{406914401})
%\textremb{for} \textaddb{with the} condition (\ref{405569401}).
%}
%}{%
%\textremb{
%Также может быть проверено то, что уравнение (\ref{436283681}) является гиперболическим на множестве функций (\ref{406914401}) при условии (\ref{405569401}).
%}
%}

\engrus{0.5ex}{0.5ex}{%
One could say that relations (\ref{375438001}) define \textadd1{the} transformation of independent variables  $\{\xi,\hconj{\xi}\}$
to $\{\mtxi_{1},\mtxi_{2}\}$ for \textadd1{the} equation (\ref{436283681}).
But the definition of \textadd1{the} transformation with differential relations is not complete.
\textadd1{The} direct connection between the variables can be obtained by \textadd1{the} path integration of relations (\ref{375438001}) in
\textadd1{a} nonsingular area, that is for \textadd1{the} condition (\ref{405569401}). At the same time we must define an initial
correspondence between the variables
$\{\xi,\hconj{\xi}\}$ and $\{\mtxi_{1},\mtxi_{2}\}$.
}{%
Можно сказать, что соотношения (\ref{375438001}) определяют преобразование независимых переменных $\{\xi,\hconj{\xi}\}$
в $\{\mtxi_{1},\mtxi_{2}\}$ для уравнения (\ref{436283681}).
Однако определение преобразования дифференциальными соотношения является неполным.
Непосредственная связь между переменными может быть получена интегрированием по пути соотношений (\ref{375438001}) в
несингулярной области, то есть при условии (\ref{405569401}). В то же время мы должны определить начальное соответствие между переменными
$\{\xi,\hconj{\xi}\}$ и $\{\mtxi_{1},\mtxi_{2}\}$.
}

\engrus{0.5ex}{0.5ex}{%
It is useful to introduce polar coordinates $\{\tilde{\rho}_1,\tilde{\varphi}_2\}$
\textaddb{and $\{\tilde{\rho}_2,\tilde{\varphi}_2\}$}
for \textadd1{the} variables $\{\mtxi_1,\mtxi_2\}$
by analogy with \textadd1{the polar coordinates $\{{\rho},{\varphi}\}$} (\ref{552514381}) for \textadd1{the} variables $\{{\xi},\hconj{{\xi}}\}$:
}{%
Удобно ввести полярные координаты $\{\tilde{\rho}_1,\tilde{\varphi}_1\}$ и $\{\tilde{\rho}_2,\tilde{\varphi}_2\}$ для переменных $\{\mtxi_1,\mtxi_2\}$
по аналогии с полярными координатами $\{{\rho},{\varphi}\}$ (\ref{552514381}) для переменных $\{{\xi},\hconj{{\xi}}\}$:
}
\begin{equation}
\label{430348741}
\mtxi_1 = \tilde{\rho}_1\,\e{\him\,\tilde{\varphi}_1}\;,\quad
\mtxi_2 = \tilde{\rho}_2\,\e{\him\,\tilde{\varphi}_2}
\;.
\end{equation}

\engrus{0.5ex}{0.5ex}{%
We can consider the simplest case by taking $\xxx = 0$ in (\ref{375438001}). In this case we can put
}{%
Мы можем рассмотреть простейший случай, взяв $\xxx = 0$ in (\ref{375438001}). В этом случае можем положить
}
\begin{equation}
\label{290961541}
\begin{pmatrix}
  \mtxi_{1} \\
 \mtxi_{2} \\
\end{pmatrix}
=
\begin{pmatrix}
  \xi \\
 \hconj{\xi} \\
\end{pmatrix}
\;,
\end{equation}
\engrus{0.5ex}{0.5ex}{%
\noindent
and expression (\ref{755070621}) \textremb{is}\textaddb{becomes} evident solution of \textaddb{the} appropriate to (\ref{436283681}) linear equation when \mbox{$\xxx = 0$}.
}{%
\noindent
и выражение (\ref{755070621}) становится очевидным решением соответствующего (\ref{436283681}) линейного уравнения, когда \mbox{$\xxx = 0$}.
}

\engrus{0.5ex}{0.5ex}{%
In general case let us consider relation (\ref{290961541}) as asymptotic for \mbox{$\rho \to \infty$}.
Then \textadd1{for an area including infinity} we can \textadd1{use the following}
\textremd{designates}\textaddd{designations}:
}{%
В общем случае будем рассматривать соотношения (\ref{290961541}) как асимптотические при \mbox{$\rho \to \infty$}.
Тогда для некоторой области, включающей бесконечность, мы можем использовать следующие обозначения:
}
\begin{equation}
\label{448940041}
 \mtxi_{1} = \mtxi
 \;,\quad
 \mtxi_{2} = \mtcxi
 \;,\quad
 \tilde{\rho}_1 = \tilde{\rho}_2 =\tilde{\rho}
 \;,\quad
 \tilde{\varphi}_1 = -\tilde{\varphi}_2 = \tilde{\varphi}
\;.
\end{equation}

\engrus{0.5ex}{0.5ex}{%
\textadd1{The} general solution in the form of lightlike soliton depending on \textadd1{the} phase $\Phase$ (\ref{426404172})
can be obtained from (\ref{406914401}) by \textadd1{the} invariant substitution (\ref{592114861}).
}{%
Общее решение в виде светоподобного солитона зависящего от фазы $\Phase$ (\ref{426404172}) может быть получено из
(\ref{406914401}) инвариантной подстановкой (\ref{592114861}).
}

\engrus{0.5ex}{0.5ex}{%
It is notable that the action density for \textadd1{the} obtained solution does not contain radical. \textadd1{The} substitution solution
(\ref{592114861}) with (\ref{755070621}) into (\ref{383322881}) with (\ref{430348001}) and (\ref{303442791})
gives \textadd1{the} expression
}{%
Примечательно, что плотность действия для полученного решения не содержит радикала.
Подстановка решения (\ref{592114861}) с (\ref{755070621}) в (\ref{383322881}) с (\ref{430348001}) и (\ref{303442791})
даёт выражение
}
\begin{equation}
\label{423072731}
\LF  = \left|\frac{1 \mp \xxx^2\,\Xi_{1}^{\prime}\,\Xi_{2}^{\prime}}{1 \pm \xxx^2\,
\Xi_{1}^{\prime}\,\Xi_{2}^{\prime}}\right|
\;.
\end{equation}
\engrus{0.5ex}{0.5ex}{%
\noindent
As we see, explicit dependence on phase $\Phase$, which we have in (\ref{592114861}), here is absent.
}{%
\noindent
Как мы видим, явная зависимость от фазы $\Phase$, которую мы имеем в (\ref{592114861}), здесь отсутствует.
}

\engrus{0.5ex}{0.5ex}{%
As a consequence of probes of various arbitrary functions $\Xi_{1} (\mtxi_{1})$ and $\Xi_{2}(\mtxi_{2})$ we have obtained that
the
\textrem1{solitons }\textadd1{solutions} under consideration have \textrem1{singular }%
 tubelike shells, where the condition (\ref{405569401}) is broken.
Note that \textrem1{one }\textadd1{the lightlike} soliton can have a set of such
\textrem1{singular }shells.
Thus the obtained solution can be called the shell lightlike soliton.
}{%
Пробы различных произвольных функций $\Xi_{1} (\mtxi_{1})$ и $\Xi_{2}(\mtxi_{2})$ показали, что рассматриваемые решения
имеют трубчатые оболочки, на которых условие (\ref{405569401}) нарушается.
Надо заметить, что светоподобный солитон может иметь несколько таких оболочек.
Таким образом полученное решение может быть названо оболочечным светоподобным солитоном.
}

\engrus{0.5ex}{0.5ex}{%
In order to identify some features in the use
of the formulas (\ref{406914401}) and (\ref{405109561}),
let us consider a simplest solution which does not depend on polar angle $\varphi$.
}{%
Для выявления некоторых особенностей применения
формул (\ref{406914401}) и (\ref{405109561}) рассмотрим простейшее решение, у которого отсутствует зависимость от полярного угла
$\varphi$.
}

\engrus{0.5ex}{0.5ex}{%
This cylindrically symmetric solution is similar to the spherically symmetric one (\ref{670504501}) and it can be
obtained directly without the use of formulas
(\ref{406914401}) and (\ref{405109561}).
}{%
Это цилиндрически симметричное решение аналогично сферически симметричному решению (\ref{670504501}) и может быть получено
непосредственно без применения формул (\ref{406914401}) и (\ref{405109561}).
}

\engrus{0.5ex}{0.5ex}{%
 Relations (\ref{383322881}) and (\ref{408139091}) written in cylindrical coordinates give in this case the following expressions for  the action density and
 \textaddd{the}
 radial derivative of the field function:
}{%
Соотношения (\ref{383322881}) и (\ref{408139091}), записанные в цилиндрических координатах, в этом случае дают следующие выражения для плотности действия и радиальной производной полевой функции решения:
}
\begin{align}
\label{52268754}
\LF  &= \sqrt{\left|1 \mp \xxx^2\,\ffun_{\rho}^{2}\right|}
\\
\label{439078951}
\ffind_{\rho} &=-\frac{2\,\qL}{\rho}   %\qL\,}{\rho}
\qquad\Longrightarrow\qquad
 \frac{\p \ffun}{\p \rho}  = -\frac{2\,\qL}{\sqrt{\left|\rho^{2}\pm \brrho^{2}\right|}}
\;,
\quad
\brrho\eqdef \left|2\,\qL\,\xxx\right|     %\qL\,\xxx\right|
\;.
\end{align}
\engrus{0.5ex}{0.5ex}{%
\noindent
where $\qL$ is a real constant representing a linear charge density.
}{%
\noindent
где $\qL$ -- действительная константа, представляющая линейную плотность заряда.
}

\begin{subequations}\label{439036491}
\engrus{0.5ex}{0.5ex}{%
An integration of the expression of derivative $\ffun_\rho$ (\ref{439078951}) for the cases of top and bottom signs gives accordingly
}{%
Интегрирование выражения производной $\ffun_\rho$ (\ref{439078951}) для случаев верхнего и нижний знаков даёт соответственно
}
\begin{align}
\label{365126681}
\ffun &= -2\,\qL\,\ln\frac{\rho + \sqrt{\rho^2 + \brrho^{2}}}{\brrho}
\;,\\
\label{365126682}
\ffun &= \left[\begin{array}{ll}
-2\,\qL\,\ln\dfrac{\rho + \sqrt{\rho^2 - \brrho^{2}}}{\brrho} &\Longleftarrow\quad\rho \geqslant \brrho\;,
\\[2ex]
-2\,\him\,\qL\,\ln\dfrac{\rho + \him\sqrt{\brrho^{2} - \rho^2  }}{\brrho}
= 2\,\qL\,\arccos\bigl(\rho/\brrho\bigr) &\Longleftarrow\quad\rho \leqslant \brrho
\;.
\end{array}
\right.
\end{align}
\end{subequations}

\engrus{0.5ex}{0.5ex}{%
In the case of the first solution in (\ref{439036491})
the field function $\ffun$ (\ref{365126681}) and its derivative $\ffun_\rho$ (\ref{439078951}, top sign) are finite and
regular at any finite point (for $\rho < \infty$).
 The appropriate plots are shown on Fig. \ref{43867823}.
 It can be shown that the energy of this solution is logarithmically divergent for $\rho\to\infty$.
}{%
В случае первого решения в (\ref{439036491})
полевая функция $\ffun$ (\ref{365126681}) и её производная $\ffun_\rho$ (\ref{439078951}, верхний знак) конечны и регулярны в любой конечной точке (при $\rho < \infty$).
Соответствующие графики показаны на Рис. \ref{43867823r}. Как можно показать, энергия этого решения логарифмически расходится при $\rho\to\infty$.
}

\begin{figure}[h]
\begin{center}
\ifpdf
  {
  \unitlength 1mm
  \begin{pspicture}(0,0)(12,3.3)
   \put(0,-0.5){\includegraphics[width=55mm]{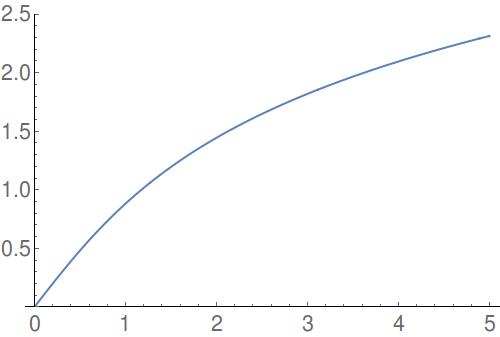}}
   \put(6.1,-0.5){\includegraphics[width=55mm]{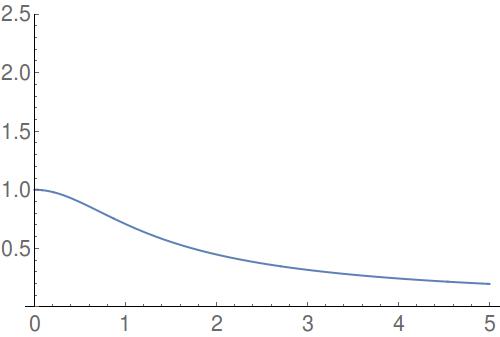}}
   \put(0.5,0){
    \put(6.1,2.7){$\dfrac{\p\ffun}{\p\rho}$}
    \put(11.2,-0.5){$\rho$}
    \put(0.05,3.0){$\ffun$}
    \put(5.2,-0.5){$\rho$}
   }
   \put(4.8,-2.6){
    \begin{picture}(3.2,3.5)
    \put(-0.5,3.8){$\qL = -1/2$}
    \put(-0.5,3.3){$\brrho = 1$}
    \end{picture}
   }
 \end{pspicture}
  }
 \else
  \begin{pspicture}(0,0)(12,3.3)
   \put(0.5,0){
    \put(6.1,2.7){$\dfrac{\p\ffun}{\p\rho}$}
    \put(11.2,-0.5){$\rho$}
    \put(0.05,3.0){$\ffun$}
    \put(5.2,-0.5){$\rho$}
   }
   \put(4.8,-2.6){
    \begin{picture}(3.2,3.5)
    \put(-0.5,3.8){$\qL = -1/2$}
    \put(-0.5,3.3){$\brrho = 1$}
    \end{picture}
   }
   \put(0,-0.5){\includegraphics[width=55mm]{LSOESTF_Fig41a.eps}}
   \put(6.1,-0.5){\includegraphics[width=55mm]{LSOESTF_Fig41b.eps}}
  \end{pspicture}
\fi
\end{center}
\engrus{0.5ex}{0.5ex}{%
 \caption{\label{43867823} Solution (\ref{365126681}) and its derivative.}
}{%
\mifeng{\addtocounter{figure}{-1}}{}
 \caption{\label{43867823r} Решение (\ref{365126681}) и его производная.}
}
\end{figure}

\engrus{0.5ex}{0.5ex}{%
In the case of the second solution in (\ref{439036491})
the field function $\ffun$ (\ref{365126682}) is finite for  \mbox{$\rho < \infty$} and it becomes zero at \mbox{$\rho = \brrho$} while remaining continuous everywhere.
But its derivative $\ffun_\rho$ (\ref{439078951}, bottom sign) becomes infinite at \mbox{$\rho = \brrho$}.
}{%
В случае второго решения в (\ref{439036491})
полевая функция $\ffun$ (\ref{365126682}) конечна при \mbox{$\rho < \infty$} и обращается в ноль при \mbox{$\rho = \brrho$}, оставаясь всюду непрерывной.
При этом её производная  $\ffun_\rho$ (\ref{439078951}, нижний знак) бесконечна при $\rho = \brrho$.
}

\engrus{0.5ex}{0.5ex}{%
Thus there is the singular surface \mbox{$\rho = \brrho$} in this case.
The appropriate plots are shown on Fig. \ref{43862378}.  The energy of this solution is logarithmically divergent for $\rho\to\infty$ only.
It can be shown that the energy in the region $0\leqslant\rho\leqslant\brrho$ is finite and positive.
}{%
Таким образом в этом случае имеется сингулярная поверхность \mbox{$\rho = \brrho$}. Соответствующие графики показаны на Рис. \ref{43862378r}.  Энергия этого решения логарифмически расходиться только при $\rho\to\infty$.
Как можно показать, энергия в области $0\leqslant\rho\leqslant\brrho$ конечна и положительна.
}

\begin{figure}[h]
\begin{center}
\ifpdf
  {
  \unitlength 1mm
  \begin{pspicture}(0,0)(12,3.3)
   \put(0,-0.5){\includegraphics[width=55mm]{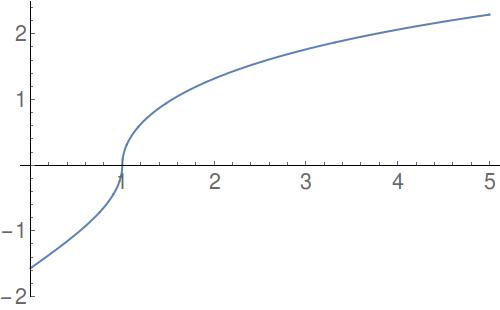}}
   \put(6.1,-0.5){\includegraphics[width=55mm]{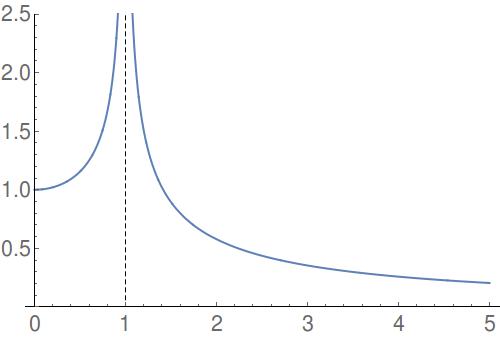}}
   \put(0.5,0){
     \put(6.05,2.8){$\dfrac{\p\ffun}{\p\rho}$}
     \put(11.15,-0.6){$\rho$}
     \put(0.0,2.7){$\ffun$}
     \put(5.0,0.7){$\rho$}
   }
   \put(4.7,-1.8){
    \begin{picture}(3.2,3.5)
    \put(-0.5,3.8){$\qL = -1/2$}
    \put(-0.5,3.3){$\brrho = 1$}
    \end{picture}
   }
 \end{pspicture}
  }
 \else
  \begin{pspicture}(0,0.0)(12,3.3)
   \put(0.5,0){
     \put(6.05,2.8){$\dfrac{\p\ffun}{\p\rho}$}
     \put(11.15,-0.6){$\rho$}
     \put(0.0,2.7){$\ffun$}
     \put(5.0,0.7){$\rho$}
   }
   \put(4.7,-1.8){
    \begin{picture}(3.2,3.5)
    \put(-0.5,3.8){$\qL = -1/2$}
    \put(-0.5,3.3){$\brrho = 1$}
    \end{picture}
   }
   \put(0,-0.5){\includegraphics[width=55mm]{LSOESTF_Fig42a.eps}}
   \put(6.1,-0.5){\includegraphics[width=55mm]{LSOESTF_Fig42b.eps}}
  \end{pspicture}
\fi
\end{center}
\engrus{0.5ex}{0.5ex}{%
 \caption{\label{43862378} Solution (\ref{365126682}) and its derivative.}
}{%
\mifeng{\addtocounter{figure}{-1}}{}
 \caption{\label{43862378r}  Решение (\ref{365126682}) и его производная.}
}
\end{figure}

\engrus{0.5ex}{0.5ex}{%
It should be noted that the choice of the top sign in (\ref{439078951}) and the appropriate solution (\ref{365126681})
correspond to the signature of metric (\ref{43842964a}) in the whole space.
}{%
Надо отметить, что выбор верхнего знака в (\ref{439078951}) и соответствующее решение (\ref{365126681}) отвечают сигнатуре метрики (\ref{43842964a}) во всём пространстве.
}

\engrus{0.5ex}{0.5ex}{%
The choice of the bottom sign in (\ref{439078951}) and the appropriate solution (\ref{365126682}) correspond to the signature of metric (\ref{43842964b})
 in the region \mbox{$\rho > \brrho$}. In the region \mbox{$0 < \rho < \brrho $}, the solution (\ref{365126682})
corresponds to the signature of metric (\ref{43842964a}). In this case, in the region  \mbox{$0 < \rho < \brrho $},
the alternative of representation of modulus in (\ref{439078951})
\mbox{$\left|\rho^{2} - \brrho^{2}\right| = \brrho^{2} - \rho^{2}$} corresponds to the action density (\ref{52268754})
in the form
\mbox{$\LF  = \sqrt{\left|1 - \xxx^2\,\ffun_{\rho}^{2}\right|} = \sqrt{\xxx^2\,\ffun_{\rho}^{2} - 1} $}.
}{%
Выбор нижнего знака в (\ref{439078951}) и соответствующее решение (\ref{365126682}) отвечают сигнатуре метрики (\ref{43842964b})
 в области \mbox{$\rho > \brrho$}. В области \mbox{$0 < \rho < \brrho $} решение (\ref{365126682})
отвечает сигнатуре метрики (\ref{43842964a}). В этом случае в области \mbox{$0 < \rho < \brrho $} способ раскрытия модуля в (\ref{439078951})
\mbox{$\left|\rho^{2} - \brrho^{2}\right| = \brrho^{2} - \rho^{2}$} соответствует плотности действия (\ref{52268754}) вида
\mbox{$\LF  = \sqrt{\left|1 - \xxx^2\,\ffun_{\rho}^{2}\right|} = \sqrt{\xxx^2\,\ffun_{\rho}^{2} - 1} $}.
}

\engrus{0.5ex}{0.5ex}{%
In consequence, as we see on Fig. \ref{43862378}, the field function of the solution (\ref{365126682})
is continuous for \mbox{$0 \leqslant \rho < \infty $} and
the appropriate configuration of the space-time film represents a smooth surface for \mbox{$0 < \rho < \infty $}.
}{%
В результате, как видно на Рис. \ref{43862378r}, полевая функция решения (\ref{365126682}) непрерывна при
\mbox{$0 \leqslant \rho < \infty $},
а соответствующая конфигурация про\-странст\-вен\-но-временной плёнки представляет собой гладкую поверхность
при  %\mifeng{}{\break}
\mbox{$0 < \rho < \infty $}.
}

\engrus{0.5ex}{0.5ex}{%
Thus in this case, the change of the signature of metric
on the singular set
allows to obtain the solution in the form of a smooth surface
for \mbox{$\rho > 0$}.
}{%
Таким образом изменение сигнатуры метрики на сингулярном множестве позволило в данном случае
получить решение в виде
гладкой поверхности при \mbox{$\rho > 0$}.
}

\begin{subequations}\label{330471621}
\engrus{0.5ex}{0.5ex}{%
Now, to obtain the solution (\ref{439036491}) with the help of the relations (\ref{406914401}), we take the arbitrary functions
of the general solution (\ref{406914401}) in the following form:
}{%
Теперь, чтобы получить решение (\ref{439036491}) при помощи соотношений (\ref{406914401}), выберем произвольные функции общего решения (\ref{406914401}) в следующем виде:
}
\begin{alignat}{2}
\label{451959281}
\Xi_{1} &= -\frac{\cbxi_1}{\xxx}\,\ln\biggl(\frac{\mtxi_1}{\cbrho}\biggr)
\;,\quad
&  \Xi_{2} &=
-\frac{\cbxi_2}{\xxx}\,\ln\biggl(\frac{\mtxi_2}{\cbrho}\biggr)
\;,\\
\label{451959281p}
\Xi_{1}^{\prime}  &=
-\frac{\cbxi_1}{\xxx\,\mtxi_1}\;,\quad  &
\Xi_{2}^{\prime} &=
-\frac{\cbxi_2}{\xxx\, \mtxi_2}
\;,
\end{alignat}
\engrus{0.5ex}{0.5ex}{%
\noindent
where $\{\cbxi_1,\cbxi_2\}$ are complex constants, $\cbrho$ is a real positive constant having the physical dimension of length
and providing the dimensionless for argument of logarithm.
}{%
\noindent
где $\{\cbxi_1,\cbxi_2\}$ -- комплексные константы, $\cbrho$ -- действительная положительная константа, имеющая физическую размерность длины и обеспечивающая безразмерность аргумента логарифма.
}
\end{subequations}

\engrus{0.5ex}{0.5ex}{%
The substitution of these functions (\ref{330471621}) into relations (\ref{375438001}) give the following dependence between the differentials
of old and new independent variables:
}{%
Подстановка этих функций в соотношения (\ref{375438001}) даёт следующие зависимости между дифференциалами старых и новых независимых переменных:
}
\begin{equation}
\label{492600841}
 \df\mxi = \df\mtxi_1 {}\pm{} \dfrac{\cbxi_2^{2}}{\mtxi_2^{2}}\,\df\mtxi_2
 \;,\quad
\df\mxic = \pm\dfrac{\cbxi_1^{2}}{\mtxi_1^{2}}\,\df\mtxi_1 + \df\mtxi_2
\;.
\end{equation}

\engrus{0.5ex}{0.5ex}{%
The condition for nonsingularity of the transformation matrix to the variables $\{\mxi,\mxic\}$ (\ref{405569401})
has the following form in the case under consideration  (\ref{492600841}):
}{%
Условие несингулярности матрицы перехода к переменным $\{\mxi,\mxic\}$ (\ref{405569401}) в рассматриваемом случае (\ref{492600841}) имеет следующий вид:
}
\begin{equation}
\label{609493561}
\frac{\cbxi_1^{2}\,\cbxi_2^{2}}{\mtxi_1^2\,\mtxi_2^2}  \neq 1
\;.
\end{equation}

\engrus{0.5ex}{0.5ex}{%
\textaddb{%
To obtain the functions $\mtxi_1 (\mxi)$ and $\mtxi_2(\mxic)$, first we must integrate relations (\ref{492600841}).
For this purpose, it is convenient to use the polar coordinates $\{\mtrho_1,\mtphi_1\}$, $\{\mtrho_2,\mtphi_2\}$ (\ref{430348741}), and $\{\rho,\varphi\}$ (\ref{552514381}).
}
}{%
Чтобы получить функции $\mtxi_1 (\mxi)$ и $\mtxi_2(\mxic)$, сначала мы должны проинтегрировать соотношения (\ref{492600841}).
С этой целью удобно использовать полярные координаты $\{\mtrho_1,\mtphi_1\}$, $\{\mtrho_2,\mtphi_2\}$ (\ref{430348741}) и $\{\rho,\varphi\}$ (\ref{552514381}).
}

\engrus{0.5ex}{0.5ex}{%
\textaddb{%
Let us take the paths of integration in the planes  $\{\mtrho_1,\mtphi_1\}$, $\{\mtrho_2,\mtphi_2\}$, and $\{\rho,\varphi\}$ with sufficiently far beginning points from coordinate origin.
Let the starting points in these planes be $\{\rho_{1\!\infty},0\}$, $\{\rho_{2\!\infty},0\}$, and $\{\rho_{\!\infty},0\}$.
In the plane $\{\mtrho_1,\mtphi_1\}$, for example, we can integrate by the following path: from $\{\rho_{1\!\infty},0\}$ to
$\{\mtrho_1,0\}$ for \mbox{$\mtphi_1=\const$} and from $\{\mtrho_1,0\}$ to $\{\mtrho_1,\mtphi_1\}$ for \mbox{$\mtrho_1 =\const$}.
Similarly, we integrate in the plane $\{\mtrho_2,\mtphi_2\}$ as well as the left parts of equalities (\ref{492600841}) in the plane $\{\rho,\varphi\}$.
}
}{%
Выберем пути интегрирования в плоскостях $\{\mtrho_1,\mtphi_1\}$, $\{\mtrho_2,\mtphi_2\}$ и $\{\rho,\varphi\}$ с начальными точками, достаточно удалёнными от координатного центра.
Пусть стартовые точки в этих плоскостях будут $\{\rho_{1\!\infty},0\}$, $\{\rho_{2\!\infty},0\}$ и $\{\rho_{\!\infty},0\}$.
В плоскости $\{\mtrho_1,\mtphi_1\}$, например, мы можем интегрировать по следующему пути: от $\{\rho_{1\!\infty},0\}$ до
$\{\mtrho_1,0\}$ для \mbox{$\mtphi_1=\const$} и  от $\{\mtrho_1,0\}$ до $\{\mtrho_1,\mtphi_1\}$ для \mbox{$\mtrho_1 =\const$}.
Аналогично интегрируем в плоскости $\{\mtrho_2,\mtphi_2\}$, а также левые части равенств (\ref{492600841}) в плоскости $\{\rho,\varphi\}$.
}

\engrus{0.5ex}{0.5ex}{%
\textaddb{%
Then we take that \mbox{$\rho_{1\!\infty}\to\infty$}, \mbox{$\rho_{2\!\infty}\to\infty$}, and \mbox{$\rho_{\!\infty}\to\infty$}.
Using again the variables $\{\mtxi_1,\mtxi_1\}$, $\{\mtxi_2,\mtxi_2\}$, and $\{\mxi,\mxic\}$,
as \textaddd{a} result we have
}
}{%
Затем мы принимаем, что \mbox{$\rho_{1\!\infty}\to\infty$}, \mbox{$\rho_{2\!\infty}\to\infty$} и \mbox{$\rho_{\!\infty}\to\infty$}.
Используя  снова переменные $\{\mtxi_1,\mtxi_1\}$, $\{\mtxi_2,\mtxi_2\}$ и $\{\mxi,\mxic\}$, в результате имеем
}
\begin{equation}
\label{40528090}
\mxi = \mtxi_1 \mp \frac{\cbxi_2^{2}}{\mtxi_2}
\;,\quad
\mxic = \mtxi_2 \mp \frac{\cbxi_1^{2}}{\mtxi_1}
\;,
\end{equation}

\begin{subequations}\label{491538841}
\engrus{0.5ex}{0.5ex}{%
Two solutions for the system of equations (\ref{40528090}) with respect to variables $\{\mtxi_1,\mtxi_2\}$ can be written
in an explicit form. But here we write these solution for the special case
}{%
Система уравнений (\ref{40528090}) относительно переменных $\{\mtxi_1,\mtxi_2\}$ имеет два решения, которые могут быть выписаны в явном виде. Здесь, однако, мы запишем эти решения для специального случая
}
\begin{alignat}{2}
\label{491841233}
 & & \cbxi_1^{2} &= \cbxi_2^{2}
\;.
\end{alignat}
\engrus{0.5ex}{0.5ex}{%
\noindent
Using the representation of the variables $\{\mxi,\mxic\}$ in polar coordinates $\{\rho,\varphi\}$ (\ref{552514381})
we have
}{%
\noindent
Используя представление переменных $\{\mxi,\mxic\}$ в полярных координатах $\{\rho,\varphi\}$ (\ref{552514381}), имеем
}
\begin{alignat}{2}
\label{491841231}
\mtxi_1 &=
\frac{\e{\him\varphi}}{2}\left(\rho + \sqrt{\rho^2 \pm 4\,\cbxi_1^{2} }\right)
\,,\quad
&  \mtxi_2 &=
\frac{\e{-\him\varphi}}{2}\left(\rho + \sqrt{\rho^2 \pm 4\,\cbxi_1^{2} }\right)
\;,\\
\label{491841232}
\mtxi_1 &=
\frac{\e{\him\varphi}}{2}\left(\rho - \sqrt{\rho^2 \pm 4\,\cbxi_1^{2} }\right)\,,\quad
&  \mtxi_2 &=
\frac{\e{-\him\varphi}}{2}\left(\rho - \sqrt{\rho^2 \pm 4\,\cbxi_1^{2} }\right)
\;.
\end{alignat}
\end{subequations}

\engrus{0.5ex}{0.5ex}{%
As can be easy seen, the solution (\ref{491841231}) satisfies the asymptotic relations \mbox{$\mtxi_1\to\mxi$} and \mbox{$\mtxi_2\to\mxic$} for \mbox{$\rho\to\infty$}.
But the solution (\ref{491841232}) gives the asymptotic relations \mbox{$\mtxi_1\to 0$} and \mbox{$\mtxi_2\to 0$} for \mbox{$\rho\to\infty$}.
}{%
Как легко заметить, решение (\ref{491841231}) удовлетворяет асимптотическим соотношениям \mbox{$\mtxi_1\to\mxi$} и \mbox{$\mtxi_2\to\mxic$} при \mbox{$\rho\to\infty$}.
Решение же (\ref{491841232}) даёт асимптотики \mbox{$\mtxi_1\to 0$} и \mbox{$\mtxi_2\to 0$} при \mbox{$\rho\to\infty$}.
}

\engrus{0.5ex}{0.5ex}{%
Let us consider the solution (\ref{491538841}) with equal and real constants $\{\cbxi_1,\cbxi_2\}$ such that
}{%
Рассмотрим решение (\ref{491538841}) с равными и действительными константами $\{\cbxi_1,\cbxi_2\}$
так, что
}
\begin{equation}
\label{660547091}
\cbxi_1 = \cbxi_2  = \qL\,\xxx\;,\qquad  %\frac{\qL\,\xxx}{2}\;,\qquad
\left|\qL\,\xxx\right| \eqdef \frac{\brrho}{2} \eqdef \cbrho  %\frac{\left|\qL\,\xxx\right|}{2} \eqdef \frac{\brrho}{2} \eqdef \cbrho
\;.
\end{equation}
\engrus{0.5ex}{0.5ex}{%
\noindent
Here $\cbrho$ is the real positive constant contained also in the formulas for the functions $\Xi_{1}$ and $\Xi_{2}$ (\ref{451959281}).
The constants $\qL$ and $\brrho$ are contained in the solutions (\ref{439036491}).
}{%
\noindent
Здесь $\cbrho$ -- действительная положительная константа, входящая также в формулы для функций $\Xi_{1}$ и $\Xi_{2}$ (\ref{451959281}). Константы $\qL$ и $\brrho$ содержатся в решениях (\ref{439036491}).
}

\engrus{0.5ex}{0.5ex}{%
In this case (\ref{660547091}), the consecutive substitution (\ref{491841231}) to (\ref{451959281}) and (\ref{451959281}) to (\ref{755070621})
gives the solution (\ref{365126681}) for the signature (\ref{43842964a}) and the solution (\ref{365126682}) in the region $\rho \geqslant \brrho$ for the signature (\ref{43842964b}). The second solution (\ref{491841232})
gives the same field function $\ffun$ but with the opposite sign.
}{%
В этом случае (\ref{660547091}) последовательная подстановка (\ref{491841231}) в
(\ref{451959281}) и (\ref{451959281}) в
(\ref{755070621})
даёт решение (\ref{365126681}) для сигнатуры (\ref{43842964a}) и решение (\ref{365126682}) в области $\rho \geqslant \brrho$ для сигнатуры (\ref{43842964b}). Второе решение (\ref{491841232})
даёт ту же полевую функцию $\ffun$, но с противоположным знаком.
}

\engrus{0.5ex}{0.5ex}{%
 We can see from (\ref{491538841}), the tilde variables $\{\mtxi_1,\mtxi_2\}$ are mutually complex conjugated
 in this case such that
 the designations without indexes $\{\mtxi,\mtcxi\}$ (\ref{448940041}) can be used. Then according to (\ref{660547091}),
 the condition for nonsingularity of the transformation matrix (\ref{609493561}) takes the form
}{%
При этом, как видно из (\ref{491538841}), тильдовые переменные $\{\mtxi_1,\mtxi_2\}$ являются взаимно комплексно сопряжёнными так, что можно использовать обозначения без индексов
$\{\mtxi,\mtcxi\}$ (\ref{448940041}). Тогда согласно (\ref{660547091}) условие несингулярности матрицы перехода (\ref{609493561}) принимает вид
}
\begin{equation}
\label{436605551}
\mtrho  \neq \cbrho\quad\Longleftrightarrow\quad \mtrho \neq \frac{\brrho}{2}
\;.
\end{equation}

\engrus{0.5ex}{0.5ex}{%
Thus the determinant of matrix for
the transformation of variables $\{\mtxi,\mtcxi\}\to\{\mxi,\mxic\}$ (\ref{375438001})
for the functions (\ref{451959281}) is zero at the ring with
\textremd{the}radius \mbox{$\tilde{\rho} = \cbrho = \brrho/2$} in the variables $\{\mtxi,\mtcxi\}$.
According to \textremd{the}equations (\ref{40528090}), in the case of the signature of metric (\ref{43842964a}) (top sign in (\ref{40528090})) this ring is reflected to the origin of the initial coordinate system $\rho=0$.
But in the case of signature of metric (\ref{43842964b}) (bottom sign in (\ref{40528090})) the ring
\mbox{$\tilde{\rho} = \cbrho = \brrho/2$} in the variables $\{\mtxi,\mtcxi\}$ is reflected to the ring
 \mbox{$\rho = 2\,\cbrho = \brrho$} in the variables $\{\mxi,\mxic\}$.
}{%
Таким образом детерминант матрицы преобразования переменных\mifeng{}{\break} $\{\mtxi,\mtcxi\}\to\{\mxi,\mxic\}$ (\ref{375438001}) для функций (\ref{451959281}) обращается в ноль на кольце радиуса \mbox{$\tilde{\rho} = \cbrho = \brrho/2$} в переменных $\{\mtxi,\mtcxi\}$. Согласно уравнениям (\ref{40528090}), в случае сигнатуры метрики (\ref{43842964a}) (верхний знак в (\ref{40528090})) это кольцо отображается в начало координат исходной координатной системы $\rho=0$.
В случае же метрики (\ref{43842964b}) (нижний знак в (\ref{40528090})) кольцо \mbox{$\tilde{\rho} = \cbrho = \brrho/2$} в переменных $\{\mtxi,\mtcxi\}$ отображается в кольцо \mbox{$\rho = 2\,\cbrho = \brrho$} в переменных $\{\mxi,\mxic\}$.
}

\begin{subequations}\label{523967371}
\engrus{0.5ex}{0.5ex}{%
The formulas (\ref{491538841}) give the appropriate  inverse mappings.
Note that for the signature of metric (\ref{43842964b}) we must use the substitution
\mbox{$\sqrt{-\brrho}\to \him\,\sqrt{\brrho}$} in (\ref{491538841}).
Thus we have the following mutually mappings of the regions for the two signatures of metric (\ref{43842964a}) and (\ref{43842964b}):
}{%
Формулы (\ref{491538841}) дают соответствующие обратные отображения.
При этом в случае метрики (\ref{43842964b}) мы должны использовать замену
\mbox{$\sqrt{-\brrho}\to \him\,\sqrt{\brrho}$} в (\ref{491538841}).
Таким образом имеем следующие взаимные отображения областей для двух сигнатур метрики (\ref{43842964a}) и (\ref{43842964b}):
}
\begin{align}
\label{523880481}
\rho=0\quad &\longleftrightarrow\quad \tilde{\rho} = \cbrho = \frac{\brrho}{2}
\;\;,\\
\label{523880482}
\rho = 2\,\cbrho = \brrho\quad  &\longleftrightarrow\quad \tilde{\rho} = \cbrho = \frac{\brrho}{2}
\;\;.
\end{align}
\end{subequations}

\engrus{0.5ex}{0.5ex}{%
To obtain the solution (\ref{365126682}) in the region \mbox{$\rho \leqslant \brrho$}, we consider the case of equal and
 purely imaginary constants in (\ref{491538841}):
}{%
Для получения решения (\ref{365126682}) в области \mbox{$\rho \leqslant \brrho$} рассмотрим случай равных чисто мнимых констант
в (\ref{491538841}):
}
\begin{equation}
\label{451959128}
\cbxi_1 = \cbxi_2  =
\him\,\qL\,\xxx    %\frac{\him\,\qL\,\xxx}{2}
\;,\qquad
\left|\qL\,\xxx\right| \eqdef \frac{\brrho}{2} \eqdef \cbrho   %\frac{\left|\qL\,\xxx\right|}{2} \eqdef \frac{\brrho}{2} \eqdef \cbrho
 \;.
\end{equation}
\engrus{0.5ex}{0.5ex}{%
\noindent
Now the tilde variables $\mtxi_{1}$ and $\mtxi_{2}$ are not mutually complex conjugated in general.
}{%
\noindent
Теперь тильдовые переменные $\mtxi_{1}$ и $\mtxi_{2}$ вообще не являются взаимно комплексно сопряжёнными.
}

\engrus{0.5ex}{0.5ex}{%
Substituting (\ref{451959128}) into (\ref{491841231}) and using relation $\sqrt{-1} = \him$, we obtain the solution
for the system  of equations (\ref{40528090}) in the following form:
}{%
Подставляя (\ref{451959128}) в (\ref{491841231}) и используя соотношение $\sqrt{-1} = \him$, получаем решение системы уравнений (\ref{40528090}) в следующем виде:
}
\begin{alignat}{2}
\label{420179871}
\mtxi_1 &=
\frac{\e{\him\varphi}}{2}\left(\rho + \him\sqrt{\pm \brrho^{2} - \rho^2 }\right)
\,,\quad
&  \mtxi_2 &=
\frac{\e{-\him\varphi}}{2}\left(\rho + \him\sqrt{\pm \brrho^{2} - \rho^2 }\right)
\;.
\end{alignat}

\engrus{0.5ex}{0.5ex}{%
The substitution (\ref{451959128}) and (\ref{420179871}) into
\textremd{the}expressions (\ref{451959281}) for the functions $\Xi_{1}$ and $\Xi_{2}$
gives the solution (\ref{755070621}) which is real for the signature of metric
(\ref{43842964a}) (top sign in (\ref{420179871})) in the region $\rho \leqslant \brrho$.
This solution coincides with (\ref{365126682}) for $\rho \leqslant \brrho$.
}{%
Подстановка (\ref{451959128}) и (\ref{420179871}) в выражения (\ref{451959281}) для функций $\Xi_{1}$ и $\Xi_{2}$
даёт решение (\ref{755070621}),
которое является действительным для сигнатуры метрики (\ref{43842964a}) (верхний знак в (\ref{420179871})) в области $\rho \leqslant \brrho$.
Это решение совпадает с (\ref{365126682}) для $\rho \leqslant \brrho$.
}

\engrus{0.5ex}{0.5ex}{%
It should be noted also that
\textremd{the change of the type for the constants $\cbxi_1$ and $\cbxi_2$ of the solution (\ref{330471621})
from (\ref{660547091}) to (\ref{451959128})
swaps the paces of}top and bottom signs in the expressions
(\ref{492600841}), (\ref{40528090}), and (\ref{491538841})
\textaddd{%
are reversed when
the type of the constants $\cbxi_1$ and $\cbxi_2$ for solution (\ref{330471621})
is changed from (\ref{660547091}) to (\ref{451959128}).%
}
}{%
Надо ещё отметить, что замена типа констант $\cbxi_1$ и $\cbxi_2$ решения (\ref{330471621}) с (\ref{660547091}) на (\ref{451959128}) меняет местами верхний и нижний знаки в выражениях
(\ref{492600841}), (\ref{40528090}) и (\ref{491538841}).
}

\engrus{0.5ex}{0.5ex}{%
Now let us obtain the energy, momentum, and angular momentum densities for a lightlike soliton.
 For this purpose we substitute solution (\ref{755070621}) with scale-rotation transformation (\ref{592114861}) to formulas
(\ref{481319051}).
}{%
Теперь получим плотности энергии, импульса и момента импульса для светоподобного солитона.
Для этой цели подставляем решение (\ref{755070621}) с масштабно-поворотным преобразованием (\ref{592114861}) в формулы
(\ref{481319051}).
}

\begin{subequations}\label{495810601}
\engrus{0.5ex}{0.5ex}{%
Using relations (\ref{430348001}) and (\ref{303442791}), we obtain the expressions for energy, momentum, and angular momentum densities
\textremb{with }\textaddb{containing} some common functions, which
\textremb{will be designate as }\textaddb{we denote by}
$\fcE_{i}$.
Then we have
}{%
Используя соотношения (\ref{430348001}) и (\ref{303442791}), получаем выражения для плотностей энергии, импульса и момента импульса,
содержащие некоторые общие функции, которые мы обозначаем как $\fcE_{i}$.
Тогда имеем
}
\begin{align}
%\nonumber
\cE    &= \fcE_{0} + \omega^2
\left(\bigl(\btap^{\prime}\bigr)^2\,\fcE_{1} +
 \bigl(\btam^{\prime}\bigr)^2\left(\fcE_{2}/\btam^2 + \fcE_{3}/\btam + \fcE_{4}\right)
 %\right.
 %\\
 %&\qquad\qquad\qquad\qquad\qquad\qquad\qquad\qquad
 %\left.
 + \btam^{\prime}\,\btap^{\prime}\left(\fcE_{5}/\btam + \fcE_{6}\right)
 \vphantom{\bigl(\btap^{\prime}\bigr)^2\,\btam^{\prime}\bigr)^2}
 \right)
\label{495868961}
\;,\\
%\nonumber
\cP^{3} &= k\, \omega
\left(\bigl(\btap^{\prime}\bigr)^2\,\fcE_{1} +
 \bigl(\btam^{\prime}\bigr)^2\left(\fcE_{2}/\btam^2 + \fcE_{3}/\btam + \fcE_{4}\right)
% \right.
% \\
% &\qquad\qquad\qquad\qquad\qquad\qquad\qquad\qquad
% \left.
 + \btam^{\prime}\,\btap^{\prime}\left(\fcE_{5}/\btam + \fcE_{6}\right)
 \vphantom{\bigl(\btap^{\prime}\bigr)^2\,\btam^{\prime}\bigr)^2}
 \right)
 \label{495868962}
\;,\\
\label{49742337}
\cM_{3} &= \omega
\left(
\btap^{\prime}\,\fcE_{1} + \btam^{\prime}\left(\fcE_{5}/\btam + \fcE_{6}\right)\!/2
\vphantom{\bigl(\btap^{\prime}\bigr)^2\,\btam^{\prime}\bigr)^2}
 \right)
\;,
\end{align}
\engrus{0ex}{0.5ex}{%
\noindent
where $k_3^2 = \omega^2$ according to (\ref{426404172}).
}{%
\noindent
где $k_3^2 = \omega^2$ в соответствии с (\ref{426404172}).
}
\end{subequations}

\engrus{0.5ex}{0.5ex}{%
Here we write explicitly only two functions:
}{%
Здесь мы выпишем явно только две функции $\fcE_{0}$ и $\fcE_{1}$:
}
\begin{equation}
\label{504942361}
\fcE_{0}\eqdef \frac{1}{2\pi}\,
 \frac{\Xi_{1}^{\prime}\,\Xi_{2}^{\prime}}{1 \pm \xxx^2\,\Xi_{1}^{\prime}\,\Xi_{2}^{\prime}}
  \;,\quad
  \fcE_{1}\eqdef -\frac{1}{4\pi}\,
 \frac{\left(
 \xi\,\e{-\him\,\btap}\,\Xi_{1}^{\prime}
  -
 \hconj{\xi}\,\e{\him\,\btap}\,\Xi_{2}^{\prime}
 \right)^2
 }{1 - \xxx^4\,\bigl( \Xi_{1}^{\prime}\bigr)^2\bigl(\Xi_{2}^{\prime}\bigr)^2}
\;.
\end{equation}
\engrus{0.5ex}{0.5ex}{\noindent
These functions play \textadd1{the} main role in the area, where the scale function $\btam (\theta)$ is almost constant: $\btam^{\prime}\to 0$.
}{\noindent
Эти функции играют главную роль в области, где масштабная функция $\btam (\theta)$ почти постоянна: $\btam^{\prime}\to 0$.
}

\engrus{0.5ex}{0.5ex}{%
We have from (\ref{495810601}) the following notable relation
for the case $\btam^{\prime}\to 0$:
}{%
Мы имеем из (\ref{495810601}) следующее примечательное соотношение для случая
$\btam^{\prime}\to 0$:
}
\begin{equation}
\label{332235821}
\cE - \fcE_{0}  = \left|\cP_{3}\right| = \omega\left|\btap^{\prime} \cM_{3}\right|
\;.
\end{equation}

\engrus{0.5ex}{0.5ex}{%
The arbitrary functions $\btam (\Phase)$ and $\btap (\Phase)$ (\ref{591083061}) define scale and rotation in the plane  $\{x^{1},x^{2}\}$ \textrem1{accordingly }\textadd1{respectively}. Using
(\ref{519662391}), (\ref{592114861}), and (\ref{426404172}), we can show that the case $\btap^{\prime} > 0$
corresponds to positive rotation by angle $\varphi$ in time $x^{0}$ and in $x^{3}$ axis for $k_{3} > 0$.
}{%
Произвольные функции $\btam (\Phase)$ и $\btap (\Phase)$ (\ref{591083061})
определяют масштаб и вращение в плоскости  $\{x^{1},x^{2}\}$ соответственно.
 Используя (\ref{519662391}), (\ref{592114861}) и (\ref{426404172}), мы можем показать, что случай $\btap^{\prime} > 0$
отвечает положительному вращению на угол $\varphi$ за время $x^{0}$ и вдоль оси $x^{3}$ для $k_{3} > 0$.
}

\engrus{0.5ex}{0.5ex}{%
Thus for right-handed coordinate system
$\{x^{1},x^{2},x^{3}\}$, the cases $\btap^{\prime} > 0$ and $\btap^{\prime} < 0$ correspond to right
and left local twist of the soliton accordingly.
}{%
Таким образом для правой координатной системы $\{x^{1},x^{2},x^{3}\}$ случаи $\btap^{\prime} > 0$ и $\btap^{\prime} < 0$ отвечают
правой и левой локальной закрученности солитона соответственно.
}

\engrus{0.5ex}{0.5ex}{%
It is interesting to consider the solitons with constant twist:
}{%
Интересно рассмотреть солитоны с постоянной закрученностью:
}
\begin{equation}
\label{667163021}
\btap^{\prime}  = \const \neq 0
\;.
\end{equation}
\engrus{0.5ex}{0.5ex}{%
Such solitons can be called the uniformly twisted ones.
For conciseness we will call them the twisted solitons.
}{%
Подобные солитоны могут быть названы равномерно закрученными.
Для краткости мы будем называть их закрученными солитонами.
}

\engrus{0.5ex}{0.5ex}{%
As we see in (\ref{332235821}), for the case (\ref{667163021}) the soliton energy density $\cE$ is proportional to its angular momentum density $\cM$ in high-frequency approximation,
that is for $\omega\left|\btap^{\prime} \cM_{3}\right| \gg \left|\fcE_{0}\right|$.
The appropriate proportionality relation between soliton energy and its angular momentum is notable property of the twisted lightlike soliton.
}{%
Как мы видим в (\ref{332235821}), для случая (\ref{667163021}) плотность энергии солитона $\cE$ пропорциональна плотности его момента импульса $\cM$ в высокочастотном приближении,
то есть для $\omega\left|\btap^{\prime} \cM_{3}\right| \gg \left|\fcE_{0}\right|$.
Соответствующее соотношение пропорциональности между энергией солитона и его моментом импульса является примечательным свойством закрученного светоподобного солитона.
}

\engrus{0.5ex}{0.5ex}{%
To obtain integral characteristics of the soliton it is necessary to integrate the functions $\{\fcE_{0},...,\fcE_{6}\}$ in the plane $\{x^{1},x^{2}\}$.
}{%
Чтобы получить интегральные характеристики солитона необходимо проинтегрировать функции $\{\fcE_{0},...,\fcE_{6}\}$ в плоскости $\{x^{1},x^{2}\}$.
}

\engrus{0.5ex}{0.5ex}{%
Considering (\ref{375438001}), \textremb{and (\ref{430348741}) }we can see that the appropriate integrands \textaddb{(\ref{504942361})} have notable simple form in the
\textaddb{tilde} variables
\textremb{$\{\tilde{\rho},\tilde{\varphi}\}$ (\ref{430348741}) }$\{\mtxi_{1},\mtxi_{2}\}$.
\textaddb{Here we obtain these expressions for the case of their mutually complex conjugation
  (\ref{448940041}) such that
 \mbox{$\{\mtxi_{1},\mtxi_{2}\} = \{\mtxi,\mtcxi\}$}.}
}{%
Рассматривая (\ref{375438001}), мы можем заметить, что соответствующие
 подынтегральные выражения (\ref{504942361}) имеют примечательно простую форму в тильдовых переменных $\{\mtxi_{1},\mtxi_{2}\}$.
 Здесь мы получим эти выражения для случая их взаимной комплексной сопряжённости (\ref{448940041}) так, что
 \mbox{$\{\mtxi_{1},\mtxi_{2}\} = \{\mtxi,\mtcxi\}$}.
}

\engrus{0.5ex}{0.5ex}{%
We must take into consideration \textaddb{also} that the functions $\{\mtxi,\mtcxi\}$ and, accordingly,
$\{\Xi_1,\Xi_2\}$ depended on arguments
$\{\xi/\bta,\hconj{\xi}/\hconj{\bta}\}$ after \textaddb{the} scale-rotation transformation (\ref{592114861}).
}{%
Мы должны учесть также, что функции $\{\mtxi,\mtcxi\}$ и,
соответственно,
$\{\Xi_1,\Xi_2\}$ зависят от аргументов
$\{\xi/\bta,\hconj{\xi}/\hconj{\bta}\}$ после масштабно-поворотного преобразования (\ref{592114861}).
}

\begin{subequations}\label{352154411}
\engrus{0.5ex}{0.5ex}{%
Thus making additional substitution $\{\xi/\bta,\hconj{\xi}/\hconj{\bta}\}\to \{\xi,\hconj{\xi}\}$ \textaddb{and using the polar coordinates
$\{\rho,\varphi\}$ and $\{\tilde{\rho},\tilde{\varphi}\}$}, we \textremb{have }\textaddb{obtain} the following integrands:
}{%
Таким образом, делая дополнительную подстановку $\{\xi/\bta,\hconj{\xi}/\hconj{\bta}\}\to \{\xi,\hconj{\xi}\}$
и используя полярные координаты $\{\rho,\varphi\}$ и $\{\tilde{\rho},\tilde{\varphi}\}$, получаем следующие
подынтегральные выражения:
}
\begin{align}
\label{352106482}
\fcE_{0}\,\btam^2\,\rho\,\df\varphi\,\df\rho &=
\frac{\btam^2}{2\pi}\,
\Xi_{1}^{\prime}\,\Xi_{2}^{\prime}\left(1 \mp \xxx^2\,\Xi_{1}^{\prime}\,\Xi_{2}^{\prime}\right)
 \tilde{\rho}\,\df\tilde{\varphi}\,\df\tilde{\rho}
\;,\\
\label{352106481}
 \fcE_{1}\,\btam^2\,\rho\,\df\varphi\,\df\rho &=
 -\frac{\btam^2}{4\pi}
 \left(\bta\,\xi\,\e{-\him\,\btap}\,\Xi_{1}^{\prime} -
 \hconj{\bta}\,\hconj{\xi}\,\e{\him\, \btap}\,\Xi_{2}^{\prime} \right)^2
  \tilde{\rho}\,\df\tilde{\varphi}\,\df\tilde{\rho}
  \;,\\
  \fcE_{2}\,\btam^2\,\rho\,\df\varphi\,\df\rho &=
\frac{\btam^2}{4\pi}
 \left(\bta\,\xi\,\e{-\him\,\btap}\,\Xi_{1}^{\prime} +
 \hconj{\bta}\,\hconj{\xi}\,\e{\him\, \btap}\,\Xi_{2}^{\prime} \right)^2
  \tilde{\rho}\,\df\tilde{\varphi}\,\df\tilde{\rho}
  \;,\\
  %\nonumber
  \fcE_{3}\,\btam^2\,\rho\,\df\varphi\,\df\rho &=
   -\frac{\btam^2}{2\pi}
   \left(\Xi_{1}+ \Xi_{2}\right)
 \left(\bta\,\xi\,\e{-\him\,\btap}\,\Xi_{1}^{\prime} +
 \hconj{\bta}\,\hconj{\xi}\,\e{\him\, \btap}\,\Xi_{2}^{\prime} \right)
 %\\
 %&\qquad\qquad\qquad\qquad\qquad
 %{\scriptstyle\times}
 \left(1 + \xxx^2\,\Xi_{1}^{\prime}\,\Xi_{2}^{\prime}\right)
  \tilde{\rho}\,\df\tilde{\varphi}\,\df\tilde{\rho}
  \;,\\
  \fcE_{4}\,\btam^2\,\rho\,\df\varphi\,\df\rho &=
  \frac{\btam^2}{4\pi}
   \left(\Xi_{1}+ \Xi_{2}\right)^2
   \left(1 \pm \xxx^2\,\Xi_{1}^{\prime}\,\Xi_{2}^{\prime}\right)^2
  \tilde{\rho}\,\df\tilde{\varphi}\,\df\tilde{\rho}
  \;,\\
  \fcE_{5}\,\btam^2\,\rho\,\df\varphi\,\df\rho &=
  \frac{\btam^2\,\him}{2\pi}
 \left(\bta^2\,\xi^2\,\e{-\him\,2\,\btap}\,(\Xi_{1}^{\prime})^2 -
 \hconj{\bta^2}\,\hconj{\xi^2}\,\e{\him\, 2\,\btap}\,(\Xi_{2}^{\prime})^2 \right)
  \tilde{\rho}\,\df\tilde{\varphi}\,\df\tilde{\rho}
  \;,\\
  %\nonumber
  \fcE_{6}\,\btam^2\,\rho\,\df\varphi\,\df\rho &=
   -\frac{\btam^2\,\him}{2\pi}
   \left(\Xi_{1}+ \Xi_{2}\right)
 \left(\bta\,\xi\,\e{-\him\,\btap}\,\Xi_{1}^{\prime} -
 \hconj{\bta}\,\hconj{\xi}\,\e{\him\, \btap}\,\Xi_{2}^{\prime} \right)
% \\
% &\qquad\qquad\qquad\qquad\qquad
% {\scriptstyle\times}
 \left(1 \pm \xxx^2\,\Xi_{1}^{\prime}\,\Xi_{2}^{\prime}\right)
  \tilde{\rho}\,\df\tilde{\varphi}\,\df\tilde{\rho}
 \;,
\end{align}
\engrus{0.5ex}{0.5ex}{%
\noindent
where top and bottom signs are appropriate to metrics (\ref{43842964a}) and (\ref{43842964b}) accordingly.
}{%
\noindent
где верхний и нижний знаки отвечают метрикам (\ref{43842964a}) и (\ref{43842964b}) соответственно.
}
\end{subequations}

\engrus{3ex}{2ex}{%
\section{Twisted lightlike soliton}
\label{twlls}
}{%
\mifeng{\addtocounter{section}{-1}}{}
\section{Закрученный светоподобный солитон}
\mifeng{}{\label{twlls}}
}
\begin{subequations}\label{462088071}
\engrus{0.5ex}{0.5ex}{%
For further calculations, we define the arbitrary functions $\Xi_{1}$ and $\Xi_{2}$.
Let us take power function with integer negative exponent
\textaddb{for the tilde variables $\{\mtxi_1,\mtxi_2\}$}.
Introducing  \textremb{necessary }\textaddb{also the appropriate} multiplicative \textaddb{complex} constants
\textaddb{$\{\cbxi_1,\cbxi_2\}$ and considering the necessity} for concordance of physical dimension,
\textremb{and for simplification of resulting formulas }we \textremb{have }\textaddb{can write}
}{%
Для дальнейших вычислений мы определяем произвольные функции $\Xi_{1}$ и $\Xi_{2}$.
Возьмём степенную функцию с целым отрицательным показателем для
тильдовых
переменных $\mtxi_1$ и $\mtxi_2$.
Вводя также соответствующие мультипликативные комплексные константы $\{\cbxi_1,\cbxi_2\}$ и учитывая необходимость согласования физической размерности,
можем написать
}
\begin{alignat}{2}
\label{462409051}
\Xi_{1} &= \frac{\cbxi_1^{m+1}}{\xxx\,m}\,\mtxi_1^{-m}\;,\quad  &  \Xi_{2} &=
\frac{\cbxi_2^{m+1}}{\xxx\,m}\, \mtxi_2^{-m}
\;,\\
\label{462409052}
\Xi_{1}^{\prime}  &=
-\frac{1}{\xxx}\biggl(\frac{\cbxi_1}{\mtxi_1}\biggr)^{\!\! m+1}
\;,\quad  &
\Xi_{2}^{\prime} &=
-\frac{1}{\xxx}\biggl(\frac{\cbxi_2}{\mtxi_2}\biggr)^{\!\! m+1}
\;,
\end{alignat}
\engrus{0.5ex}{0.5ex}{%
\noindent
where $m$ is natural number.
}{%
\noindent
где $m$ -- натуральное число.
}
\end{subequations}

\engrus{0.5ex}{0.5ex}{%
Then \textadd1{the} formula
(\ref{755070621}) representing the solution of equation (\ref{436283681})
\textremb{has the form }\textaddb{gives the following expression:}
}{%
Тогда формула (\ref{755070621}), представляющая решение уравнения (\ref{436283681}), даёт следующее выражение:
}
\begin{equation}
\label{519662391}
\ffun  = \frac{1}{\xxx\,m}\left(\cbxi_1^{m+1}\;\mtxi_1^{-m} + \cbxi_2^{m+1}\;\mtxi_2^{-m}\right)
\;.
\end{equation}

\engrus{0.5ex}{0.5ex}{%
By analogy with derivation of formulas (\ref{492600841}), (\ref{609493561}), and (\ref{40528090}), we obtain the following relations:
}{%
По аналогии с выводом формул (\ref{492600841}), (\ref{609493561}) и (\ref{40528090}) получаем следующие соотношения:
}
\begin{alignat}{2}
\label{369128871}
 \df\mxi &= \df\mtxi_1 {}\pm{} \biggl(\dfrac{\cbxi_2}{\mtxi_2}\biggr)^{\!\!2(m+1)}\!\!\df\mtxi_2
 \;,\quad   &
\df\mxic  &= \pm\biggl(\dfrac{\cbxi_1}{\mtxi_1}\biggr)^{\!\!2(m+1)}\!\!\df\mtxi_1 + \df\mtxi_2
\;,\\
\label{369128870}
&
\phantom{......................} \biggl(\frac{\cbxi_1\,\cbxi_2}{\mtxi_1\,\mtxi_2}\biggr)^{\!\!2(m+1)}  &\neq 1
&
\;,\\
\label{369128872}
\mxi &= \mtxi_1 \mp \frac{\cbxi_2^{2(m+1)}}{\left(2\, m + 1\right)\mtxi_2^{2\,m+1}}
\;,\quad   &
\mxic    &= \mtxi_2 \mp \frac{\cbxi_1^{2(m+1)}}{\left(2\, m + 1\right)\mtxi_1^{2\,m+1}}
\;.
\end{alignat}
\engrus{0.5ex}{0.5ex}{%
It should be noted that the formulas (\ref{462409052}) and (\ref{369128871})--(\ref{369128872}) for \mbox{$m = 0$} give the formulas
(\ref{451959281p}) and (\ref{492600841})--(\ref{40528090}) accordingly.
}{%
Надо заметить, что при \mbox{$m = 0$} формулы (\ref{462409052}) и (\ref{369128871})--(\ref{369128872}) переходят в формулы
(\ref{451959281p}) и (\ref{492600841})--(\ref{40528090}) соответственно.
}

\begin{subequations}\label{323085231}
\engrus{0.5ex}{0.5ex}{%
Without restricting generality we can assume that the complex constants $\{\cbxi_1,\cbxi_2\}$ have identical modulus $\cbrho$ with physical dimension of length.
Let us represent it in the following form:
}{%
Без ограничения общности можем считать, что комплексные константы $\{\cbxi_1,\cbxi_2\}$ имеют одинаковый модуль $\cbrho$ с физической размерностью длины.
Представим их в следующем виде:
}
\begin{alignat}{2}
\label{43196058}
\cbxi_1  &= \cbrho\;\e{\him\,\cbphi_1}
\;,\quad  &
\cbxi_2  &= \cbrho\;\e{\him\,\cbphi_2}
\;.
\end{alignat}

\engrus{0.5ex}{0.5ex}{%
According to the asymptotic condition (\ref{290961541}), the tilde variables are mutually complex conjugated
\mbox{$\{\mtxi_1,\mtxi_2\} = \{\mtxi,\hconj{\mtxi}\}$} (\ref{448940041}) in the region including infinity.
For this case the phase constants $\{\cbphi_1,\cbphi_2\}$ in (\ref{43196058}) obey the relation
}{%
Согласно асимптотическому условию (\ref{290961541}), в области, включающей бесконечность,
тильдовые переменные являются взаимно комплексно-сопря\-жён\-ными \mbox{$\{\mtxi_1,\mtxi_2\} = \{\mtxi,\hconj{\mtxi}\}$} (\ref{448940041}), а фазовые константы $\{\cbphi_1,\cbphi_2\}$ в (\ref{43196058}) подчиняются соотношению
}
\begin{equation}
\label{322598351}
 \cbphi_1 = -\cbphi_2 = \cbphi
\;.
\end{equation}
\end{subequations}

\engrus{0.5ex}{0.5ex}{%
Taking into account the scale-rotation transformation (\ref{592114861}) we must make the following changes in the formulas \mbox{(\ref{462088071})--(\ref{369128872})}:
}{%
Учёт масштабно-поворотного преобразования (\ref{592114861}) требует следующих замен в формулах \mbox{(\ref{462088071})--(\ref{369128872})}:
}
\begin{equation}
\label{458615411}
\ffun\;\to\;\btam\,\ffun
\;,\quad
\mxi\;\to\;\frac{\e{-\him\,\btap}}{\btam}\;\mxi
\;,\quad
\mxic\;\to\;\frac{\e{\him\,\btap}}{\btam}\,\mxic
\;.
\end{equation}

\engrus{0.5ex}{0.5ex}{%
Because \mbox{$\xi\sim \mtxi$} and \mbox{$\hconj{\xi}\sim \mtcxi$} at
\textaddb{infinity}
\mbox{$\rho\to\infty$}, we have from (\ref{519662391}),
\textremb{with the scale-rotation transformation (\ref{592114861}) }\textaddb{(\ref{323085231}), and (\ref{458615411})}
the following asymptotic solution:
}{%
Поскольку \mbox{$\xi\sim \mtxi$} и \mbox{$\hconj{\xi}\sim \mtcxi$} на бесконечности
\mbox{$\rho\to\infty$}, имеем из (\ref{519662391}), (\ref{323085231}) и (\ref{458615411}) следующее асимптотическое решение:
}
\begin{equation}
\label{549399221}
\ffun \, \sim\, \frac{2\left(\btam\,\cbrho\right)^{m+1}}{\xxx\,m\,\rho^{m}}\,
\cos\bigl(m\left(\varphi - \btap\right) - \left(m + 1\right)\cbphi\bigr)
\quad\text{\mifeng{at}{при}}\quad
{\rho}\to\infty
\;.
\end{equation}

\engrus{0.5ex}{0.5ex}{%
In view of dependence on phase  $\btam (\theta)$ and $\btap (\theta)$, the formula (\ref{549399221})
describes the propagating wave along the $x^{3}$ axis. The dependence $\btap (\theta)$
in (\ref{549399221}) describes also the twist of this wave about the propagation direction.
}{%
Ввиду зависимости от фазы $\btam (\theta)$ и $\btap (\theta)$ формула (\ref{549399221})
описывает волну, распространяющуюся вдоль оси $x^{3}$. Зависимость $\btap (\theta)$
в (\ref{549399221}) описывает также закрученность этой волны относительно направления распространения.
}

\engrus{0.5ex}{0.5ex}{%
Let us consider the twisted lightlike soliton with
\textremb{condition }\textaddb{constant twist} (\ref{667163021}).
We \textaddb{can} put for this case
}{%
Рассмотрим закрученный светоподобный солитон с постоянной закрученностью (\ref{667163021}).
Мы можем положить для этого случая
}
\begin{equation}
\label{519019131}
 \btap = \pm\frac{\Phase}{m}
\;,
\end{equation}
\engrus{0.5ex}{0.5ex}{%
\noindent
where the signs '$+$' and '$-$' correspond to right and left twisted soliton accordingly.
}{%
\noindent
где знаки '$+$' и '$-$' отвечают правому и левому закрученному солитону соответственно.
}

\engrus{0.5ex}{0.5ex}{%
In addition,
let us \textremb{consider }\textaddb{assume} that the scale function $\btam (\theta)$
is almost constant: $\btam^{\prime} \sim 0$. As we can see in (\ref{549399221}) with (\ref{519019131}),
in this case  $\omega$ in (\ref{426404172}) is \textadd1{a} radian frequency of the soliton wave \textremb{and }\textaddb{with the wave length}
$2\pi/\left|k_{3}\right|$. \textremb{is the appropriate wave length.}
}{%
Кроме того предположим, что масштабная функция $\btam (\theta)$
почти постоянна: $\btam^{\prime} \sim 0$. Как мы можем видеть в  (\ref{549399221}) с (\ref{519019131}),
в этом случае  $\omega$ в (\ref{426404172}) представляет собой круговую частоту солитонной волны
с длиной волны $2\pi/\left|k_{3}\right|$.
}

\engrus{0.5ex}{0.5ex}{%
Now let us consider the projection of the singular surface of the lightlike soliton  to the transverse plane, that is a line on which the condition (\ref{369128870}) is violated.
}{%
Рассмотрим теперь проекцию сингулярной поверхности солитона на поперечную
плоскость, то есть линию, на которой нарушается условие (\ref{369128870}).
}

\engrus{0.5ex}{0.5ex}{%
In the plane of the tilde variables \mbox{$\{\mtxi,\hconj{\mtxi}\}$}, the singular line is a circle of radius $\cbrho$,
 just as for the cylindrically symmetric solution considered in the previous section (\ref{523967371}):
}{%
В плоскости тильдовых переменных сингулярная линия представляет собой окружность радиуса
$\cbrho$,
также как и для для цилиндрически
симметричного решения, рассмотренного в предыдущей секции (\ref{523967371}):
}
\begin{equation}
\label{473309241}
\tilde{\rho}  = \cbrho
\;.
\end{equation}

\engrus{0.5ex}{0.5ex}{%
It can be shown that, the action density (\ref{423072731}) vanishes or becomes infinite  on this line for the signatures of metric (\ref{43842964a}) and (\ref{43842964b})
\textrem1{accordingly }\textadd1{respectively}.
Also, accordingly, the vector components $\ffind^{\mu}$ (\ref{421005381}) or $\dffun^{\mu}$ (\ref{450528581}) become infinite on it.
}{%
Можно показать, что плотность действия (\ref{423072731}) обращается в ноль или
 бесконечность на этой линии для сигнатур метрики (\ref{43842964a}) и (\ref{43842964b}) соответственно.
Также, соответственно, векторные компоненты $\ffind^{\mu}$ (\ref{421005381}) или $\dffun^{\mu}$ (\ref{450528581}) обращаются на ней в бесконечность.
}

\engrus{0.5ex}{0.5ex}{%
In the plane $\{\mxi,\mxic\}$,
in conformity with \mbox{(\ref{369128872})--(\ref{458615411})} and (\ref{473309241}),
 the singular line
is described by \textadd1{the} formula
}{%
В плоскости $\{\mxi,\mxic\}$,
в соответствии с \mbox{(\ref{369128872})--(\ref{458615411})} и (\ref{473309241}),
   сингулярная линия описывается формулой
}
\begin{equation}
\label{402041591}
\mxis  = \cbrho\;\e{\him\left(\tilde{\varphi} + \btap\right)}
\Biggl( 1 \mp \frac{\e{\him\,2\left(\vphantom{I^{I}}m\,\tilde{\varphi}\, {}-{}\,
(m + 1)\,
\cbphi\right)}}{2\, m + 1}
\Biggr)
\;.
\end{equation}
\engrus{0ex}{0.5ex}{%
\noindent
Here the function $\mxis = \mxis (\tilde{\varphi})$ represents a parametric expression for the singular line in the complex plane
of \textaddb{the} variable $\mxi$.
The phases $\btap$ and $\cbphi$ define a rotating of this closed curve
as a whole.
}{%
\noindent
Здесь функция $\mxis = \mxis (\tilde{\varphi})$ представляет параметрическое выражение сингулярной линии на комплексной плоскости переменной  $\mxi$.
Фазы $\btap$ и $\cbphi$ задают поворот этой замкнутой кривой как целого.
}

\engrus{0.5ex}{0.5ex}{%
\textremb{Expression }\textaddb{The curve $\mxis (\tilde{\varphi})$} (\ref{402041591}) \textremb{represents }\textaddb{is an}  epicycloid with $2\,m$ cusps.
For $m=1$ this line is shown in Fig. \ref{37607901}.
 These figures was obtained also by R.~Ferraro \cite{FerraroRafaelHepTh0309185,FerraroRafael1007p2651}
for mathematically similar but another problem.
}{%
Кривая $\mxis (\tilde{\varphi})$ (\ref{402041591}) представляет собой эпициклоиду с  $2\,m$ остриями.
Для $m=1$ эта линия показана на Рис. \ref{37607901r}.
Эти фигуры были получены также Р.~Ферраро \cite{FerraroRafaelHepTh0309185,FerraroRafael1007p2651} для математически сходной, но
другой задачи.
}

\begin{figure}[h]
\begin{center}
\ifpdf
  {
  \unitlength 1mm
  \begin{pspicture}(0,0)(12,6.2)
    \put(0.3,-0.7){\includegraphics[width=45mm]{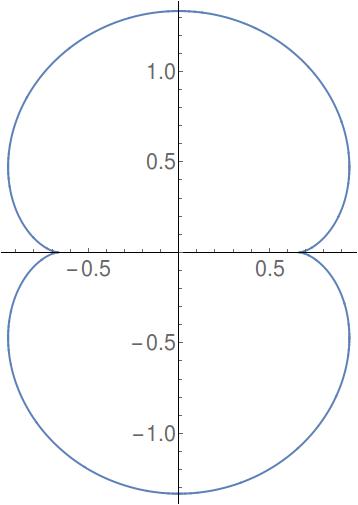}}
   \put(5.4,0.25){\includegraphics[width=63mm]{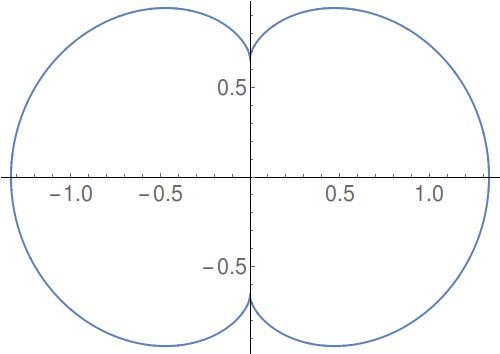}}
      \put(0.5,0){
    \put(0.4,2.6){${\mathcal B}_{1}$}
    \put(3.2,2.6){${\mathcal A}_{1}$}
    \put(3.7,-0.2){$\metrp^{00} = 1$}
    \put(10.6,2.1){${\mathcal C}_1$}
    \put(5.2,2.6){${\mathcal D}_{1}$}
    \put(7.4,-0.2){$\metrp^{00} = -1$}
    \put(2.2,5.6){$x^{2}$}
    \put(4.1,2.55){$x^{1}$}
    \put(8.15,4.65){$x^{2}$}
    \put(11.15,2.55){$x^{1}$}
   }
   \put(-3.5,2.2){
    \begin{picture}(3,2.5)
    \put(8,3.75){$m = 1$}
    \put(8,3.3){$\cbrho = 1,$}
    \put(9.1,3.3){$\xxx = 1$}
    \put(8,2.85){$\btam = 1,$}
    \put(9.1,2.85){$\btap = 0,$}
    \put(10.25,2.85){$\cbphi = 0$}
    \end{picture}
   }
 \end{pspicture}
  }
 \else
  \begin{pspicture}(0,0)(12,6.2)
   \put(0.5,0){
    \put(0.4,2.6){${\mathcal B}_{1}$}
    \put(2.9,2.6){${\mathcal A}_{1}$}
    \put(3.7,-0.2){$\metrp^{00} = 1$}
    \put(10.7,2.1){${\mathcal C}_1$}
    \put(5.3,2.6){${\mathcal D}_{1}$}
    \put(7.4,-0.2){$\metrp^{00} = -1$}
    \put(2.2,5.6){$x^{2}$}
    \put(4.1,2.55){$x^{1}$}
    \put(8.15,4.65){$x^{2}$}
    \put(11.15,2.55){$x^{1}$}
   }
   \put(-3.5,2.2){
    \begin{picture}(3,2.5)
    \put(8,3.75){$m = 1$}
    \put(8,3.3){$\cbrho = 1,$}
    \put(9.1,3.3){$\xxx = 1$}
    \put(8,2.85){$\btam = 1,$}
    \put(9.1,2.85){$\btap = 0,$}
    \put(10.25,2.85){$\cbphi = 0$}
    \end{picture}
   }
  \put(0.3,-0.7){\includegraphics[width=45mm]{LSOESTF_Fig1a.eps}}
  \put(5.4,0.25){\includegraphics[width=63mm]{LSOESTF_Fig1b.eps}}
  \end{pspicture}
\fi
\end{center}
\engrus{0.5ex}{0.5ex}{%
 \caption{\label{37607901} Singular line on the plane $\{x^{1},x^{2}\}$ with \textaddb{the} parameter \mbox{$m = 1$}.}
}{%
\mifeng{\addtocounter{figure}{-1}}{}
 \caption{\label{37607901r} Сингулярная линия на плоскости $\{x^{1},x^{2}\}$ со значением параметра \mbox{$m = 1$}.}
}
\end{figure}

\begin{figure}[h]
\begin{center}
\ifpdf
  {
  \unitlength 1mm
  \begin{pspicture}(0,0)(12,7)
   \put(0,-0.4){\includegraphics[width=55mm]{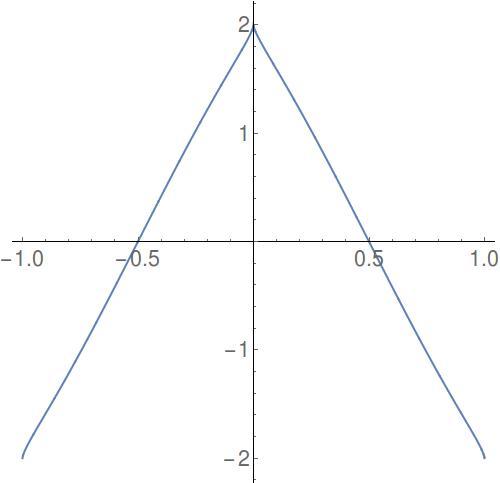}}
   \put(6.2,-0.4){\includegraphics[width=55mm]{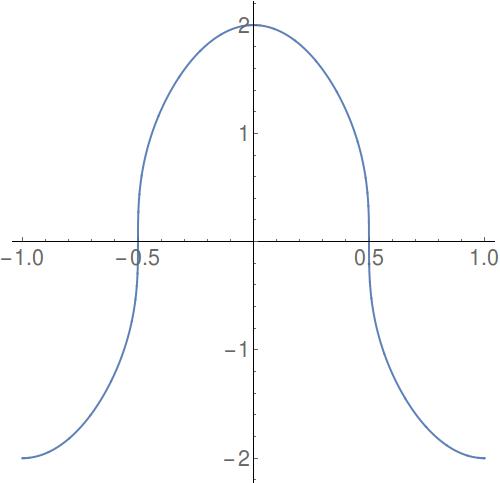}}
   \put(2.9,4.95){$\ffun (\mxis)$}
   \put(3.1,4.4){${\mathcal A}_{1}$}
   \put(4.8,-0.15){${\mathcal B}_{1}$}
   \put(0.4,-0.15){${\mathcal B}_{1}$}
   \put(9.1,4.95){$\ffun (\mxis)$}
   \put(9.1,4.3){${\mathcal C}_{1}$}
   \put(11.3,0.1){${\mathcal D}_{1}$}
   \put(6.3,0.1){${\mathcal D}_{1}$}
   \put(3.0,-0.6){$\metrp^{00} = 1$}
   \put(9.3,-0.6){$\metrp^{00} = -1$}
   \put(-3.8,0.5){
    \begin{picture}(3,2.5)
    \put(8,3.75){$m = 1$}
    \put(8,3.3){$\cbrho = 1,$}
    \put(9.1,3.3){$\xxx = 1$}
    \put(8,2.85){$\btam = 1,$}
    \put(9.1,2.85){$\btap = 0,$}
    \put(10.25,2.85){$\cbphi = 0$}
    \end{picture}
   }
   \put(11,2.5){$\varphi/\pi$}
   \put(4.8,2.5){$\varphi/\pi$}
   \end{pspicture}
   }
 \else
  \begin{pspicture}(0,0)(12,5)
   \put(0,-0.4){\includegraphics[width=55mm]{LSOESTF_Fig2a.eps}}
   \put(6.2,-0.4){\includegraphics[width=55mm]{LSOESTF_Fig2b.eps}}
   \put(2.9,4.95){$\ffun (\mxis)$}
   \put(3.1,4.4){${\mathcal A}_{1}$}
   \put(4.8,-0.15){${\mathcal B}_{1}$}
   \put(0.4,-0.15){${\mathcal B}_{1}$}
   \put(9.1,4.95){$\ffun (\mxis)$}
   \put(9.1,4.3){${\mathcal C}_{1}$}
   \put(11.3,0.1){${\mathcal D}_{1}$}
   \put(6.3,0.1){${\mathcal D}_{1}$}
   \put(3.0,-0.6){$\metrp^{00} = 1$}
   \put(9.3,-0.6){$\metrp^{00} = -1$}
   \put(-3.8,0.5){
  \begin{picture}(3,2.5)
   \put(8,3.75){$m = 1$}
   \put(8,3.3){$\cbrho = 1,$}
   \put(9.1,3.3){$\xxx = 1$}
   \put(8,2.85){$\btam = 1,$}
   \put(9.1,2.85){$\btap = 0,$}
   \put(10.25,2.85){$\cbphi = 0$}
  \end{picture}
  }
   \put(11,2.5){$\varphi/\pi$}
   \put(4.8,2.5){$\varphi/\pi$}
  \end{pspicture}
\fi
\end{center}
  \engrus{0.5ex}{0.5ex}{%
  \caption{\label{83116575} The field function $\ffun$ on the singular line of the plane $\{x^{1},x^{2}\}$ for \mbox{$m = 1$}.}
  }{%
  \mifeng{\addtocounter{figure}{-1}}{}
  \caption{\label{83116575r} Полевая функция $\ffun$ на сингулярной линии в плоскости $\{x^{1},x^{2}\}$ для \mbox{$m = 1$}.}
  }
\end{figure}

\engrus{0.5ex}{0.5ex}{%
\textremd{A}\textaddd{The} change of the signature of metric in (\ref{402041591}) from (\ref{43842964a}) (top sign, \mbox{$\metrp^{00} = 1$})
to (\ref{43842964b}) (bottom sign, \mbox{$\metrp^{00} = -1$}) implies a turn of the whole curve to the angle $\pi/(2\,m)$.
}{%
Изменение сигнатуры метрики в (\ref{402041591}) с (\ref{43842964a}) (верхний знак, \mbox{$\metrp^{00} = 1$})
на (\ref{43842964b}) (нижний знак, \mbox{$\metrp^{00} = -1$}) влечёт за собой поворот
всей кривой на угол $\pi/(2\,m)$.
}

\engrus{0.5ex}{0.5ex}{%
In the present investigation, the system (\ref{369128872}) with \mbox{$\cbrho = 1$} for given values of the parameter $m$ and the variables $\{\mxi,\mxic\}$
is solved numerically with respect to the variables $\{\mtxi_1,\mtxi_2\}$ in all characteristic areas of the plane $\{x^1,x^2\}$.
}{%
В настоящем исследовании система уравнений (\ref{369128872}) с \mbox{$\cbrho = 1$} для заданных значений параметра $m$ и переменных $\{\mxi,\mxic\}$
решалась численно относительно переменных $\{\mtxi_1,\mtxi_2\}$ во всех характерных областях плоскости $\{x^1,x^2\}$.
}

\engrus{0.5ex}{0.5ex}{%
In the \textremb{area }\textaddb{region} of the plane $\{x^1,x^2\}$ outside of the singular line (\ref{402041591}),  we have one-to-one mapping
$\{\mtxi,\mtcxi\}\Longleftrightarrow\{\mxi,\mxic\}$ with the condition (\ref{290961541}) at infinity \mbox{$\rho\to\infty$}.
}{%
В области плоскости $\{x^1,x^2\}$ вне сингулярной линии (\ref{402041591}) имеется вза\-имно-однозначное отображение
\mbox{$\{\mtxi,\mtcxi\}\Longleftrightarrow\{\mxi,\mxic\}$} при условии (\ref{290961541}) на бесконечности \mbox{$\rho\to\infty$}.
}

\engrus{0.5ex}{0.5ex}{%
\textaddd{But} the solutions of equations (\ref{369128872}) give a multivalued mapping $\{\mxi,\mxic\}\Longrightarrow\{\mtxi_1,\mtxi_2\}$
in the interior of the singular line (\ref{402041591}) on the plane $\{x^1,x^2\}$.
}{%
Однако в области плоскости $\{x^1,x^2\}$ внутри сингулярной линии (\ref{402041591}) решения уравнений (\ref{369128872}) дают
многозначное отображение  $\{\mxi,\mxic\}\Longrightarrow\{\mtxi_1,\mtxi_2\}$.
}

\engrus{0.5ex}{0.5ex}{%
The class of solutions under consideration (\ref{462088071}) is characterized by zero value of the field function $\ffun$ at the
origin of the coordinates $\{x^1,x^2\}$.
\textremd{Using}\textaddd{In the scope of relations} (\ref{462088071}) -- (\ref{369128872}), it is possible to make a composite \textaddd{continuous everywhere} solution
\textremd{with}\textaddd{consisting of}
sectorial pieces and satisfying
this condition \textaddd{$\ffun (0,0) = 0$}.
\textremd{This composite solution is continuous everywhere}But its absolute value increases at infinity ($\rho\to\infty$).
}{%
Класс рассматриваемых решений (\ref{462088071}) характеризуется нулевым значением
полевой функции $\ffun$ в начале координат $\{x^1,x^2\}$. В рамках соотношений (\ref{462088071}) -- (\ref{369128872}) оказывается возможным составить из секториальных кусков
всюду непрерывное решение, удовлетворяющее этому условию $\ffun (0,0) = 0$.
Однако оно возрастает по абсолютной величине на бесконечности ($\rho\to\infty$).
}

\engrus{0.5ex}{0.5ex}{%
The values of the field function $\ffun$ on the line (\ref{402041591}) for this
\textaddd{inner}
solution are distinct from
ones for the \textaddd{external} solution with asymptotics (\ref{549399221}).
Thus the \textremd{appropriate}composite solution
\textremd{has}\textaddd{consisting of these external and inner ones would have} a finite discontinuity on the singular line (\ref{402041591}).
}{%
Значения полевой функции $\ffun$ на линии (\ref{402041591}) для этого внутреннего решения  отличаются от значений полевой функции для внешнего решения
с асимптотикой (\ref{549399221}).
Таким образом
%соответствующее составное
решение,
составленное из этих внешнего и внутреннего,
 имело бы конечный разрыв на сингулярной линии (\ref{402041591}).
}

\engrus{0.5ex}{0.5ex}{%
But in this case we \textremd{can}\textaddd{could} take also the trivial solution \textaddd{$\ffun = 0$} for the inner area of the singular line (\ref{402041591}).
}{%
Но в таком случае мы могли бы взять также тривиальное решение $\ffun = 0$ для внутренней
области сингулярной линии (\ref{402041591}).
}

\engrus{0.5ex}{0.5ex}{%
To obtain an everywhere continuous solution, the various kinds of composite solutions can be used.
One of them will be considered below (see Fig. \ref{37607999} and Fig. \ref{4755348855}).
}{%
Для получения всюду непрерывного решения могут использоваться различные типы составных решений. Одно из них будет рассмотрено ниже (см. Рис. \ref{37607999r} и Рис. \ref{4755348855r}).
}

\engrus{0.5ex}{0.5ex}{%
Apart from this one case we shall consider the solutions of the
exterior domain with respect to the singular line (\ref{402041591}).
The region of space in the interior of the singular line will be excluded from the space of the problem.
In this case we have a soliton with an appropriate cavity.
Such soliton has all characteristic properties of \textremd{a}\textaddd{the} twisted soliton.
It is possible to generate of such solitons, in particular, by sources having cavities also.
}{%
Кроме этого одного случая мы будем рассматривать только решения внешней по отношению к линии (\ref{402041591}) области.
Область пространства внутри сингулярной линии будет исключена из пространства задачи.
В этом случае мы имеем солитон с соответствующей полостью.
Такой солитон обладает всеми характерными особенностями вращающегося солитона.
Образование таких солитонов возможно, в частности, источниками, также имеющими полости.
}

\engrus{0.5ex}{0.5ex}{%
Thus,
according to (\ref{473309241}) and (\ref{402041591}),
we have the following
condition for the space of the solution:
}{%
Таким образом
согласно (\ref{473309241}) и (\ref{402041591})
имеем следующее условие для пространства решения:
}
\begin{equation}
\label{433253941}
 \rho \geqslant \left|\mxis\bigl(\btap (\theta)\bigr)\right|\,\btam (\theta)
 \;,\quad
 \tilde{\rho} \geqslant \cbrho
\;,
\end{equation}
\engrus{0.5ex}{0.5ex}{%
\noindent
where \textadd1{the} dependence \mbox{$\mxis = \mxis\bigl(\btap (\theta)\bigr)$} corresponds
to \textadd1{the} rotation of the
singular contour in the plane $\{x^{1},x^{2}\}$ by the angle $\btap (\theta)$ (\ref{591083061}).
}{%
\noindent
где зависимость \mbox{$\mxis = \mxis\bigl(\btap (\theta)\bigr)$} соответствует вращению сингулярного контура в плоскости\break
 $\{x^{1},x^{2}\}$ на угол $\btap (\theta)$ (\ref{591083061}).
}

\engrus{0.5ex}{0.5ex}{%
Thus, according to (\ref{433253941}), we have the soliton with an inner shell.
}{%
Таким образом согласно (\ref{433253941}) имеем солитон с внутренней оболочкой.
}

\engrus{0.5ex}{0.5ex}{%
\textaddd{It must be emphasized that the expression under modulus in the action density (\ref{383322881})
is non-negative for these solutions.
Therefore they are free from the features connected with a
%transition of this expression through zero value.
change of the sign of this expression.
}
}{%
Надо отметить, что для этих решений модели выражение под знаком модуля в плотности
действия (\ref{383322881}) неотрицательно.
Следовательно они свободны от особенностей, связанных с
%переходом этого выражения через нулевое значение.
изменением знака этого выражения.
}

\engrus{0.5ex}{0.5ex}{%\mifrus{}{}%\noindent%
\textaddd{In what follows we consider just such solitons and they will be correlated
with real photons in section \ref{relphot}.
}
}{%\mifeng{}{}%\noindent%
Далее именно подобные солитоны мы рассматриваем, а в секции \ref{relphot}
они будут сопоставлены с реальными фотонами.
}

\engrus{0.5ex}{0.5ex}{%
The results of numerical calculations for the field function $\ffun$ (\ref{519662391})
for \mbox{$m=1$} on the singular line are shown in Fig. \ref{83116575}.
The appropriate results for the field function $\ffun$ on the plane
$\{x^{1},x^{2}\}$ are shown in Fig. \ref{53247958} and Fig. \ref{53247959}.
The points $\{\mathcal{A}_{1},\,\mathcal{B}_{1},\,\mathcal{C}_{1},\,\mathcal{D}_{1}\}$
are corresponding in Figures  \ref{519662391} -- \ref{53247959}.
}{%
Результаты численных расчётов для полевой функции $\ffun$ (\ref{519662391})
при \mbox{$m=1$} на сингулярной линии  показаны на Рис. \ref{83116575r}.
Соответствующие результаты для полевой функции $\ffun$ на плоскости
$\{x^{1},x^{2}\}$ показаны на Рис. \ref{53247958r} и Рис. \ref{53247959r}.
Точки $\{\mathcal{A}_{1},\,\mathcal{B}_{1},\,\mathcal{C}_{1},\,\mathcal{D}_{1}\}$
являются соответственными на Рис. \ref{83116575r} -- \ref{53247959r}.
}

\engrus{0.5ex}{0.5ex}{%
It should be noted that the values of the parameters $\cbphi$ and $\btap$,
which are contained in the formulas
(\ref{323085231}) and (\ref{458615411}), define a
turning
of the whole field configuration only and do not change the solution qualitatively.
}{%
Надо отметить, что значения параметров $\cbphi$ и $\btap$, входящих формулы
(\ref{323085231}) и (\ref{458615411}), задают только поворот всей полевой конфигурации как целого и качественно не меняют решение.
}

\engrus{0.5ex}{0.5ex}{%
To construct an everywhere continuous solution, let us consider two solutions with shifted singular line.
In these cases the constants are added to the variables $\{\mxi,\mxic\}$ in the left-hand parts of the equations (\ref{369128872}). The appropriate singular lines are shown in Fig. \ref{37607999}.
The two parts of the composed solution are taken from those half-planes, where the singular line is absent
(see Fig. \ref{37607999}).
The appropriate field function in the whole plane $\{x^{1},x^{2}\}$ is shown in Fig. \ref{4755348855}.
}{%
Для построения всюду непрерывного решения рассмотрим два решения со смещённой сингулярной линией. В этих случаях
к переменным  $\{\mxi,\mxic\}$ в левых частях уравнений (\ref{369128872}) добавляются константы.
Соответствующие сингулярные линии показаны на Рис. \ref{37607999r}. Две части составного решения берутся из
тех полуплоскостей, в которых нет сингулярной линии (см. Рис. \ref{37607999r}). Соответствующая
полевая функция на всей плоскости $\{x^{1},x^{2}\}$ показана на  Рис. \ref{4755348855r}.
}

\engrus{0.5ex}{0.5ex}{%
The field functions of the shell soliton for $m = 2$ in the plane $\{x^{1},x^{2}\}$ are shown in Fig. \ref{35836648}  and Fig. \ref{35836659}.
}{%
Полевые функции оболочечного солитона для $m = 2$ в плоскости $\{x^{1},x^{2}\}$ показаны на Рис. \ref{35836648r}  и Рис. \ref{35836659r}.
}

\begin{figure}[h]
\begin{center}
\ifpdf
  {
  \begin{pspicture}(0,0)(12,5)
\put(-0.3,-1.2){\includegraphics[width=125mm]{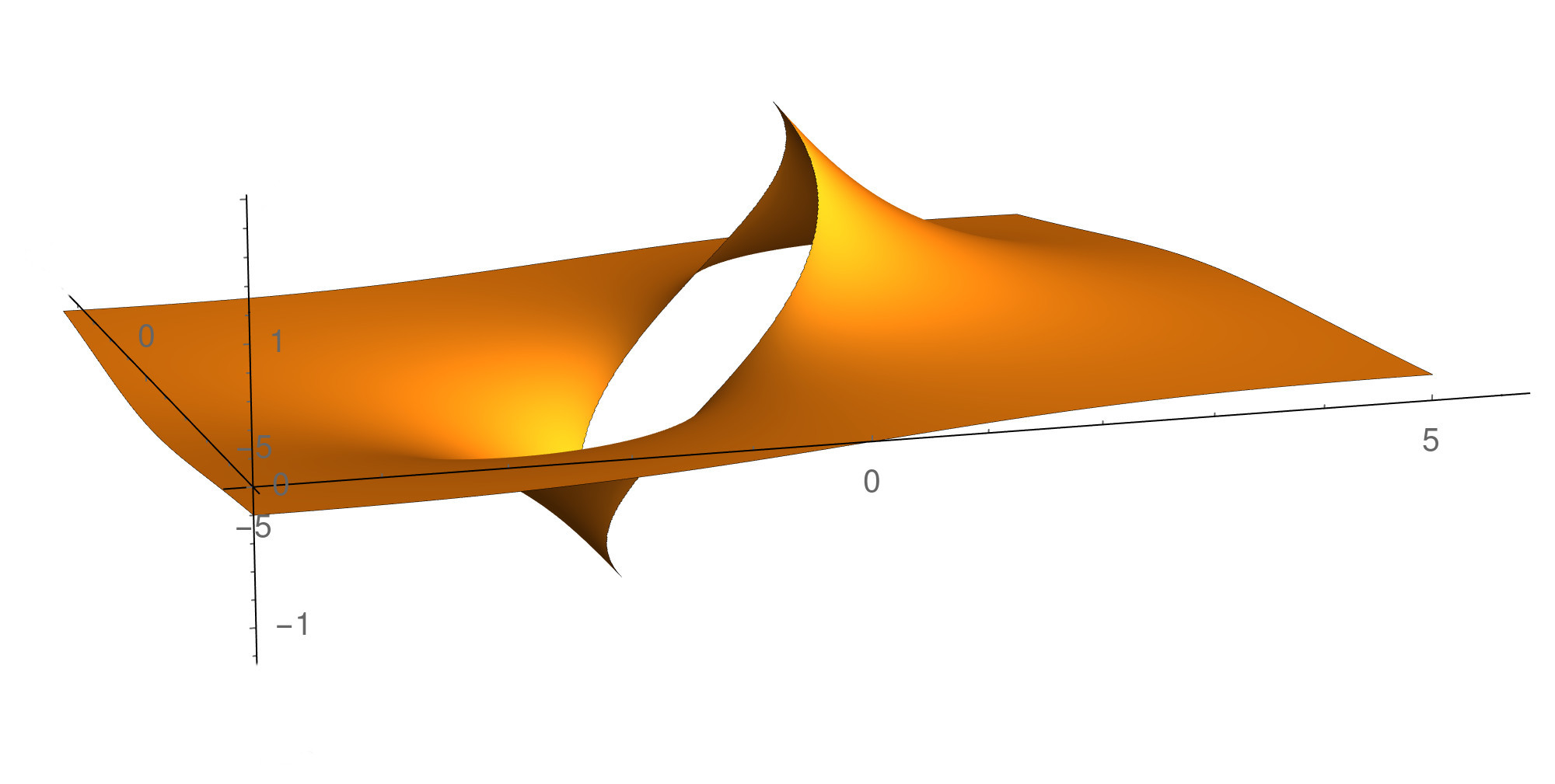}}
  \put(1.8,3.4){$\ffun$}
  \put(5.45,3.95){${\mathcal A}_{1}$}
  \put(4.85,0.35){${\mathcal B}_{1}$}
 \put(0.4,-3.0){
  \begin{picture}(2,2.5)
   \put(8,3.75){$m = 1,$}
   \put(9.2,3.75){$\metrp^{00} = 1$}
   \put(8,3.3){$\cbrho = 1,$}
   \put(9.1,3.3){$\xxx = 1$}
   \put(8,2.85){$\btam = 1,$}
   \put(9.1,2.85){$\btap = 0,$}
   \put(10.25,2.85){$\cbphi = 0$}
  \end{picture}
  }
  \put(0.15,2.9){$x^{2}$}
  \put(11.5,2.0){$x^{1}$}
  \end{pspicture}
  }
 \else
  \begin{pspicture}(0,0)(12,5)
  \put(1.8,3.4){$\ffun$}
  \put(5.45,3.95){${\mathcal A}_{1}$}
  \put(4.85,0.35){${\mathcal B}_{1}$}
  \put(0.4,-3.0){
  \begin{picture}(2,2.5)
   \put(8,3.75){$m = 1,$}
   \put(9.2,3.75){$\metrp^{00} = 1$}
   \put(8,3.3){$\cbrho = 1,$}
   \put(9.1,3.3){$\xxx = 1$}
   \put(8,2.85){$\btam = 1,$}
   \put(9.1,2.85){$\btap = 0,$}
   \put(10.25,2.85){$\cbphi = 0$}
  \end{picture}
  }
  \put(0.15,2.9){$x^{2}$}
  \put(11.5,2.0){$x^{1}$}
   \put(-0.3,-1.2){\includegraphics[width=125mm]{LSOESTF_Fig53p.eps}}
  \end{pspicture}
\fi
\end{center}
  \engrus{0.5ex}{0.5ex}{%
  \caption{\label{53247958}The field function $\ffun$ on the plane $\{x^{1},x^{2}\}$ for $m = 1$ and
  \mbox{$\metrp^{00} = 1$}.}
  }{%
  \mifeng{\addtocounter{figure}{-1}}{}
  \caption{\label{53247958r}Полевая функция $\ffun$ в плоскости $\{x^{1},x^{2}\}$ для $m = 1$ и \mbox{$\metrp^{00} = 1$}.}
  }
\end{figure}

\begin{figure}[h]
\begin{center}
\ifpdf
  {
  \begin{pspicture}(0,0)(12,5)
\put(-0.3,-1.2){\includegraphics[width=125mm]{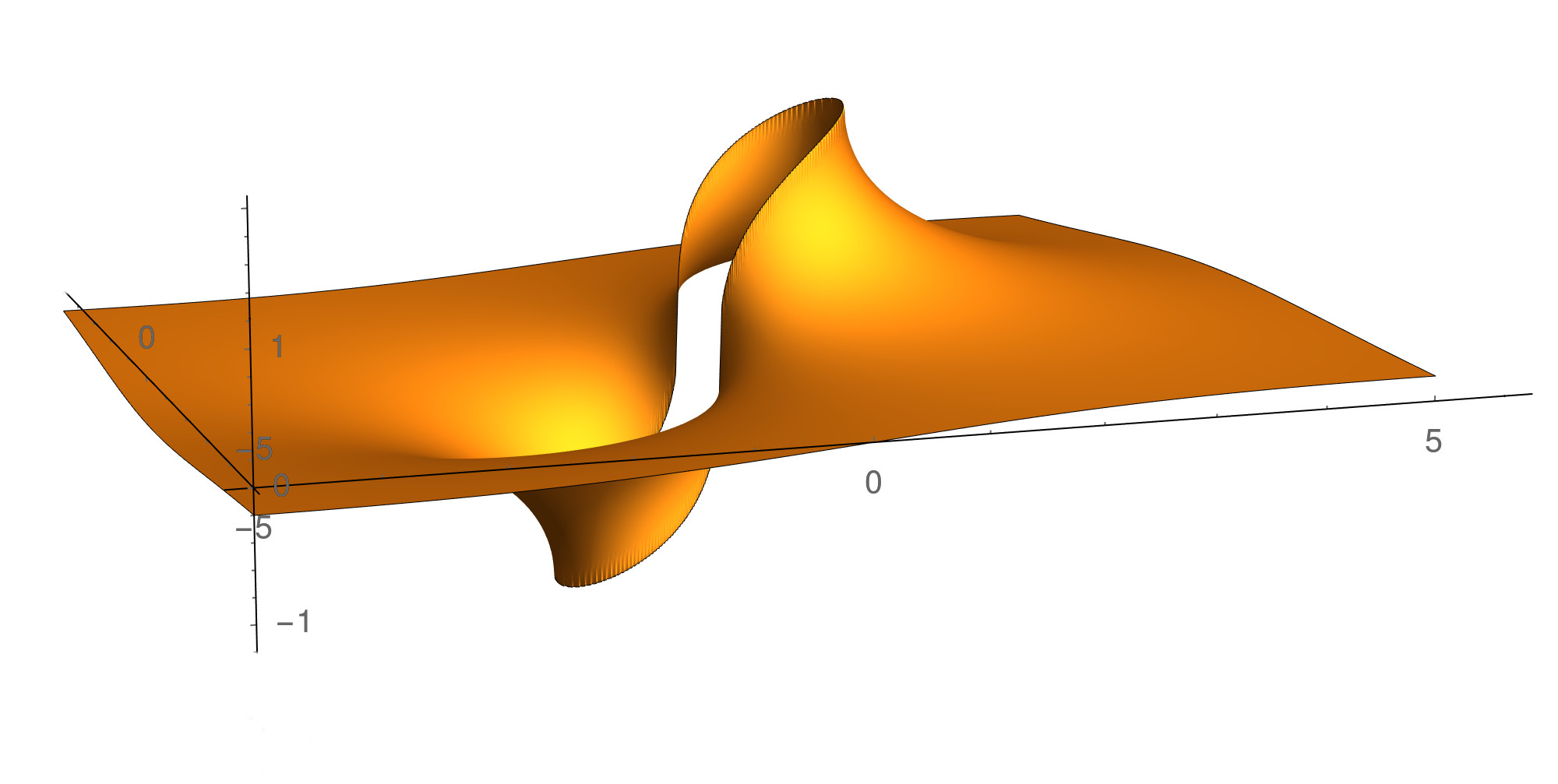}}
  \put(1.8,3.3){$\ffun$}
  \put(6.6,4.1){${\mathcal C}_{1}$}
  \put(3.7,0.2){${\mathcal D}_{1}$}
  \put(0.4,-3.0){
  \begin{picture}(2,2.5)
   \put(8,3.75){$m = 1,$}
   \put(9.2,3.75){$\metrp^{00} = -1$}
   \put(8,3.3){$\cbrho = 1,$}
   \put(9.1,3.3){$\xxx = 1$}
   \put(8,2.85){$\btam = 1,$}
   \put(9.1,2.85){$\btap = 0,$}
   \put(10.25,2.85){$\cbphi = 0$}
  \end{picture}
  }
  \put(0.15,3.1){$x^{2}$}
  \put(11.5,2.0){$x^{1}$}
  \end{pspicture}
  }
 \else
  \begin{pspicture}(0,0)(12,5)
  \put(1.8,3.3){$\ffun$}
  \put(6.6,4.1){${\mathcal C}_{1}$}
  \put(3.7,0.2){${\mathcal D}_{1}$}
  \put(0.4,-3.0){
  \begin{picture}(2,2.5)
   \put(8,3.75){$m = 1,$}
   \put(9.2,3.75){$\metrp^{00} = -1$}
   \put(8,3.3){$\cbrho = 1,$}
   \put(9.1,3.3){$\xxx = 1$}
   \put(8,2.85){$\btam = 1,$}
   \put(9.1,2.85){$\btap = 0,$}
   \put(10.25,2.85){$\cbphi = 0$}
  \end{picture}
  }
  \put(0.15,3.1){$x^{2}$}
  \put(11.5,2.0){$x^{1}$}
   \put(-0.3,-1.2){\includegraphics[width=125mm]{LSOESTF_Fig54p.eps}}
  \end{pspicture}
\fi
\end{center}
  \engrus{0.5ex}{0.5ex}{%
  \caption{\label{53247959}The field function $\ffun$ on the plane $\{x^{1},x^{2}\}$ for $m = 1$ and
  \mbox{$\metrp^{00} = -1$}.}
  }{%
  \mifeng{\addtocounter{figure}{-1}}{}
  \caption{\label{53247959r}Полевая функция $\ffun$ в плоскости $\{x^{1},x^{2}\}$ для $m = 1$ и \mbox{$\metrp^{00} = -1$}.}
  }
\end{figure}

\begin{figure}[h]
\begin{center}
\ifpdf
  {
  \unitlength 1mm
  \begin{pspicture}(0,0)(12,6.2)
  \put(0.5,-0.5){\includegraphics[width=40mm]{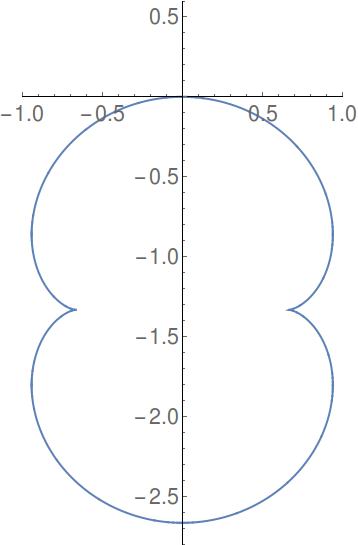}}
  \put(7.0,-0.5){\includegraphics[width=40mm]{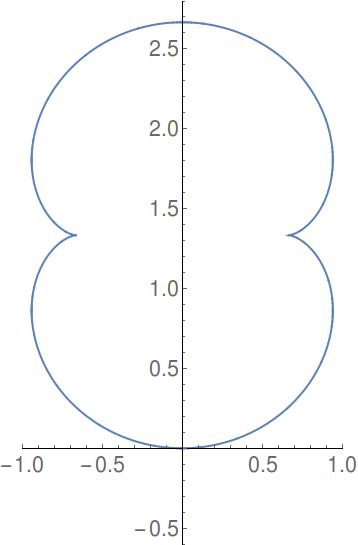}}
   \put(0.5,0){
   \put(2.2,5.5){$x^{2}$}
   \put(3.7,4.7){$x^{1}$}
   \put(8.66,5.5){$x^{2}$}
   \put(10.2,0.7){$x^{1}$}
   }
    \put(-3.7,-3.2){
  \begin{picture}(3,2.5)
   \put(8,3.75){$m = 1,$}
   \put(9.2,3.75){$\metrp^{00} = 1$}
   \put(8,3.3){$\cbrho = 1,$}
   \put(9.1,3.3){$\xxx = 1$}
   \put(8,2.85){$\btam = 1,$}
   \put(9.1,2.85){$\btap = 0,$}
   \put(10.25,2.85){$\cbphi = 0$}
  \end{picture}
  }
  \end{pspicture}
  }
 \else
  \begin{pspicture}(0,0)(12,6.2)
   \put(0.5,0){
   \put(2.2,5.5){$x^{2}$}
   \put(3.7,4.7){$x^{1}$}
   \put(8.66,5.5){$x^{2}$}
   \put(10.2,0.7){$x^{1}$}
   }
    \put(-3.7,-3.2){
  \begin{picture}(3,2.5)
   \put(8,3.75){$m = 1,$}
   \put(9.2,3.75){$\metrp^{00} = 1$}
   \put(8,3.3){$\cbrho = 1,$}
   \put(9.1,3.3){$\xxx = 1$}
   \put(8,2.85){$\btam = 1,$}
   \put(9.1,2.85){$\btap = 0,$}
   \put(10.25,2.85){$\cbphi = 0$}
  \end{picture}
  }
  \put(0.5,-0.5){\includegraphics[width=40mm]{LSOESTF_Fig55a.eps}}
  \put(7.0,-0.5){\includegraphics[width=40mm]{LSOESTF_Fig55b.eps}}
  \end{pspicture}
\fi
\end{center}
\engrus{0.5ex}{0.5ex}{%
 \caption{\label{37607999} Shifted singular lines on the plane $\{x^{1},x^{2}\}$ with \textaddb{the} parameter \mbox{$m = 1$}.}
}{%
\mifeng{\addtocounter{figure}{-1}}{}
 \caption{\label{37607999r} Смещённые сингулярные линии на плоскости $\{x^{1},x^{2}\}$
 со значением параметра \mbox{$m = 1$}.}
}
\end{figure}

\begin{figure}[h]
\begin{center}
\ifpdf
  \begin{pspicture}(0,0)(12,5)
    \put(0,-1.2){\includegraphics[width=120mm]{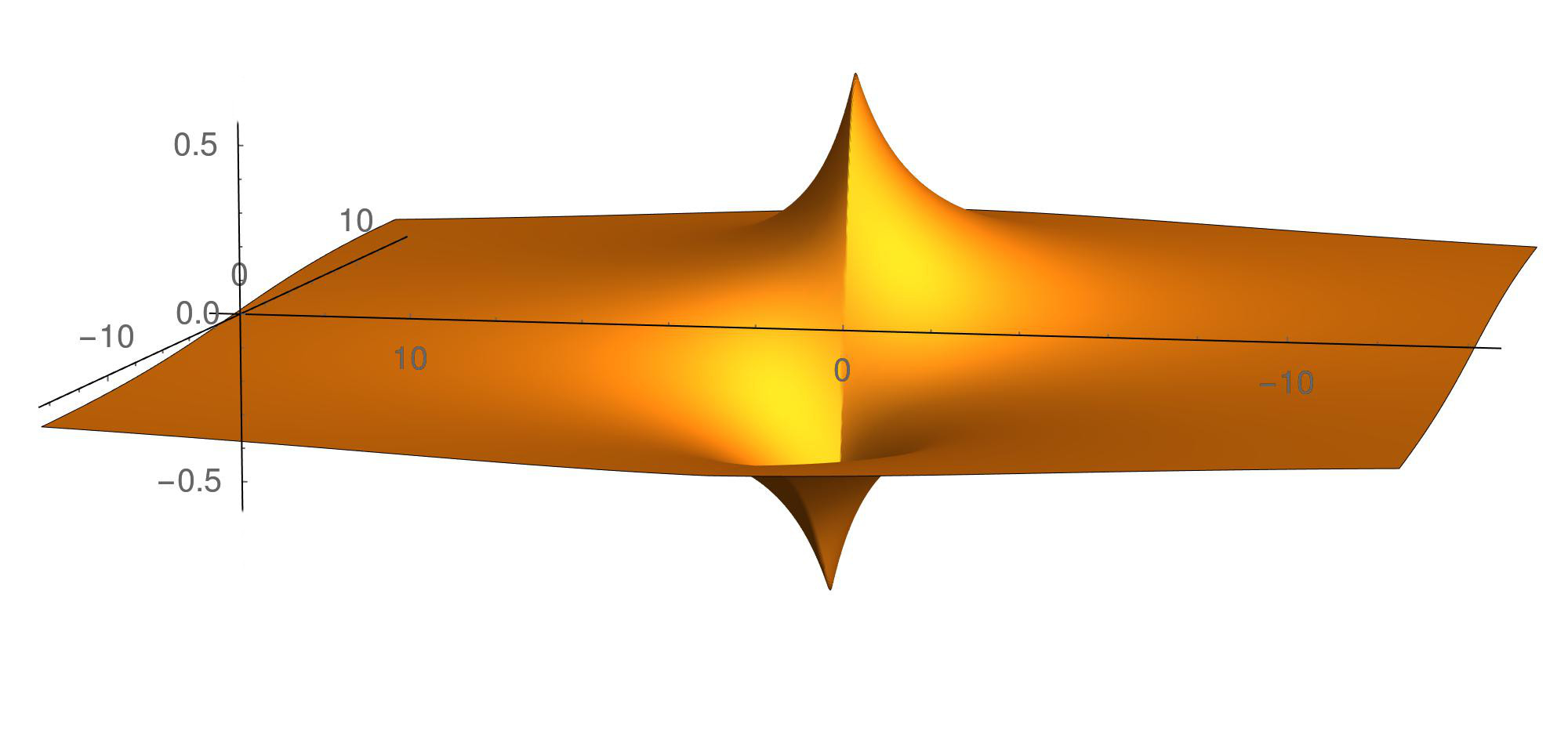}}
   \put(2.5,0.5){$m = 1$}
   \put(2,3.3){$\ffun$}
  \put(3.3,2.9){$x^{1}$}
  \put(0.9,2.1){$x^{2}$}
  \end{pspicture}
 \else
  \begin{pspicture}(0,0)(12,5)
    \put(0,-1.2){\includegraphics[width=120mm]{LSOESTF_Fig56p.eps}}
   \put(2.5,0.5){$m = 1$}
   \put(2,3.3){$\ffun$}
  \put(3.3,2.9){$x^{1}$}
  \put(0.9,2.1){$x^{2}$}
  \end{pspicture}
\fi
\end{center}
 \engrus{0.5ex}{0.5ex}{%
 \caption{\label{4755348855} The field function $\ffun$ of the composed solution for the soliton without a cavity when
 \mbox{$m = 1$} and \mbox{$\metrp^{00} = 1$} in the plane  $\{x^1,x^2\}$.
 }
 }{%
\mifeng{\addtocounter{figure}{-1}}{}
  \caption{\label{4755348855r} Полевая функция $\ffun$ составного решения в плоскости $\{x^1,x^2\}$ для солитона без полости при \mbox{$m = 1$} и  \mbox{$\metrp^{00} = 1$}.}
 }
\end{figure}

\begin{figure}[h]
\begin{center}
\ifpdf
  {
  \begin{pspicture}(0,0)(12,5)
   \put(0,-1.2){\includegraphics[width=125mm]{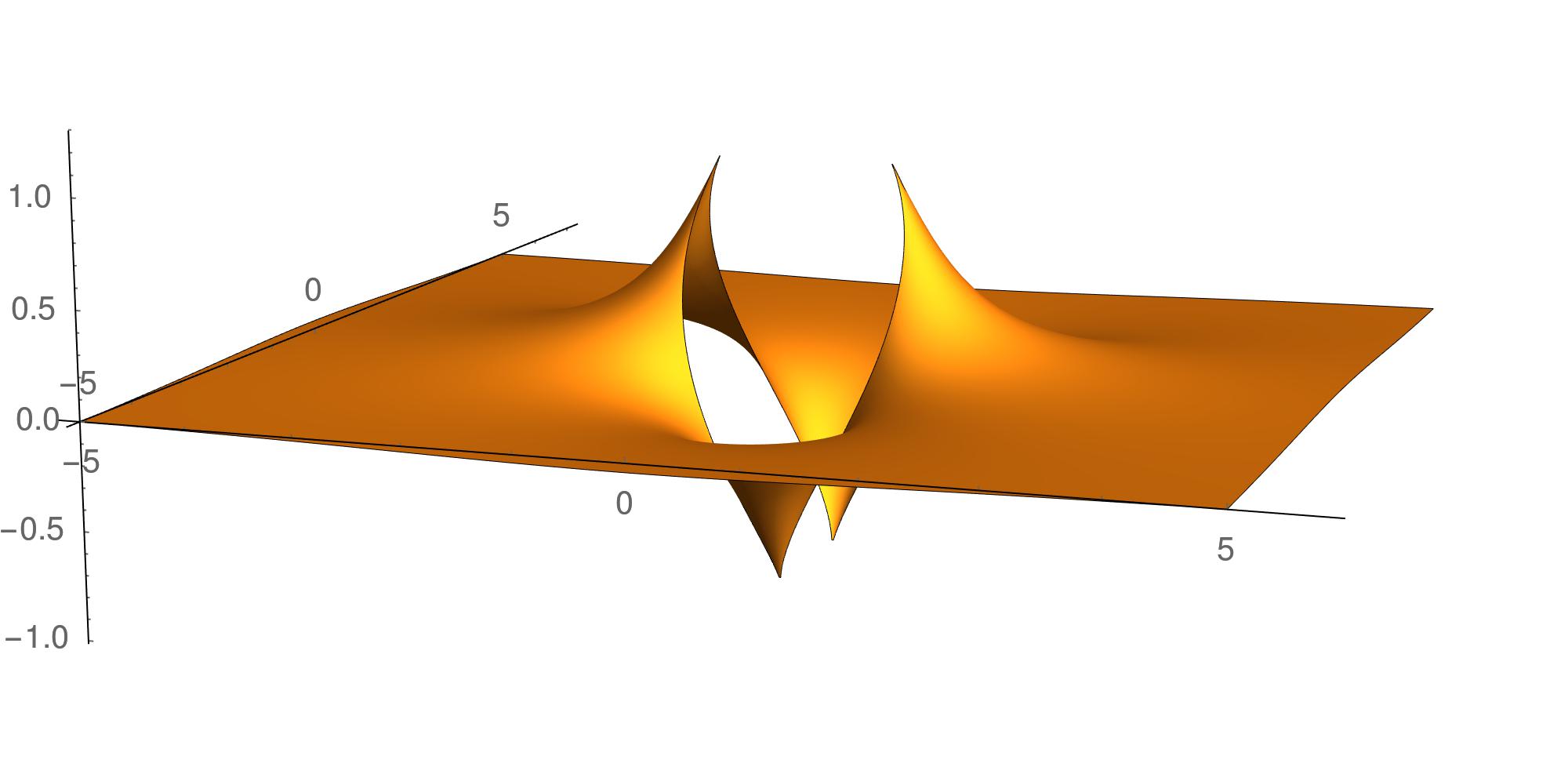}}
  \put(0.7,3.7){$\ffun$}
  \put(0,0.4){
 \begin{picture}(3,2.5)
   \put(8,3.75){$m = 2$}
   \put(8,3.3){$\cbrho = 1,$}
   \put(9.1,3.3){$\xxx = 1$}
   \put(8,2.85){$\btam = 1,$}
   \put(9.1,2.85){$\btap = 0,$}
   \put(10.25,2.85){$\cbphi = 0$}
   \put(8,2.4){$\metrp^{00} = 1$}
  \end{picture}
  }
  \put(4.7,3.15){$x^{2}$}
  \put(10.65,0.7){$x^{1}$}
  \end{pspicture}
  }
 \else
  \begin{pspicture}(0,0)(12,5)
  \put(0.7,3.7){$\ffun$}
  \put(0,0.4){
  \begin{picture}(3,2.5)
   \put(8,3.75){$m = 2$}
   \put(8,3.3){$\cbrho = 1,$}
   \put(9.1,3.3){$\xxx = 1$}
   \put(8,2.85){$\btam = 1,$}
   \put(9.1,2.85){$\btap = 0,$}
   \put(10.25,2.85){$\cbphi = 0$}
   \put(8,2.4){$\metrp^{00} = 1$}
  \end{picture}
  }
  \put(4.7,3.15){$x^{2}$}
  \put(10.75,0.7){$x^{1}$}
   \put(0,-1.2){\includegraphics[width=125mm]{LSOESTF_Fig57.eps}}
  \end{pspicture}
\fi
\end{center}
  \engrus{0.5ex}{0.5ex}{%
  \caption{\label{35836648}The field function $\ffun$ on the plane $\{x^{1},x^{2}\}$ for $m = 2$.}
  }{%
  \mifeng{\addtocounter{figure}{-1}}{}
  \caption{\label{35836648r}Полевая функция $\ffun$ в плоскости $\{x^{1},x^{2}\}$ для $m = 2$ и \mbox{$\metrp^{00} = -1$}.}
  }
\end{figure}

\begin{figure}[h]
\begin{center}
\ifpdf
  {
  \begin{pspicture}(0,0)(12,5)
   \put(0,-1.2){\includegraphics[width=125mm]{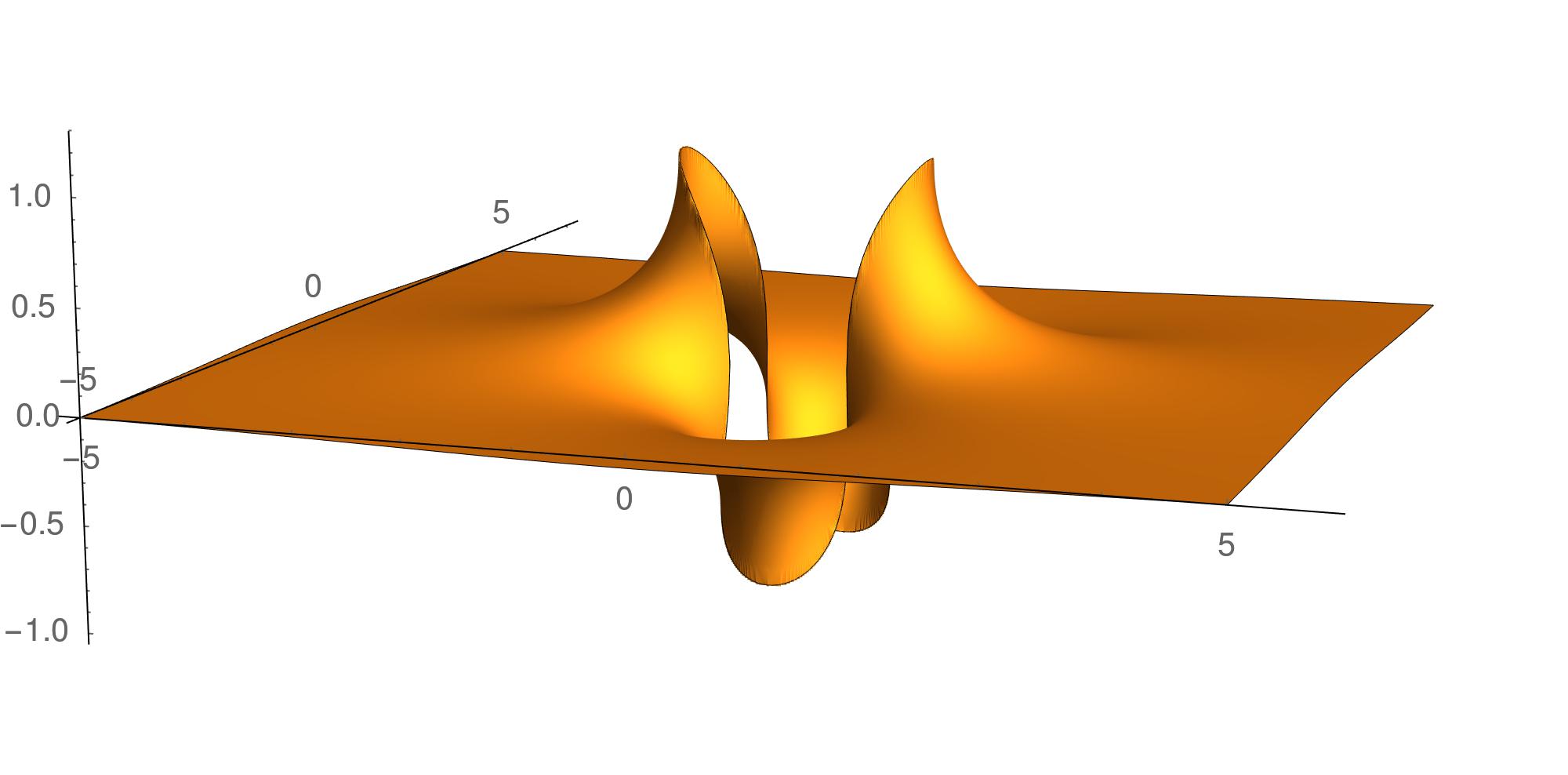}}
  \put(0.7,3.7){$\ffun$}
  \put(0,0.4){
 \begin{picture}(3,2.5)
   \put(8,3.75){$m = 2$}
   \put(8,3.3){$\cbrho = 1,$}
   \put(9.1,3.3){$\xxx = 1$}
   \put(8,2.85){$\btam = 1,$}
   \put(9.1,2.85){$\btap = 0,$}
   \put(10.25,2.85){$\cbphi = 0$}
   \put(8,2.4){$\metrp^{00} = -1$}
  \end{picture}
  }
  \put(4.7,3.15){$x^{2}$}
  \put(10.65,0.7){$x^{1}$}
  \end{pspicture}
  }
 \else
  \begin{pspicture}(0,0)(12,5)
  \put(0.7,3.7){$\ffun$}
  \put(0,0.4){
  \begin{picture}(3,2.5)
   \put(8,3.75){$m = 2$}
   \put(8,3.3){$\cbrho = 1,$}
   \put(9.1,3.3){$\xxx = 1$}
   \put(8,2.85){$\btam = 1,$}
   \put(9.1,2.85){$\btap = 0,$}
   \put(10.25,2.85){$\cbphi = 0$}
   \put(8,2.4){$\metrp^{00} = -1$}
  \end{picture}
  }
  \put(4.7,3.15){$x^{2}$}
  \put(10.75,0.7){$x^{1}$}
   \put(0,-1.2){\includegraphics[width=125mm]{LSOESTF_Fig58.eps}}
  \end{pspicture}
\fi
\end{center}
  \engrus{0.5ex}{0.5ex}{%
  \caption{\label{35836659}The field function $\ffun$ on the plane $\{x^{1},x^{2}\}$ for $m = 2$ and
  \mbox{$\metrp^{00} = -1$}.}
  }{%
  \mifeng{\addtocounter{figure}{-1}}{}
  \caption{\label{35836659r}Полевая функция $\ffun$ в плоскости $\{x^{1},x^{2}\}$ для $m = 2$ и
   \mbox{$\metrp^{00} = -1$}.}
  }
\end{figure}

\engrus{0.5ex}{0.5ex}{%
Now let us obtain the expressions of full energy, momentum, and angular momentum \textaddd{(\ref{576969441})} for the
\textremd{solution under consideration}\textaddd{twisted lightlike shell soliton}
in bounded three-dimensional volume.
For convenience we consider the tubular volume in coordinates
$\{\tilde{\rho},\tilde{\varphi},x^{3}\}$. Its internal radius is defined in (\ref{433253941}).
Let its external radius and length be designated as $\trhoinf$ and $\zdlina$ accordingly. Thus in addition to
condition (\ref{433253941}) we have
}{%
Теперь получим выражения для полной энергии, импульса и момента импульса (\ref{576969441}) для
вращающегося светоподобного оболочечного солитона в ограниченном
трёхмерном объёме. Для удобства мы рассматриваем трубчатый объём в координатах $\{\tilde{\rho},\tilde{\varphi},x^{3}\}$. Его внутренний радиус определён в (\ref{433253941}).
Обозначим его внешний радиус и длину как $\trhoinf$ и $\zdlina$ соответственно. Таким образом в дополнение к условиям
 (\ref{433253941}) имеем
}
\begin{equation}
\label{328475681}
\tilde{\rho}\leqslant \trhoinf
\;,\quad
-\frac{\zdlina}{2}\leqslant x^{3}\leqslant \frac{\zdlina}{2}
\;.
\end{equation}

\engrus{0.5ex}{0.5ex}{%
First we calculate the integrals on right-hand parts of relations (\ref{352154411})
by variables $\{\tilde{\rho},\tilde{\varphi}\}$ in area
$\{[\cbrho,\trhoinf],[-\pi,\pi]\}$.
This integration corresponds to
\textremd{the integration}\textaddd{one}
of the left-hand parts for the relations (\ref{352154411}) by the variables $\{\rho,\varphi\}$ in the outside area of the singular line $\mxis$ \textaddd{(\ref{402041591}) and} bounded by the line
$\mxi (\trhoinf\,\e{-\him\,\tilde{\varphi}},\trhoinf\,\e{\him\,\tilde{\varphi}})$,
\textaddd{where the function $\mxi (\mtxi,\hconj{\mtxi})$ is defined in (\ref{369128872})}.
}{%
Сначала мы вычисляем интегралы от правых частей соотношений
(\ref{352154411}) по переменным $\{\tilde{\rho},\tilde{\varphi}\}$
в области $\{[\cbrho,\trhoinf],[-\pi,\pi]\}$.
 Это интегрирование соответствует интегрированию левых частей соотношений (\ref{352154411}) по переменным $\{\rho,\varphi\}$ во внешней области сингулярной линии $\mxis$ (\ref{402041591}) и ограниченной линией
$\mxi (\trhoinf\,\e{-\him\,\tilde{\varphi}},\trhoinf\,\e{\him\,\tilde{\varphi}})$,
где функция $\mxi (\mtxi,\hconj{\mtxi})$ определена в (\ref{369128872}).
}

\engrus{0.5ex}{0.5ex}{%
Making the integration in the plane $\{x^{1},x^{2}\}$ we can put the rotation parameter
\textaddd{(\ref{519019131})} be zero: $\btap = 0$.
\textaddd{Also we put $\cbphi = 0$.}
Let us substitute (\ref{462088071})
and (\ref{369128872}) with (\ref{430348741}) and (\ref{448940041}) to the right-hand parts of (\ref{352154411}).
We change the integration by variable $x^{3}$  to
one by phase $\Phase$ (\ref{426404172}).
}{%
Осуществляя интегрирование в плоскости $\{x^{1},x^{2}\}$,  мы можем положить параметр поворота
(\ref{519019131}) равным нулю: $\btap = 0$.
Также мы полагаем  $\cbphi = 0$.
Подставим (\ref{462088071}) и (\ref{369128872}) с (\ref{430348741}) и (\ref{448940041}) в правые части (\ref{352154411}).
Мы заменяем интегрирование по переменной $x^{3}$  на интегрирование по фазе $\Phase$ (\ref{426404172}).
}

\begin{subequations}\label{338555621}
\engrus{0.5ex}{0.5ex}{%
As \textaddd{a} result, we have the following expressions for energy and absolute values of momentum and angular momentum:
}{%
В результате имеем следующие выражения для энергии и абсолютных значений импульса и момента импульса:
}
\begin{align}
\label{338678741}
\Energy   &= \EMV + \underline{\Energy}
\;,\\
\label{338678749}
&\phantom{{}={}}\quad\;\;\;\underline{\Energy} = \frac{\cbrho^{2}}{\omega\,\xxx^2}\,\pcoef_{0}\,\pint_{0}
= \EMV\,\frac{1}{\cbrho^2\,\omega^2}\,\frac{\pcoef_{0}\,\pint_{0}}{\pcoef_{1}\,(\pcoef_{2}\,\pint_{1} + \pcoef_{3}\,\pint_{2})}
\;,\\
\label{338678742}
\EMV &= \frac{\omega\,\cbrho^{4}}{\xxx^2}\,\pcoef_{1}\left(\pcoef_{2}\,\pint_{1} + \pcoef_{3}\,\pint_{2}\right)
\;,\\
\label{34148764}
\AMV  &= \frac{\cbrho^{4}}{\xxx^2}\,\pcoef_{1}\,\pcoef_{2}\,|\pint_{3}|
\;,
\end{align}
\engrus{0.5ex}{0.5ex}{%
\noindent
where $\underline{\Energy}$ is the part of soliton energy obtained
from the part $\fcE_{0}$ of energy density $\cE $ (\ref{495868961}),
}{%
\noindent
где $\underline{\Energy}$ -- часть энергии солитона полученная из части $\fcE_{0}$ плотности энергии $\cE $ (\ref{495868961}),
}
\end{subequations}
\begin{alignat}{2}
\nonumber
\pint_{0}&\eqdef \int\limits_{-\omega\,\zdlina/2}^{\omega\,\zdlina/2}\btam^2\df\Phase\;,   &  \qquad
\pint_{1} &\eqdef \int\limits_{-\omega\,\zdlina/2}^{\omega\,\zdlina/2}\btam^4\left(\btap^{\prime}\right)^2\df\Phase
\;,\\
\label{374782791}
\pint_{2} &\eqdef \int\limits_{-\omega\,\zdlina/2}^{\omega\,\zdlina/2}\btam^2\left(\btam^{\prime}\right)^2\df\Phase\;,   &
\pint_{3} &\eqdef \int\limits_{-\omega\,\zdlina/2}^{\omega\,\zdlina/2}\btam^4\,\btap^{\prime}\,\df\Phase
\;,
\end{alignat}
\vspace{-2ex}
\begin{subequations}\label{354384451}
\begin{align}
\label{462322731}
\pcoef_{0} &\eqdef
\frac{1}{2}\,\biggl(\frac{1}{m} \mp \frac{1}{2\,m + 1}\biggr)
-\frac{(\cbrho/\trhoinf)^{2\,m}}{2\,m}
\pm \frac{(\cbrho/\trhoinf)^{2\,(2\,m+1)}}{2\,(2\,m+1)}
\;,\\[1ex]
\label{35855292}
\pcoef_{1} &\eqdef \ln\biggl(\frac{\trhoinf}{\cbrho}\biggr)
\pm \frac{1}{6}\,\biggl(1 - \frac{\cbrho^{\,4}}{\trhoinf^{\,4}}\biggr)
 + \frac{1}{72} \,\biggl(1 - \frac{\cbrho^{\,8}}{\trhoinf^{\,8}}\biggr)
\qquad\text{for}\quad m = 1
\;,\\[1ex]
\pcoef_{1} &\eqdef \frac{1}{2\,m^{2}}
\left(
\frac{2}{2\,m+1}\left(\frac{2\,m\,(3\,m^2 + 4\,m + 2)}{(3\, m +1)\,(2\, m + 1)\,(m-1)} \pm \frac{1}{2\,m}\right)
- \frac{(\cbrho/\trhoinf)^{2\,(m-1)}}{m-1}
\right.
\nonumber
\\
& \left.
\mp \frac{(\cbrho/\trhoinf)^{4\,m}}{m\,(2\,m+1)}
- \frac{(\cbrho/\trhoinf)^{2\,(3\,m+1)}}{(2\,m+1)^2\,(3\,m+1)}
\right)
\quad\quad\quad\,\text{for}\quad m \geqslant 2\;,
\label{462322732}
\\[1ex]
\label{46972542}
\pcoef_{2} &\eqdef m^2\;,\quad \pcoef_{3} \eqdef (m+1)^2
\;.
\end{align}
\end{subequations}

\engrus{0.5ex}{0.5ex}{%
Here in (\ref{354384451}) we have the different expressions for the two metric signatures (\ref{43842964a}) and (\ref{43842964b})
(top and bottom signs \textrem1{accordingly}\textadd1{respectively}).
}{%
Здесь в (\ref{354384451}) имеем различные выражения для двух сигнатур метрики (\ref{43842964a}) и (\ref{43842964b})
(верхние и нижние знаки соответственно).
}

\engrus{0.5ex}{0.5ex}{%
The value of \textadd1{the} $x^{3}$ momentum projection is defined by the sign of wave vector projection $k_{3}$ (\ref{426404172}): \mbox{$\EMV_{3}=\pm\EMV$}.
}{%
Значение проекции импульса на ось $x^{3}$ определяется знаком проекции волнового вектора $k_{3}$ (\ref{426404172}): \mbox{$\EMV_{3}=\pm\EMV$}.
}

\engrus{0.5ex}{0.5ex}{%
In general case the $x^{3}$ angular momentum projection
is defined by \textadd1{the} integral
$\pint_{3}$ (\ref{374782791}), which can be called the integral twist of the soliton with  weight $\btam^4$:
$\AMV_{3} = \pm\AMV$.
}{%
В общем случае проекция момента импульса на ось $x^{3}$ определяется интегралом
$\pint_{3}$ (\ref{374782791}), который может быть назван интегральной закрученностью солитона с весом $\btam^4$:
$\AMV_{3} = \pm\AMV$.
}

\engrus{0.5ex}{0.5ex}{%
Let us write the appropriate to (\ref{338555621}) expressions for the twisted soliton with constant twist (\ref{519019131}). Using \textadd1{the} condition
(\ref{519019131}) and formulas (\ref{374782791}), we have
}{%
Запишем соответствующее (\ref{338555621}) выражение для закрученного солитона с постоянной закрученностью (\ref{519019131}). Используя условие
(\ref{519019131}) и формулы (\ref{374782791}), имеем
}
\begin{equation}
\label{817526421}
 |\pint_{3}| = \frac{1}{m}\,\tilde{\pint}_{1}
 \;,\quad
 \pint_{1} = \frac{1}{m^2}\,\tilde{\pint}_{1}
 \;,\quad
 \tilde{\pint}_{1} \eqdef \int\limits_{-\omega\,\zdlina/2}^{\omega\,\zdlina/2}\btam^4\df\Phase
\;.
\end{equation}

\begin{subequations}\label{388299621}
\engrus{0.5ex}{0.5ex}{%
Using (\ref{817526421}) and (\ref{354384451}), we obtain from (\ref{338555621}) the following expressions  for the twisted soliton:
}{%
Используя (\ref{817526421}) и (\ref{354384451}), получаем из (\ref{338555621}) следующие выражения для закрученного
солитона:
}
\begin{align}
\label{38887087}
\Energy   &=
\EMV
\left(1 + \frac{1}{\cbrho^2\,\omega^2}\,\frac{\pcoef_{0}\,\pint_{0}}{\pcoef_{1}\,(\tilde{\pint}_{1} + \pcoef_{3}\,\pint_{2})}\right)
\;,\\
\label{38890960}
\EMV &= \omega\,\AMV\,\frac{1}{m}\left(1 + \frac{\pcoef_{3}\,\pint_{2}}{\tilde{\pint}_{1}}\right)
\;,\\
\label{38894581}
\AMV  &= \frac{\cbrho^{4}}{\xxx^2}\,m\,\pcoef_{1}\,\tilde{\pint}_{1}
\;,
\end{align}
\end{subequations}

\begin{subequations}\label{417457381}
\engrus{0.5ex}{0.5ex}{%
For the twisted soliton let us consider the case for
slowly varying scale function $\btam (\Phase)$, such that $\pint_{2}\to 0$ (\ref{374782791}). Also we
suppose that the frequency $\omega$ is sufficiently high, such that $\cbrho\,\omega\to\infty$.
According to expressions (\ref{388299621}), in this case
we have the following relations:
}{%
Для закрученного солитона рассмотрим случай медленно меняющейся масштабной функции $\btam (\Phase)$
так, что $\pint_{2}\to 0$ (\ref{374782791}). Также мы предполагаем, что частота
 $\omega$ достаточно высока, так что $\cbrho\,\omega\to\infty$.
Согласно выражениям (\ref{388299621}), в этом случае имеем следующие соотношения:
}
\begin{equation}
\label{832897461}
\Energy  = \EMV  = \romega\,\AMV
\;,
\end{equation}
\engrus{0.5ex}{0.5ex}{%
\noindent
where
}{%
\noindent
где
}
\begin{equation}
\label{57899949}
\romega  \eqdef \frac{\omega}{m}
\end{equation}
\engrus{0.5ex}{0.5ex}{%
\noindent
is the angular velocity of the twisted soliton.
}{%
\noindent
представляет собой угловую скорость закрученного солитона.
}
\end{subequations}

\engrus{0.5ex}{0.5ex}{%
Let us consider the twisted soliton with scale function in the form of Gaussian curve:
}{%
Рассмотрим закрученный солитон с масштабной функцией в виде Гауссовой кривой:
}
\begin{equation}
\label{494507011}
\btam = \exp\!\left(-\frac{\Phase^2}{2\,\bPhase^2}\right)
 \;,
\end{equation}
\engrus{0.5ex}{0.5ex}{%
\noindent
where $\bPhase$  is \textadd1{the} characteristic length of the soliton measured in radians and numerically equals
to a total angle of twist on the characteristic length of the soliton along $x^{3}$ axis.
A  twist angle $2\pi/m$ corresponds to soliton wave-length along $x^{3}$ axis.
}{%
\noindent
где $\bPhase$  -- характерная длина солитона измеренная в радианах и численно равная полному углу закрученности на характерной длине солитона вдоль оси  $x^{3}$.
Угол закрученности $2\pi/m$ соответствует длине волны солитона вдоль оси $x^{3}$.
}

\engrus{0.5ex}{0.5ex}{%
Let us consider the case of infinite space with the conditions
}{%
Рассмотрим случай бесконечного пространства с условием
}
\begin{equation}
\label{350512411}
 \frac{\trhoinf}{\cbrho}\to\infty\;,\quad\omega\,\zdlina\to\infty
\;.
\end{equation}
\engrus{0.5ex}{0.5ex}{%
Then the calculation of the essential integrals  in (\ref{374782791}) and (\ref{817526421}) for the function (\ref{494507011}) gives
}{%
Тогда вычисление существенных интегралов в (\ref{374782791}) и (\ref{817526421}) для функции (\ref{494507011}) даёт
}
\begin{equation}
\label{499888381}
\pint_{0}  = \bPhase\,\sqrt{\pi}
\;,\quad
\tilde{\pint}_{1}  = \bPhase\,\sqrt{\frac{\pi}{2}}
\;,\quad
\pint_{2}  =
\frac{1}{4\,\bPhase}\,\sqrt{\frac{\pi}{2}}
\;.
\end{equation}

\begin{subequations}\label{389638041}
\engrus{0.5ex}{0.5ex}{%
As we see in (\ref{338555621}) and (\ref{354384451}) with (\ref{350512411}) and (\ref{499888381}), for the case of infinite space we have the finite values of energy, momentum, and angular momentum if $m\geqslant 2$.
Using (\ref{388299621}) with (\ref{354384451}), (\ref{817526421}), (\ref{350512411}), and (\ref{499888381}), let us write the appropriate expressions for $m = 2$
and the metric signature (\ref{43842964a}):
}{%
Как видно в (\ref{338555621}) и (\ref{354384451}) с (\ref{350512411}) и (\ref{499888381}),
в случае бесконечного пространства мы имеем конечное значение энергии, импульса и момента импульса при
 $m\geqslant 2$.
 Используя (\ref{388299621}) с (\ref{354384451}), (\ref{817526421}), (\ref{350512411}) и (\ref{499888381}), запишем
  соответствующее выражение для $m = 2$ и сигнатуры метрики (\ref{43842964a}):
}
\begin{align}
\label{388617411}
\Energy  &=
\EMV\left( 1 + \frac{560\,\sqrt{2}\;\bPhase^2}{129\,(9 + 4\,\bPhase^2)\,\cbrho^{2}\,\omega^2}\right)
\;,\\
\EMV &= \omega\,\AMV\,\frac{1}{2}\left(1 + \frac{9}{4\,\bPhase^2}\right)
\label{388617412}
\;,\\
\AMV &=
\frac{387}{1400}\,\sqrt{\frac{\pi}{2}}
\,
\frac{\cbrho^{4}\,\bPhase}{\xxx^2}
\;.
\end{align}
\end{subequations}

\engrus{0.5ex}{0.5ex}{%
It is evident that the case $\bPhase \gg 1$ and $\cbrho\;\omega \gg 1$ for expressions (\ref{389638041}) gives relations (\ref{417457381}).
}{%
Очевидно, что случай $\bPhase \gg 1$ и $\cbrho\;\omega \gg 1$ для выражений (\ref{389638041}) даёт выражения (\ref{417457381}).
}

\engrus{0.5ex}{0.5ex}{%
The shell of the twisted soliton with Gaussian scale phase functions is shown on Fig. \ref{76297469} and Fig. \ref{47553488}.
}{%
Оболочка закрученного солитона с Гауссовой масштабной фазовой функцией показана на Рис. \ref{76297469r} и Рис. \ref{47553488r}.
}

\begin{figure}[h]
\begin{center}
\ifpdf
  {
  \unitlength 1mm
  \begin{picture}(120,45)
 \put(-6,-6){\includegraphics[width=131mm]{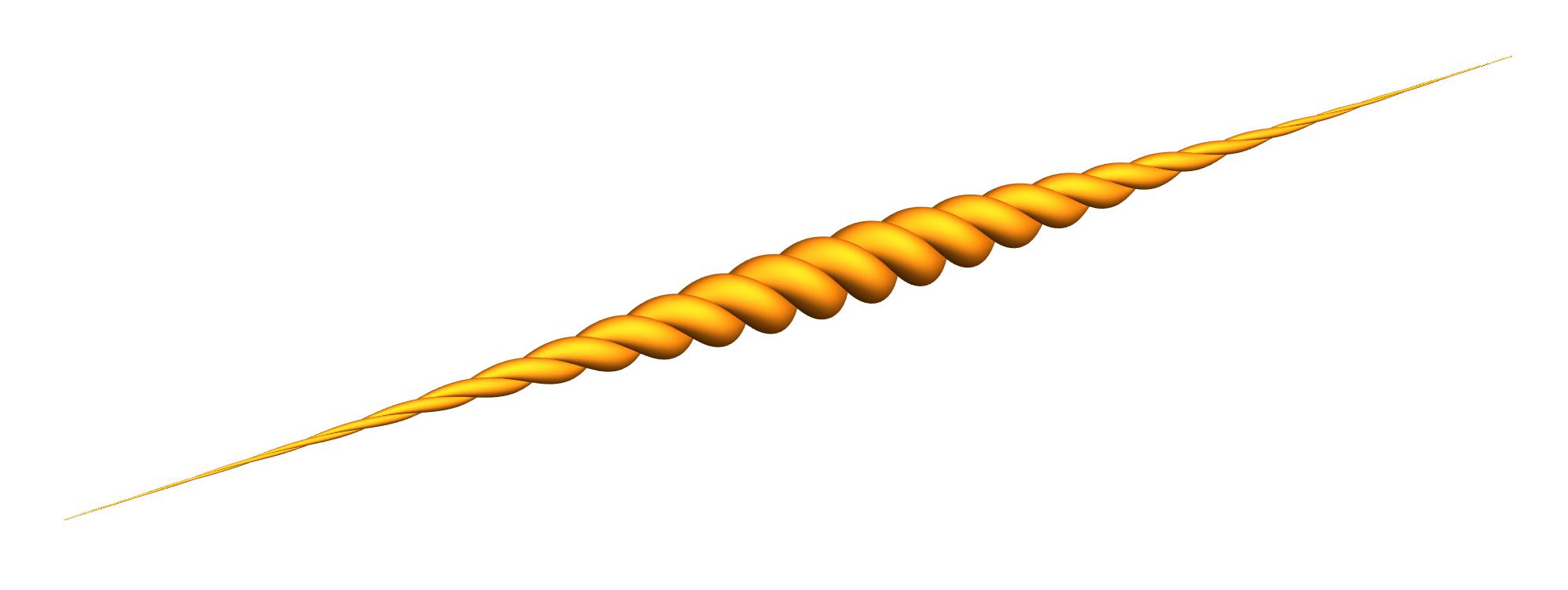}}
  \put(61,14){\line(1,-3){2}}
  \put(70,16.4){\line(1,-3){2}}
  \put(66.8,11.5){\vector(3,1){4.3}}
  \put(66.8,11.5){\vector(-3,-1){4.3}}
  \put(66.5,8){$\lambda$}
   \put(90,10){$m = 1$}
   \put(90,5){$\bPhase = 10$}
  \end{picture}
  }
 \else
  \begin{pspicture}(0,0)(12,4.5)
  \put(6.1,1.4){\line(1,-3){0.2}}
  \put(7.0,1.64){\line(1,-3){0.2}}
  \put(6.68,1.15){\vector(3,1){0.43}}
  \put(6.68,1.15){\vector(-3,-1){0.43}}
  \put(6.65,0.8){$\lambda$}
   \put(-0.6,-0.5){\includegraphics[width=130mm]{LSOESTF_Fig59.eps}}
   \put(9,1){$m = 1$}
   \put(9,0.5){$\bPhase = 10$}
  \end{pspicture}
\fi
\end{center}
\engrus{0.5ex}{0.5ex}{%
\caption{\label{76297469}The shell of Gaussian twisted soliton for $m = 1$.}
}{%
\mifeng{\addtocounter{figure}{-1}}{}
  \caption{\label{76297469r}Оболочка Гауссова вращающегося солитона для $m = 1$.}
}
\end{figure}

\begin{figure}[h]
\begin{center}
\ifpdf
  {
  \unitlength 1mm
  \begin{picture}(120,45)
   \put(-6,-6){\includegraphics[width=131mm]{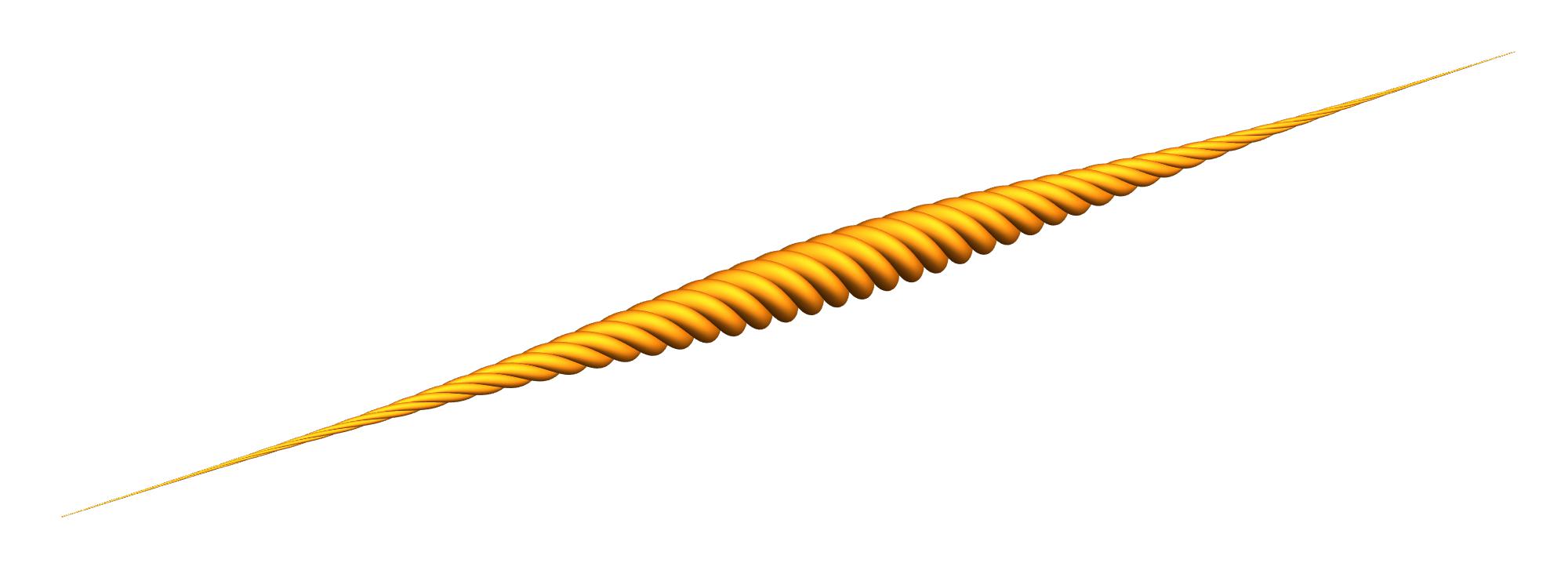}}
   \put(63,15.3){\line(1,-3){2}}
   \put(67.5,16.4){\line(1,-3){2}}
   \put(66.8,11.5){\vector(3,1){2.15}}
   \put(66.8,11.5){\vector(-3,-1){2.15}}
   \put(66.5,7.5){$\lambda$}
   \put(90,10){$m = 2$}
   \put(90,5){$\bPhase = 10$}
  \end{picture}
  }
 \else
  \begin{pspicture}(0,0)(12,4.5)
   \put(-0.6,-0.5){\includegraphics[width=130mm]{LSOESTF_Fig510.eps}}
   \put(9,1){$m = 2$}
   \put(9,0.5){$\bPhase = 10$}
  \put(6.30,1.53){\line(1,-3){0.2}}
  \put(6.75,1.64){\line(1,-3){0.2}}
  \put(6.68,1.15){\vector(3,1){0.215}}
  \put(6.68,1.15){\vector(-3,-1){0.215}}
  \put(6.65,0.75){$\lambda$}
  \end{pspicture}
\fi
\end{center}
  \engrus{0.5ex}{0.5ex}{%
  \caption{\label{47553488}The shell of Gaussian twisted soliton for $m = 2$.}
  }{%
  \mifeng{\addtocounter{figure}{-1}}{}
  \caption{\label{47553488r}Оболочка Гауссова вращающегося солитона для $m = 2$.}
  }
\end{figure}

\engrus{0.5ex}{0.5ex}{%
It is significant that the twist parameter $m$ is a topological invariant for diffeomorphism.
The shell of twisted lightlike soliton is diffeomorphic to a cylindrical surface with
threads by multifilar helix, where the number of continuous threads is $2\,m$.
These threads correspond to the singular lines on the shell, which we can see on
Fig. \ref{76297469} and Fig. \ref{47553488}.
}{%
Важно, что параметр закрученности $m$ представляет собой топологический инвариант диффеоморфизма.
Оболочка закрученного светоподобного солитона диффеоморфна цилиндрической поверхности с нарезами
многозаходной спиралью, где $2\,m$ -- число непрерывных нарезов.
Эти нарезы соответствуют сингулярным линиям на оболочке, которые мы можем видеть на Рис. \ref{76297469r} и Рис. \ref{47553488r}.
}

\engrus{0.5ex}{0.5ex}{%
The field function $\ffun$ of the Gaussian twisted soliton in the plane section $\{x^1,x^3\}$ for $x^2 = 0$ is shown on  Fig. \ref{762974691} and Fig. \ref{475534881}.
}{%
Полевая функция $\ffun$ Гауссова закрученного солитона в плоском сечении $\{x^1,x^3\}$ для $x^2 = 0$
показана на Рис. \ref{762974691r} и Рис. \ref{475534881r}.
}

\begin{figure}[h]
\begin{center}
\ifpdf
  \begin{pspicture}(0,0)(12,6.0)
    \put(-0.3,-2.5){\includegraphics[width=127mm]{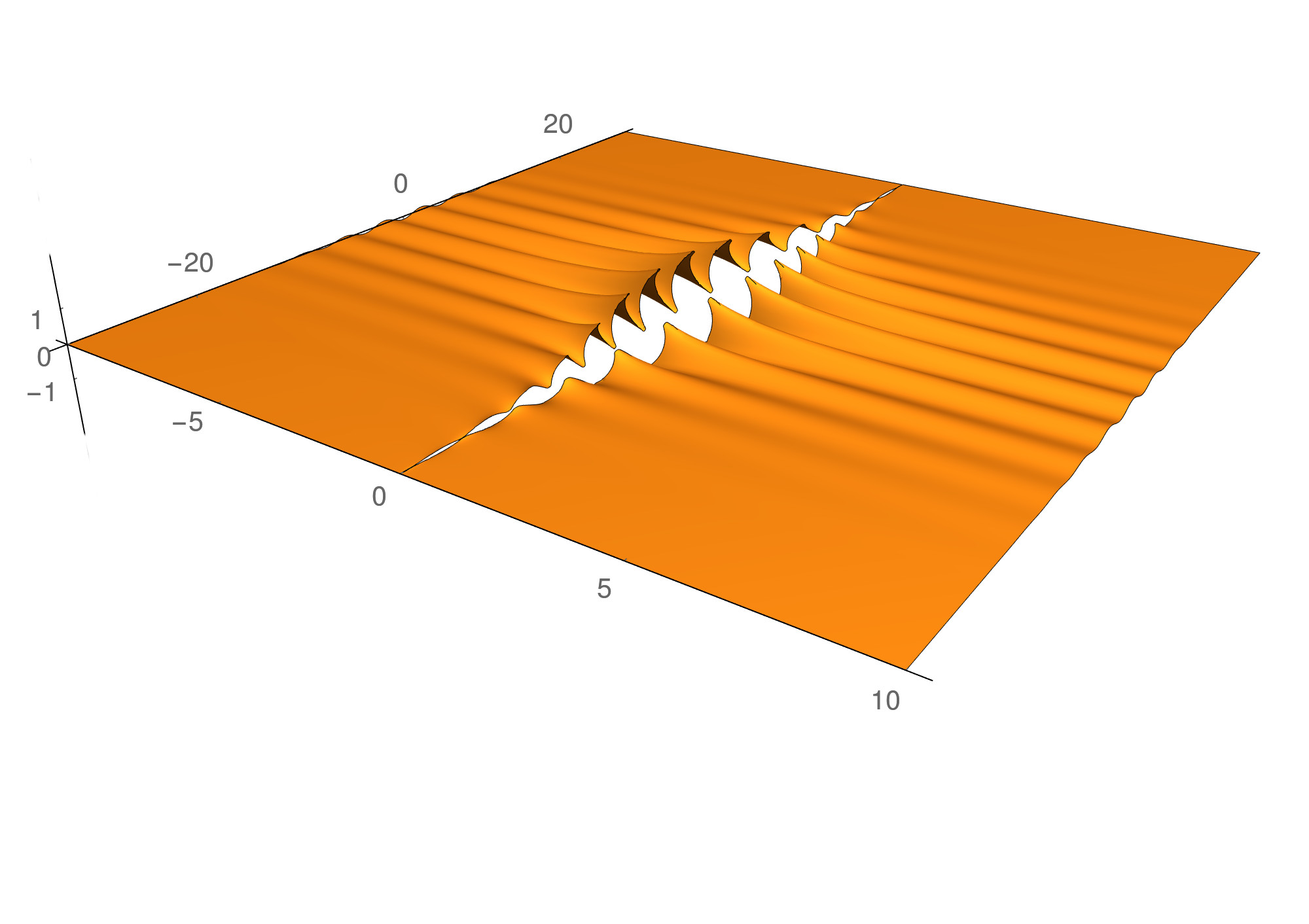}}
   \put(1.9,1){$m = 1,$}
   \put(3.2,1){$\metrp^{00} = 1$}
   \put(1.9,0.5){$\bPhase = 10$}
  \put(0.3,3.9){$\ffun$}
  \put(6,5.3){$x^{3}$}
  \put(8.9,-0.3){$x^{1}$}
  \end{pspicture}
 \else
  \begin{pspicture}(0,0)(12,6.0)
    \put(-0.3,-2.5){\includegraphics[width=127mm]{LSOESTF_Fig511p.eps}}
   \put(1.9,1){$m = 1,$}
   \put(3.2,1){$\metrp^{00} = 1$}
   \put(1.9,0.5){$\bPhase = 10$}
  \put(0.3,3.9){$\ffun$}
  \put(6,5.3){$x^{3}$}
  \put(8.9,-0.3){$x^{1}$}
  \end{pspicture}
\fi
\end{center}
\engrus{0.5ex}{0.5ex}{%
\caption{\label{762974691} The field function $\ffun$ of Gaussian twisted soliton for \mbox{$m = 1$}
 and \mbox{$\metrp^{00} = 1$} on the plane $\{x^{1},x^{3}\}$ for $x^{2} = 0$.}
}{%
  \mifeng{\addtocounter{figure}{-1}}{}
  \caption{\label{762974691r} Полевая функция $\ffun$ Гауссова вращающегося солитона для \mbox{$m = 1$}
 и \mbox{$\metrp^{00} = 1$} в плоскости $\{x^{1},x^{3}\}$ для $x^{2} = 0$.}
}
\end{figure}

\begin{figure}[h]
\begin{center}
\ifpdf
  \begin{pspicture}(0,0)(12,6.0)
   \put(-0.3,-2.5){\includegraphics[width=127mm]{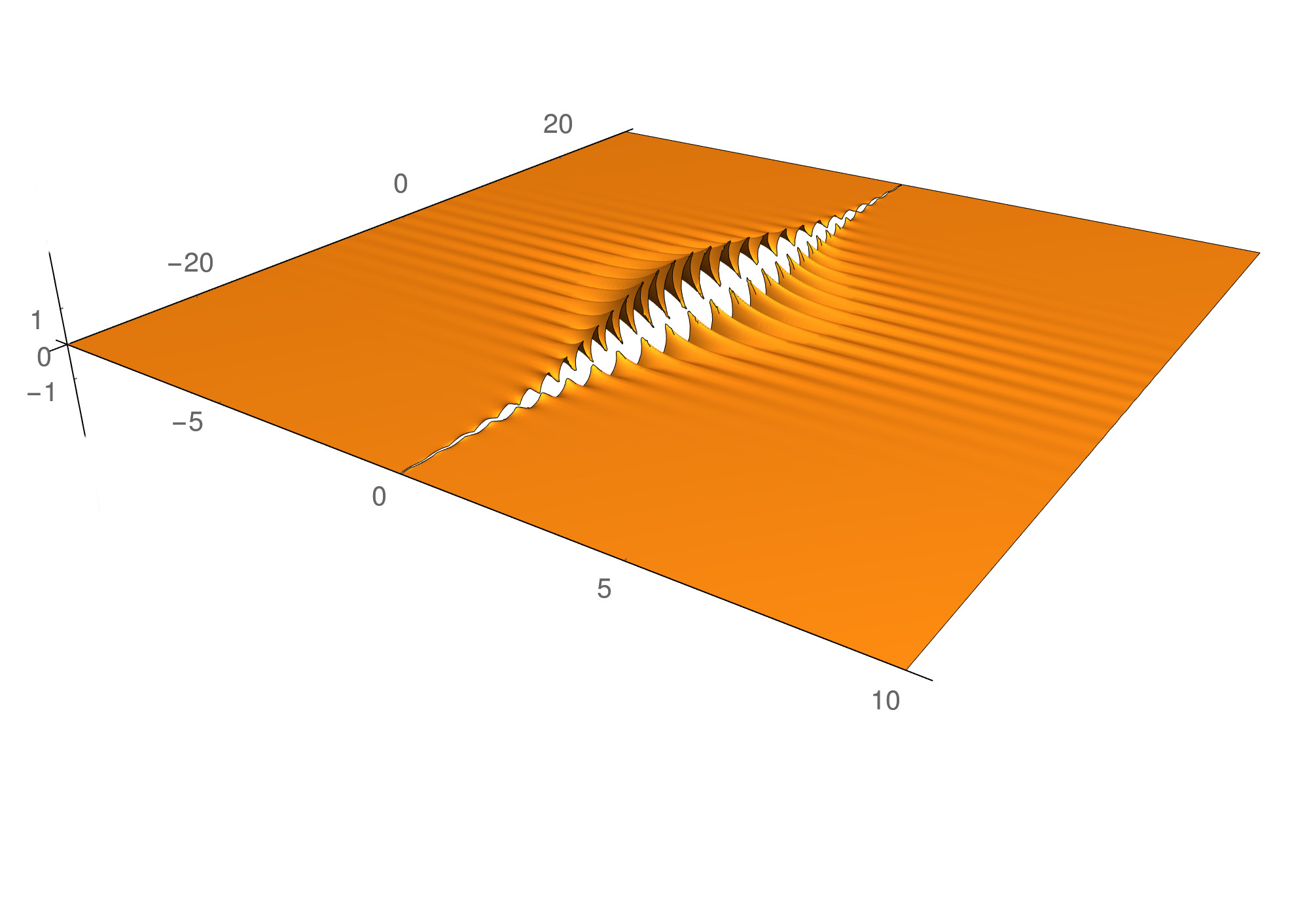}}
  \put(1.9,1){$m = 2,$}
   \put(3.2,1){$\metrp^{00} = 1$}
   \put(1.9,0.5){$\bPhase = 10$}
  \put(0.3,3.9){$\ffun$}
  \put(6,5.3){$x^{3}$}
  \put(8.9,-0.3){$x^{1}$}
  \end{pspicture}
 \else
  \begin{pspicture}(0,0)(12,6.0)
   \put(-0.3,-2.5){\includegraphics[width=127mm]{LSOESTF_Fig512p.eps}}
  \put(1.9,1){$m = 2,$}
   \put(3.2,1){$\metrp^{00} = 1$}
   \put(1.9,0.5){$\bPhase = 10$}
  \put(0.3,3.9){$\ffun$}
  \put(6,5.3){$x^{3}$}
  \put(8.9,-0.3){$x^{1}$}
  \end{pspicture}
\fi
\end{center}
 \engrus{0.5ex}{0.5ex}{%
 \caption{\label{475534881} The field function $\ffun$ of Gaussian twisted soliton for \mbox{$m = 2$}
  and \mbox{$\metrp^{00} = 1$} on the plane $\{x^{1},x^{3}\}$ for $x^{2} = 0$.}
 }{%
 \mifeng{\addtocounter{figure}{-1}}{}
  \caption{\label{475534881r} Полевая функция $\ffun$ Гауссова закрученного солитона для \mbox{$m = 2$}
   и \mbox{$\metrp^{00} = 1$} в плоскости $\{x^{1},x^{3}\}$ для $x^{2} = 0$.}
 }
\end{figure}

\engrus{0.5ex}{0.5ex}{%
At last we show zero level surfaces of the field function $\ffun$ for the Gaussian twisted soliton with $m = 1$ (Fig. \ref{7629746911}) and $m = 2$ (Fig. \ref{4755348811}).
The twist of the solitons is well seen also on these figures.
We have two-sheeted
helical surface with excluded cavity for $m=1$  and we have four-sheeted one for $m=2$.
}{%
Наконец мы показываем поверхность нулевого уровня полевой функции $\ffun$ для Гауссова закрученного солитона с $m = 1$ (Рис. \ref{7629746911r}) и $m = 2$ (Рис. \ref{4755348811r}).
Закрученность также хорошо видна на этих фигурах.
Мы имеем двулистную спиральную поверхность с исключённой полостью для $m=1$  и четырёхлистную для  $m=2$.
}

\begin{figure}[h]
\begin{center}
\ifpdf
  {
  \unitlength 1mm
  \begin{picture}(120,50)
  \put(0,-7){\includegraphics[width=120mm]{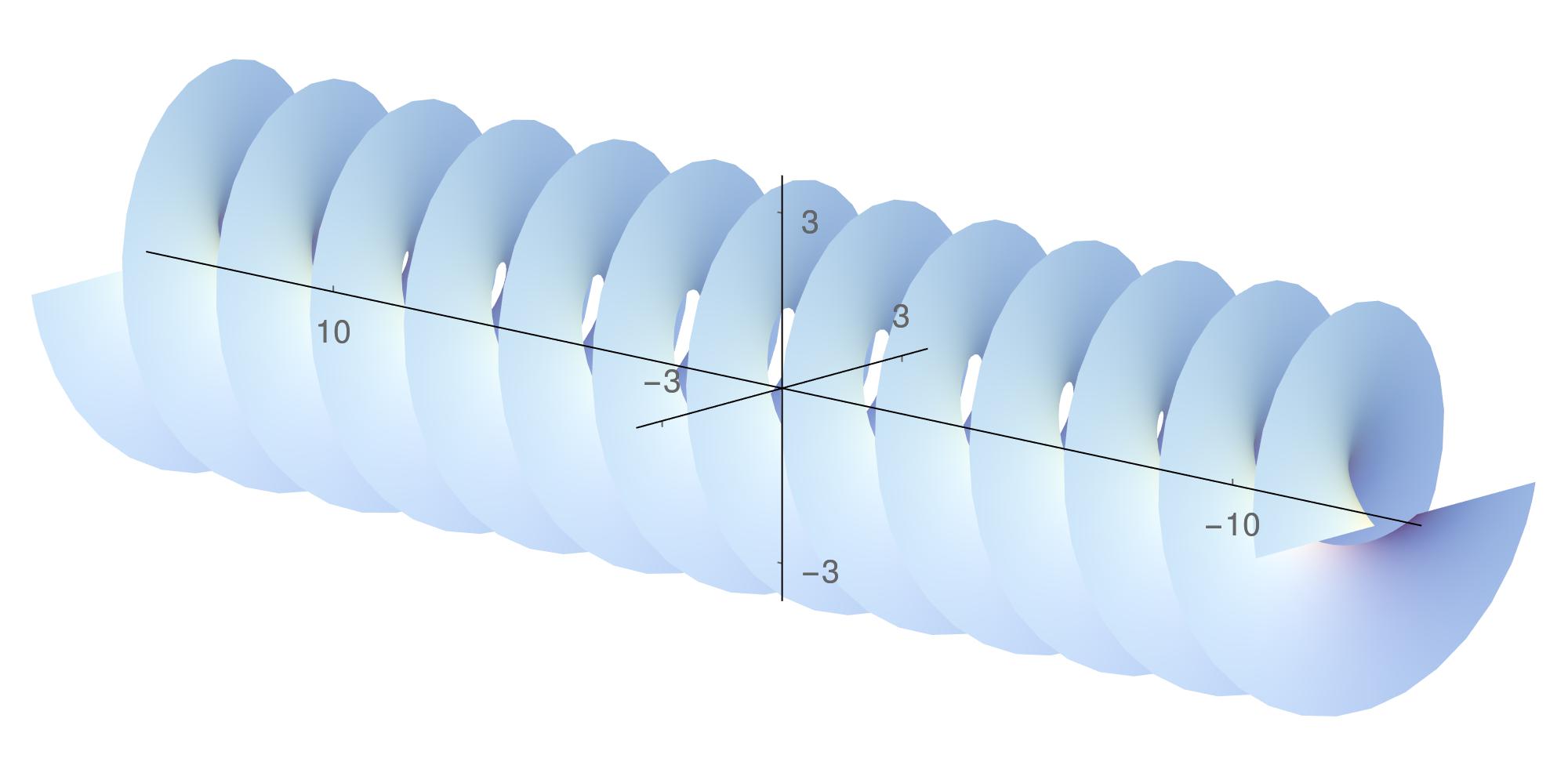}}
  \put(10,5){$m = 1$}
  \put(10,0){$\bPhase = 10$}
  \put(10,10){$\ffun = 0$}
  \put(5,34){$x^{3}$}
  \put(59,41){$x^{1}$}
  \put(90,34){$x^{2}$}
  \end{picture}
  }
 \else
  \begin{pspicture}(0,0)(12,5.0)
    \put(0,-0.7){\includegraphics[width=120mm]{LSOESTF_Fig513.eps}}
   \put(1,0.5){$m = 1$}
  \put(1,0.0){$\bPhase = 10$}
  \put(1,1){$\ffun = 0$}
  \put(0.5,3.4){$x^{3}$}
  \put(5.9,4.1){$x^{1}$}
  \put(9.0,3.4){$x^{2}$}
  \end{pspicture}
\fi
\end{center}
  \engrus{0.5ex}{0.5ex}{%
  \caption{\label{7629746911}Zero level surfaces of the field function $\ffun$ for the Gaussian twisted soliton with $m = 1$.}
  }{%
  \mifeng{\addtocounter{figure}{-1}}{}
  \caption{\label{7629746911r}Поверхность нулевого уровня полевой функции $\ffun$ для Гауссова вращающегося солитона с $m = 1$.}
  }
\end{figure}

\begin{figure}[h]
\begin{center}
\ifpdf
  {
  \unitlength 1mm
  \begin{picture}(120,50)
  \put(0,-5){\includegraphics[width=120mm]{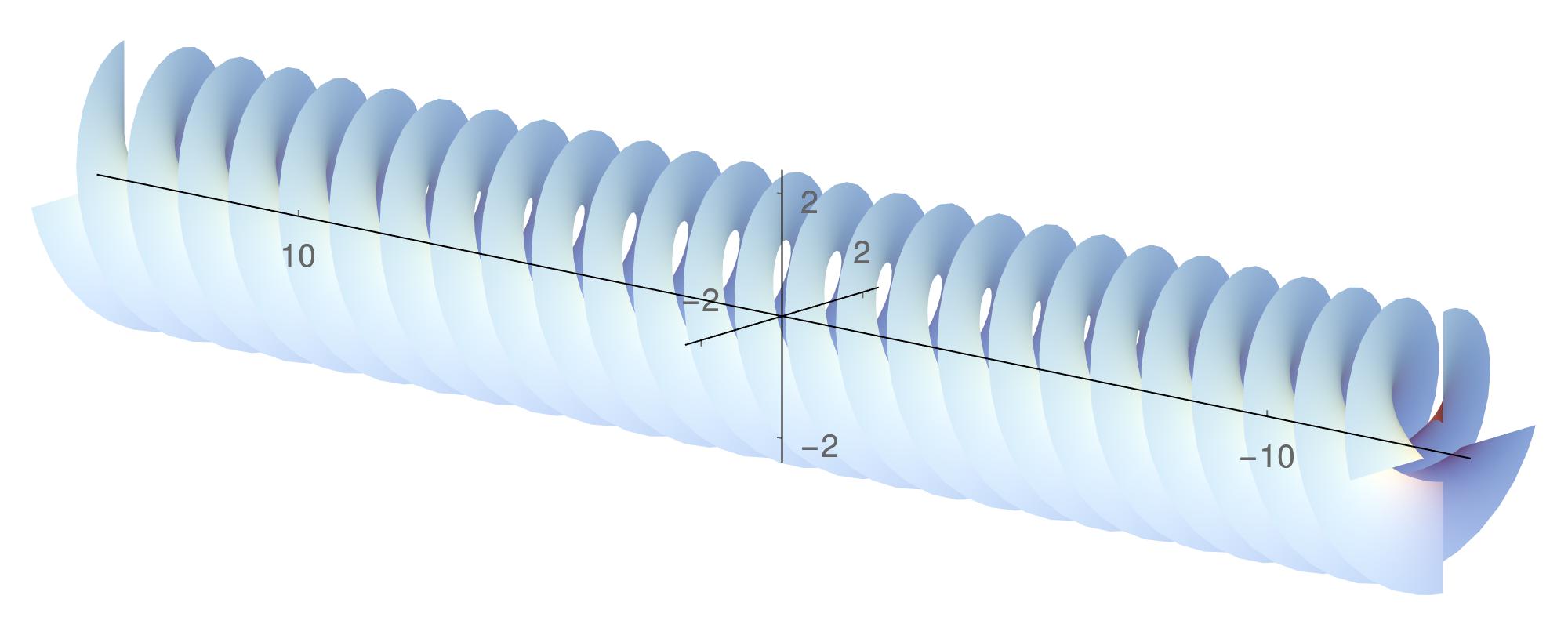}}
  \put(10,5){$m = 2$}
  \put(10,0){$\bPhase = 10$}
  \put(10,10){$\ffun = 0$}
  \put(2,30){$x^{3}$}
  \put(59,32){$x^{1}$}
  \put(81,27){$x^{2}$}
  \end{picture}
  }
 \else
  \begin{pspicture}(0,0)(12,5)
    \put(0,-0.5){\includegraphics[width=120mm]{LSOESTF_Fig514.eps}}
   \put(1,0.5){$m = 2$}
  \put(1,0.0){$\bPhase = 10$}
  \put(1,1){$\ffun = 0$}
  \put(0.2,3){$x^{3}$}
  \put(5.9,3.2){$x^{1}$}
  \put(8.1,2.7){$x^{2}$}
  \end{pspicture}
\fi
\end{center}
 \engrus{0.5ex}{0.5ex}{%
 \caption{\label{4755348811} Zero level surface of the field function $\ffun$ for the Gaussian twisted soliton with $m = 2$.}
 }{%
\mifeng{\addtocounter{figure}{-1}}{}
  \caption{\label{4755348811r} Поверхность нулевого уровня полевой функции $\ffun$ для Гауссова вращающегося солитона с $m = 2$.}
 }
\end{figure}

\engrus{0.5ex}{0.5ex}{%
All images  \mbox{\ref{76297469} -- \ref{4755348811}} are appropriate to the solitons twisted on the right.
}{%
Все изображения \mbox{\ref{76297469r} -- \ref{4755348811r}} соответствуют солитонам закрученным вправо.
}

\engrus{0.5ex}{0.5ex}{%
Here we have considered the simplest arbitrary functions $\Xi_{1}$ and $\Xi_{2}$, which give the twisted shell lightlike soliton with one cavity.
For more complicated cases we can have the appropriate solitons with a set of cavities.
But we will have the notable asymptotic relation between energy, momentum, and angular momentum
(\ref{417457381}) for these cases, because of the appropriate relation for the densities (\ref{332235821}).
}{%
Здесь мы рассмотрели простейшие произвольные функции $\Xi_{1}$ и $\Xi_{2}$, которые дают закрученный оболочечный светоподобный солитон с одной полостью.
Для более сложных случаев могут существовать соответствующие солитоны со множеством полостей.
Однако мы будем иметь примечательное асимптотическое соотношение между энергией, импульсом и моментом импульса
(\ref{417457381}) для этих случаев ввиду соответствующего соотношения для плотностей (\ref{332235821}).
}

\mifengrus{\newpage}{}
\engrus{3ex}{2ex}{%
\section{Relation to photons}
\label{relphot}
}{%
\mifeng{\addtocounter{section}{-1}}{}
\section{Отношение к фотонам}
\mifeng{}{\label{relphot}}
}

\engrus{0.5ex}{0.5ex}{%
\textremb{Because}\textaddb{In view} of \textaddb{the obtained} notable connection (\ref{417457381}) between energy, momentum, and angular momentum \textremb{of}\textaddb{for} the twisted
lightlike solitons, it is reasonable to consider their relation to photons.
}{%
Ввиду полученной примечательной связи (\ref{417457381}) между энергией, импульсом и моментом импульса для вращающихся светоподобных солитонов,
резонно рассмотреть их отношение к фотонам.
}

\engrus{0.5ex}{0.5ex}{%
For this purpose first we consider an ideal gas of these solitons in bounded three-dimensional volume $\Vols$.
}{%
Для этой цели мы рассматриваем сначала идеальный газ этих солитонов в ограниченном трёхмерном объёме.
}

\engrus{0.5ex}{0.5ex}{%
As \textadd1{it} is known, the ideal gas \textremb{behaviour}\textaddb{model} is characterized by \textadd1{the} zero interaction between the particles. But
an interaction of the particles with the \textremb{volume} walls \textaddb{of the bounding volume} provides
thermodynamic equilibrium of the ideal gas.
}{%
Как известно, модель идеального газа характеризуется отсутствием взаимодействия между частицами.
 При этом взаимодействие частиц со стенками вмещающего объёма обеспечивает термодинамическое равновесие идеального газа.
}

\engrus{0.5ex}{0.5ex}{%
\textadd1{
Thus we consider a great number of the twisted lightlike solitons in the three-volume $\Vols$ for the case of negligible their interaction with each other
because of the adequate mutual remoteness.}
}{%
Таким образом мы рассматриваем большое число вращающихся светоподобных солитонов в трёхмерном объёме $\Vols$ для случая пренебрежимого их взаимодействием друг с другом
ввиду достаточной взаимной удалённости.
}

\engrus{0.5ex}{0.5ex}{%
Let us suppose that absorptive
and emissive capacities
of the walls are provided by soliton-particles having the following constant absolute value of angular momentum
}{%
Допустим, что поглощательная и испускательная способности стенок обеспечивается со\-лито\-нами-частицами, имеющими следующее постоянное абсолютное значение момента импульса:
}
\begin{equation}
\label{539834181}
\AMV_{e}  = \frac{\hbar}{2}
\;,
\end{equation}
\engrus{0.5ex}{0.5ex}{%
\noindent
where $\hbar$ is Planck constant.
}{%
\noindent
где $\hbar$ -- постоянная Планка.
}

\engrus{0.5ex}{0.5ex}{%
We suppose also that each lightlike soliton can interact simultaneously with only one soliton-particle of the wall.
We assume \textadd1{the} angular momentum conservation for the combination of lightlike soliton with \textadd1{the} soliton-particle
of the wall in \textadd1{an} absorption \textremb{or}\textaddb{and} \textadd1{an} emission event.
}{%
Мы предполагаем также, что каждый светоподобный солитон может одновременно взаимодействовать только с одним солитоном-частицей стенки.
Мы принимаем сохранение момента импульса для комбинации светоподобного солитона с солитоном-частицей стенки в акте поглощения и испускания.
}

\engrus{0.5ex}{0.5ex}{%
Then, because of the angular momentum conservation, \textadd1{an} absorption or \textadd1{an} emission of a twisted lightlike soliton is possible only when the angular momentum of \textadd1{the} soliton-particle in the wall
is oppositely directed to the angular momentum of the lightlike soliton.
The soliton-particle angular momentum is reversed in \textremd{an}\textaddd{the} absorption
\textremd{or}\textaddd{and} \textremd{an}\textaddd{the} emission event.
}{%
Тогда, ввиду сохранения момента импульса, поглощение или испускание вращающегося светоподобного солитона возможно только в том случае, если момент импульса солитона-частицы стенки направлен противоположно к
моменту импульса светоподобного солитона.
Момент импульса солитона-частицы стенки переворачивается (меняет направление на противоположное) в акте поглощения или испускания.
}

\engrus{0.5ex}{0.5ex}{%\mifrus{}{}%\noindent%
\textaddd{%
According to known theses of quantum physics,
these conditions correspond exactly to absorption and emission processes of photons by electrons of real walls.
The fixity of the angular momentum absolute value of the electron (\ref{539834181}) is the principal feature here.
The electrons interacting with photons can be conduction ones or it can be bound in atoms.
In any case the electrons interacting also with a lattice take part in a very complicated motion.
As \textaddd{a} result of such complex interactions we have that any incident photon is absorbed by some electron for
a sufficient thickness of the wall. The electrons, in one's turn, emit the photons.
We suppose the same situation for the twisted lightlike solitons interacting with the wall.%
}
}{%\mifeng{}{}%\noindent%
Согласно известным положениям квантовой физики эти условия в точности соответствуют процессам поглощения и испускания фотонов электронами реальных стенок.
Принципиальным здесь является фиксированность модуля момента импульса электрона (\ref{539834181}).
Взаимодействующие с фотонами электроны могут быть электронами проводимости или быть связаны в атомах.
В любом случае электроны, взаимодействуя также с решёткой, участвуют в весьма сложном движении.
В результате таких сложных взаимодействий получается, что при достаточной толщине стенок любой падающий на неё
фотон поглощается каким-либо электроном. Электроны в свою очередь испускают фотоны. Такую же ситуацию
мы предполагаем и для взаимодействующих со стенкой вращающихся светоподобных солитонов.
}

\engrus{0.5ex}{0.5ex}{%
Thus the absolute value of angular momentum of \textaddb{any} twisted lightlike soliton in the volume $\Vols$
must be equal to $\hbar$ \textaddb{in the case of thermodynamic equilibrium}.
}{%
Таким образом абсолютное значение момента импульса любого вращающегося светоподобного солитона в объёме $\Vols$ должно быть равно $\hbar$
в случае термодинамического равновесия.
}

\engrus{0.5ex}{0.5ex}{%
The structure of twisted lightlike solitons
depends on \textadd1{the} structure and states of emissive and absorbent
soliton-particles. We must define the value of twist parameter $m$ and the scale phase function $\btam(\Phase)$ for the twisted lightlike solitons in the volume $\Vols$.
}{%
Структура вращающихся светоподобных солитонов зависит от структуры и состояния испускающих и поглощающих солитонов-частиц. Мы должны определить значение параметра вращения
$m$ и масштабную фазовую функцию $\btam(\Phase)$ для вращающихся светоподобных солитонов в объёме $\Vols$.
}

\begin{subequations}\label{485500821}
\engrus{0.5ex}{0.5ex}{%
Let us consider the case
}{%
Рассмотрим случай
}
\begin{equation}
\label{374045971}
m=1
\;.
\end{equation}
\engrus{0.5ex}{0.5ex}{%
As we see in (\ref{35855292}), in this case the energy of the soliton is logarithmically divergent in
infinite space. But here we consider the finite volume, where its energy is finite.
}{%
Как мы видим в (\ref{35855292}), в этом случае энергия солитона логарифмически расходится в бесконечном пространстве.
Однако здесь мы рассматриваем конечный объём, где его энергия конечна.
}

\engrus{0.5ex}{0.5ex}{%
Strictly speaking, the obtained soliton solutions must be modified  for \textadd1{the} finite volume.
But here we consider the integral characteristics of the solitons only.
Thus we can consider the soliton solutions of infinite space for the finite volume
in some approximation.
}{%
Строго говоря, полученные солитонные решения должны быть модифицированы для конечного объёма.
Однако здесь мы рассматриваем только интегральные характеристики солитонов.
Таким образом мы можем рассматривать солитонные решения бесконечного пространства для конечного объёма в некотором приближении.
}

\engrus{0.5ex}{0.5ex}{%
Let us suppose also that the scale phase function $\btam(\Phase)$ is slow variable:
}{%
Предположим также, что масштабная фазовая функция $\btam(\Phase)$ является медленно меняющейся величиной:
}
\begin{equation}
\label{371726911}
\btam^{\prime} \to 0
\;.
\end{equation}
\end{subequations}

\begin{subequations}\label{490939841}
\engrus{0.5ex}{0.5ex}{%
Thus, taking into account (\ref{338555621}) -- (\ref{817526421}) and (\ref{485500821}),
we have the following relations for \textremb{the}\textaddb{any} twisted lightlike soliton in the volume $\Vols$:
}{%
Таким образом, учитывая (\ref{338555621}) -- (\ref{817526421}) и (\ref{485500821}),
мы имеем следующие соотношения для любого вращающегося светоподобного солитона в объёме $\Vols$:
}
\begin{align}
\label{491083471}
\Energy &= \EMV + \underline{\Energy}
\;,\\
\label{491083472}
\EMV  &= \omega\,\hbar
\;,\\
\label{76338054}
\AMV &= \hbar
\;,
\end{align}
\engrus{0.5ex}{0.5ex}{%
\noindent
where
}{%
\noindent
где
}
\begin{equation}
\label{387320521}
 \underline{\Energy} = \int\limits_{\Vols}\fcE_{0}\,\dVols
\;,
\end{equation}
\engrus{0.5ex}{0.5ex}{%
\noindent
$\fcE_{0}$ is \textaddb{the} static part of energy density $\cE$ for \textaddb{the} lightlike soliton in expression (\ref{495868961}).
}{%
\noindent
$\fcE_{0}$ статическая часть плотности энергии $\cE$ для светоподобного солитона в выражении (\ref{495868961}).
}
\end{subequations}

\engrus{0.5ex}{0.5ex}{%
As we see in (\ref{495868961}), the static part of energy $\underline{\Energy}$ is independent explicitly of the
soliton frequency $\omega$. But the condition (\ref{76338054}) with expressions (\ref{38894581}) and (\ref{817526421}) gives a dependence of the soliton transversal size
\textaddb{(}characterized by the parameter $\cbrho$\textaddb{)} from the soliton frequency $\omega$.
Thus according to (\ref{338678749}) and (\ref{374782791}) the static energy $\underline{\Energy}$ is implicitly dependent
on the frequency $\omega$.
}{%
Как мы видим в (\ref{495868961}), статическая часть энергии $\underline{\Energy}$ не зависит явно от частоты солитона
$\omega$. Однако условие (\ref{76338054}) с выражениями (\ref{38894581}) и (\ref{817526421}) даёт зависимость поперечного размера солитона (характеризуемого параметром
$\cbrho$) от частоты солитона $\omega$.
Таким образом в соответствии с (\ref{338678749}) и (\ref{374782791}) статическая энергия $\underline{\Energy}$ неявно зависит от частоты $\omega$.
}

\engrus{0.5ex}{0.5ex}{%
An estimation for this dependence will be made below. But at first for simplicity we consider that the static energy $\underline{\Energy}$ is approximately constant:
}{%
Оценка этой зависимости будет сделана ниже. Однако сначала для простоты мы считаем, что статическая энергия $\underline{\Energy}$ приближённо постоянна:
}
\begin{equation}
\label{460697971}
\underline{\Energy}  \approx \const
\;.
\end{equation}

\engrus{0.5ex}{0.5ex}{%
The finiteness of the volume under consideration confines the set of possible
frequencies of the solitons.
As it is known, the field in any finite volume can be represented by
the appropriate mode expansion.
In the case of cuboid we have the simple space-time Fourier components, which
satisfies the periodic boundary conditions.
}{%
Конечность рассматриваемого объёма ограничивает множество возможных частот солитонов.
Как известно, поле в любом конечном объёме может быть представлено соответствующим разложением по модам.
В случае прямоугольного параллелепипеда мы имеем простые пространственно-временные Фурье-компоненты, которые удовлетворяют периодическим граничным условиям.
}

\engrus{0.5ex}{0.5ex}{%
In the case of arbitrary volume with cavities, the finding of volume modes looks very complicated.
Here we \textremb{consider}\textaddb{assume} that the cavities inside the soliton shells are sufficiently small to
neglect of their influence. Also we take that each soliton in the volume has
one of its allowed frequencies.
}{%
В случае произвольного объёма с полостями нахождение мод объёма выглядит
очень сложным. Здесь мы предполагаем, что полости внутри солитонных оболочек
достаточно малы, чтобы пренебречь их влиянием. Также мы принимаем, что каждый солитон в объёме имеет одну из его допустимых частот.
}

\engrus{0.5ex}{0.5ex}{%
Hereafter up to formulas (\ref{558433091}) we obtain the equilibrium distribution function
by soliton frequencies. The appropriate derivation of formulas is similar to
ones represented in \textaddd{the} classical works by S.~Bose \cite{BoseSN1924E}, A.~Einstein \cite{Einstein1924aE,Einstein1925aE}, and contained in monographs
(see, for example, \cite{BrillouinL1920Fr}).
}{%
Далее вплоть до формул (\ref{558433091}) мы получаем равновесную функцию распределения по солитонным частотам. Соответствующий вывод формул сходен с выводами, представленными в классических работах С.~Бозе \cite{BoseSN1924E}, А.~Эйнштейна \cite{Einstein1924aE,Einstein1925aE} и входящими в монографии
(см., например, \cite{BrillouinL1920Fr}).
}

\engrus{0.5ex}{0.5ex}{%
As distinct from \textaddb{the} cited works,
here we use the natural energy cells instead of finite phase space cells.
Complete deduction is expounded to show that all assumptions are in the framework of \textaddb{the}
real soliton dynamics only.
}{%
В отличие от цитированных работ, здесь мы используем естественные энергетические ячейки вместо конечных ячеек фазового пространства.
Для того, чтобы показать, что все предположения
находятся только в рамках реальной солитонной динамики, представлен полный вывод.
}

\engrus{0.5ex}{0.5ex}{%
For simplicity let us consider the volume
$\Vols$ in cubic form with side $\vdlina$.
Then the allowed frequencies are defined by formula
}{%
Для простоты рассмотрим объём $\Vols$ в виде куба со стороной $\vdlina$.
Тогда допустимые частоты определяются формулой
}
\begin{equation}
\label{387676861}
\omega_i  = \frac{2\pi}{\lambda_i} = \frac{2\pi}{\vdlina}\,\bar{n}_{i} = \frac{2\pi}{\vdlina}\,\sqrt{n_1^2+n_2^2+n_3^2}
\;,
\end{equation}
\engrus{0.5ex}{0.5ex}{%
\noindent
where $\{n_1,n_2,n_3\}$ are integer numbers, excepting the case when all number are zero,
$i$ is the index for different frequencies.
}{%
\noindent
где $\{n_1,n_2,n_3\}$ -- целые числа, за исключением случая, когда все числа равны нулю,
$i$ -- индекс различных частот.
}

\engrus{0.5ex}{0.5ex}{%
According to (\ref{387676861}) we have the following minimal frequency in the volume:
}{%
В соответствии с (\ref{387676861}) имеем следующую минимальную частоту в объёме:
}
\begin{equation}
\label{692670071}
\omega_{\mathrm{min}}  = \frac{2\pi}{\vdlina}
\;.
\end{equation}

\begin{subequations}\label{434323461}
\engrus{0.5ex}{0.5ex}{%
If there are $\NVols_{i}$ solitons with frequency $\omega_{i}$ in the volume $\Vols$,
then the full energy of \textaddb{the} solitons \textremb{in it}is given by \textadd1{the} formula
}{%
Если в объёме $\Vols$ имеется $\NVols_{i}$ солитонов с частотой $\omega_{i}$, то полная энергия солитонов даётся формулой
}
\begin{equation}
\label{431673021}
\Ugas  = \sum\limits_{i=1}^{\infty} \NVols_{i}\,\Energy_{i}
\;,
\end{equation}
\engrus{0.5ex}{0.5ex}{%
\noindent
where $\Energy_{i}$ is \textadd1{the} energy of the soliton with frequency $\omega_{i}$,
}{%
\noindent
где $\Energy_{i}$ -- энергия солитона с частотой $\omega_{i}$,
}
\begin{align}
\label{421593311}
 \Energy_{i} &= \omega_{i}\,\hbar + \underline{\Energy}
\;,\\
\label{421593312}
\NVols &= \sum\limits_{i=1}^{\infty} \NVols_{i}
\;,
\end{align}
\engrus{0.5ex}{0.5ex}{%
\noindent
$\NVols$ is a total number of solitons in the volume $\Vols$.
}{%
\noindent
$\NVols$ -- полное количество солитонов в объёме $\Vols$.
}
\end{subequations}

\engrus{0.5ex}{0.5ex}{%
Because there is the minimal frequency $\omega_{\mathrm{min}}$ (\ref{692670071}) for
the solitons in the volume $\Vols$, then according to (\ref{434323461})
we have the following
expression for their maximal quantity:
}{%
Поскольку имеется минимальная частота $\omega_{\mathrm{min}}$ (\ref{692670071}) для солитонов в объёме $\Vols$, то в соответствии с (\ref{434323461}) мы имеем  следующее выражение для их максимального
количества:
}
\begin{equation}
\label{435015271}
 \NVols_{\mathrm{max}} = \frac{\Ugas }{\omega_{\mathrm{min}}\,\hbar +\underline{\Energy}}
\;.
\end{equation}

\engrus{0.5ex}{0.5ex}{%
If we suppose that a full angular momentum as well as a full momentum of the
soliton gas in the volume
$\Vols$ are zero, then the total number of solitons must be even. Thus their minimal
quantity is $2$ and we have from (\ref{434323461}) the
maximal value for frequency:
}{%
Если мы предполагаем, что полный момент импульса также как и полный импульс
солитонного газа в объёме $\Vols$ есть ноль, то полное число солитонов должно быть чётным. Таким образом их минимальное количество $2$ и мы имеем из (\ref{434323461}) максимальное значение частоты:
}
\begin{equation}
\label{700370271}
\NVols_{\mathrm{min}}  = 2\;,\quad
\omega_{\mathrm{max}}  = \frac{\Ugas  - 2\,\underline{\Energy}}{2\,\hbar}
\;.
\end{equation}

\engrus{0.5ex}{0.5ex}{%
Among all the possible distributions by soliton frequencies
$\{\NVols_i\}$ there is \textremb{a piece}\textaddb{their part}
 providing an identical total energy
\textaddb{of the gas} $\Ugas$.
According to general principles of statistical physics such distributions are considered as equally probable.
}{%
Среди всех возможных распределений по солитонным частотам $\{\NVols_i\}$ имеется их часть, обеспечивающая одинаковую полную энергию газа $\Ugas$. Согласно основным принципам статистической физики такие распределения рассматриваются как равновероятные.
}

\engrus{0.5ex}{0.5ex}{%
Let us introduce the size of energy cell $\gVols_{i}$, which are the quantity
of solitons
having the energy $\Energy_{i}$ and the corresponding frequency $\omega_{i}$.
Different states in the sell are defined with the set of numbers $\{n_1,n_2,n_3\}$ in (\ref{387676861})
for frequency $\omega_{i}$ and two directions of twist (right and left).
}{%
Введём размер энергетической ячейки $\gVols_{i}$, который представляет собой количество солитонов, имеющих энергию $\Energy_{i}$ и соответствующую частоту $\omega_{i}$. Различные состояния в ячейке определяются набором чисел $\{n_1,n_2,n_3\}$ в (\ref{387676861})
для частоты $\omega_{i}$ и двумя направлениями закрученности (правой и левой).
}

\begin{subequations}\label{705021911}
\engrus{0.5ex}{0.5ex}{%
Let us count \textremb{up}the number of ways to provide the part of total energy $\Ugas$
produced by the solitons with energy $\Energy_{i}$ that is $\NVols_{i}\,\Energy_{i}$  (\ref{431673021}).
According to known representation we line up
$\NVols_{i}$ solitons ($\circ$) and $(\gVols_{i} -1)$ dividing walls ($|$)
in random order:
}{%
Подсчитаем количество способов,
обеспечивающих часть полной энергии $\Ugas$, порождённой солитонами с энергией $\Energy_{i}$, то есть $\NVols_{i}\,\Energy_{i}$  (\ref{431673021}).
Согласно известному представлению мы выстраиваем в ряд
$\NVols_{i}$ солитонов ($\circ$) и $(\gVols_{i} -1)$ перегородок ($|$)
в случайном порядке:
}
\begin{equation}
\label{601699131}
\circ\circ|\circ\circ\circ|\phantom{\circ}|\circ|\circ\circ\circ\circ|\circ\circ|
\cdots
\circ\circ \cdots|\circ\circ\circ|\phantom{\circ}|\circ
\;.
\end{equation}
\engrus{0.5ex}{0.5ex}{%
\noindent
Here the dividing walls ($|$) separate the different soliton states (\textaddc{numbers} $\{n_1,n_2,n_3\}$ and twist direction).
}{%
\noindent
Здесь перегородки ($|$) отделяют различные солитонные состояния (числа $\{n_1,n_2,n_3\}$ и направление закрученности).
}

\engrus{0.5ex}{0.5ex}{%
In that case, the permutation number $(\NVols_{i} + \gVols_{i} -1)!$ is a total number of distributions
for solitons with energy $\Energy_{i}$. Then we take into account that
$\NVols_{i}!$ permutations of solitons and
$(\gVols_{i} -1)!$ permutations of dividing walls
correspond to one state. As \textadd1{a} result, we have the sought number of ways
to provide the part $\NVols_{i}\,\Energy_{i}$ of total energy $\Ugas$:
}{%
В таком случае число перестановок $(\NVols_{i} + \gVols_{i} -1)!$
есть полное число распределений для солитонов с энергией $\Energy_{i}$.
Затем мы учитываем, что
$\NVols_{i}!$ перестановок солитонов и
$(\gVols_{i} -1)!$ перестановок перегородок соответствуют
одному состоянию. В результате имеем искомое число способов обеспечения
части $\NVols_{i}\,\Energy_{i}$ полной энергии $\Ugas$:
}
\begin{equation}
\label{443160701}
\Wspos_{i}  = \frac{(\NVols_{i} + \gVols_{i} -1)!}{\NVols_{i}!\,(\gVols_{i} - 1)!}
\;.
\end{equation}

\engrus{0.5ex}{0.5ex}{%
We obtain the total number of ways providing the energy $\Ugas$ by multiplication
of the numbers $\Wspos_{i}$:
}{%
Полное число способов, обеспечивающих энергию $\Ugas$, получаем
перемножением чисел $\Wspos_{i}$:
}
\end{subequations}
\begin{equation}
\label{443160702}
\Wspos = \prod\limits_{i=1}^{\infty}\Wspos_{i} = \prod\limits_{i=1}^{\infty}\frac{(\NVols_{i} + \gVols_{i} -1)!}{\NVols_{i}!\,(\gVols_{i} - 1)!}
\;.
\end{equation}

\engrus{0.5ex}{0.5ex}{%
According to \textadd1{the} usual method, we take \textadd1{into account} that the most probable distribution provided \textremb{with}\textadd1{by the}
maximum number of the ways $\Wspos$ corresponds to equilibrium.
The total number of solitons $\NVols$ is not fixed here.
}{%
Согласно обычному методу мы принимаем, что равновесию соответствует наиболее вероятное распределение, обеспечивающееся максимальным числом способов. Общее число солитонов здесь не фиксируется.
}

\engrus{0.5ex}{0.5ex}{%
Let us solve the problem for \textadd1{the} maximization of number $\Wspos$ with \textadd1{the} fixed total energy $\Ugas$. (\ref{431673021}).
For this purpose the method of Lagrange multipliers is used.
For \textadd1{the} convenience we maximize the natural logarithm of number $\Wspos$.
Thus the problem for finding of the equilibrium distribution $\{\NVols_{i}\}$ take the form:
}{%
Решим задачу максимизации числа $\Wspos$ с фиксированной полной энергией $\Ugas$. (\ref{431673021}).
Для этой цели используется метод множителей Лагранжа. Для удобства мы максимизируем натуральный логарифм числа
$\Wspos$. Таким образом задача нахождения равновесного распределения $\{\NVols_{i}\}$ принимает вид:
}
\begin{equation}
\label{347928321}
\underline{\Sentr} =
\ln\Wspos -\Tenerg^{-1}\,\Ugas
\;,\quad
\underline{\Sentr}\;  \to\;\; \mathrm{max}
\;,
\end{equation}
\engrus{0.5ex}{0.5ex}{%
\noindent
where $\Tenerg^{-1}$ is Lagrange multiplier, the parameter $\Tenerg$ has
a physical dimension of energy.
}{%
\noindent
где $\Tenerg^{-1}$ -- множитель Лагранжа, параметр $\Tenerg$ имеет физическую размерность энергии.
}

\engrus{0.5ex}{0.5ex}{%
Let us consider the case when the numbers $\NVols_{i}$ and $\gVols_{i}$ are
sufficiently great. In this case we use the Stirling formula for factorial of number.
Thus for $\NVols_{i}\gg 1$ and $\gVols_{i}\gg 1$ we have
}{%
Рассмотрим случай, когда числа $\NVols_{i}$ и $\gVols_{i}$ достаточно велики. В этом случае мы используем формулу Стирлинга для факториала числа.
Таким образом для $\NVols_{i}\gg 1$ и $\gVols_{i}\gg 1$ имеем
}
\begin{equation}
\label{319083391}
\ln\Wspos   \approx \sum\limits_{i=1}^{\infty}\Bigl(\left(\NVols_{i} + \gVols_{i}\right)
\ln\left(\NVols_{i} + \gVols_{i}\right) - \NVols_{i}\,\ln\NVols_{i} - \gVols_{i}\ln\gVols_{i}\Bigr)
\;.
\end{equation}

\engrus{0.5ex}{0.5ex}{%
Considering the sequence of numbers $\NVols_{i}$ as quasicontinuous, we have the following
necessary conditions for maximum of the function $\underline{\Sentr}$:
}{%
Рассматривая последовательность чисел $\NVols_{i}$ как квазинепрерывную, имеем следующее необходимое условие
максимума функции $\underline{\Sentr}$:
}
\begin{equation}
\label{327726911}
\frac{\p\underline{\Sentr}}{\p\NVols_{i}}  = 0
\;.
\end{equation}

\engrus{0.5ex}{0.5ex}{%
From (\ref{327726911})  with (\ref{347928321}), (\ref{319083391}), and (\ref{431673021})
we have the following equilibrium distribution:
}{%
Из (\ref{327726911})  с (\ref{347928321}), (\ref{319083391}) и (\ref{431673021})
имеем следующее равновесное распределение:
}
\begin{equation}
\label{361607831}
 \NVols_{i}  = \frac{\gVols_{i}}{\e{\Energy_{i}/\Tenerg} - 1}
\;.
\end{equation}
\engrus{0.5ex}{0.5ex}{%
Here the constant $\Tenerg$ can be expressed through the total energy $\Ugas$ by using the
condition (\ref{431673021}). Thus the physical quantity $\Tenerg$ is an energy parameter of the distribution (\ref{361607831}).
}{%
Здесь константа $\Tenerg$ может быть выражена через полную энергию $\Ugas$ с использованием условия (\ref{431673021}). Таким образом физическая величина $\Tenerg$ представляет собой энергетический параметр распределения (\ref{361607831}).
}

\engrus{0.5ex}{0.5ex}{%
Let us use the representation of quasicontinuous soliton energy spectrum
to obtain the size of energy cell $\gVols_{i}$.
In this case the energy cell $\gVols_{i}$ is characterized by energy gap from $\Energy_{i}$ to $\Energy_{i} + \Delta\Energy_{i}$.
}{%
Воспользуемся представлением квазинепрерывного спектра энергии солитона для получения размера энергетической ячейки
$\gVols_{i}$.
В этом случае энергетическая ячейка $\gVols_{i}$ характеризуется энергетическим интервалом от $\Energy_{i}$ до $\Energy_{i} + \Delta\Energy_{i}$.
}

\begin{subequations}\label{558433091}
\engrus{0.5ex}{0.5ex}{%
\textrem1{Having in view}\textadd1{Keeping in mind the} one-to-one correspondence
between the number $\bar{n}$ in (\ref{387676861}), frequency, and energy (\ref{490939841}),
we can obtain the quantity of different soliton states with
frequencies from $\omega_{i}$ to $\omega_{i} + \Delta\omega_{i}$.
A spherical layer in the space of numbers $\{n_1,n_2,n_3\}$ corresponds to
the frequency interval $\Delta\omega_{i}$.
Taking into account also the two directions of twist and proceeding to the limit
$\Delta\omega_{i} \to 0$, we obtain
}{%
Имея ввиду взаимно-однозначное соответствие между числом $\bar{n}$ в (\ref{387676861}), частотой и энергией (\ref{490939841}),
мы можем получить количество различных солитонных состояний с частотами от $\omega_{i}$ до $\omega_{i} + \Delta\omega_{i}$.
Сферический слой в пространстве чисел $\{n_1,n_2,n_3\}$ соответствует интервалу частот $\Delta\omega_{i}$.
Учитывая также два направления закрученности и переходя к пределу
$\Delta\omega_{i} \to 0$, получаем
}
\begin{equation}
\label{715477301}
 \gVols_{\omega,\Delta\omega}  \approx 2\cdot4\pi\,n^2\,\Delta n \approx \frac{\vdlina^{3}}{\pi^2}\,\omega^2\,\Delta\omega
\quad\to\quad
\gVols_{\omega}\,\df\omega  =  \frac{\vdlina^{3}}{\pi^2}\,\omega^2\,\df\omega
\;.
\end{equation}
\begin{equation}
\label{716526671}
  \NVols_{\omega} =  \frac{\vdlina^{3}}{\pi^2}\frac{\omega^2}{\e{\Energy_{\omega}/\Tenerg} - 1}
\;,
\end{equation}
\engrus{0.5ex}{0.5ex}{%
\noindent
where
}{%
\noindent
где
}
\begin{equation}
\label{716794521}
  \Energy_{\omega} = \hbar\,\omega + \underline{\Energy}
\;.
\end{equation}
\end{subequations}

\begin{subequations}\label{571394211}
\engrus{0.5ex}{0.5ex}{%
Then we integrate the expressions $\Energy_{\omega}\, \NVols_{\omega}$, and $\NVols_{\omega}$
with substitution $\NVols_{\omega}$ and $\Energy_{\omega}$ from (\ref{558433091})
over frequency from $\omega = 0$ to infinity.
As \textaddd{a} result, we obtain the following expressions for total energy and number of solitons
in the volume $\Vols$:
}{%
Затем мы интегрируем выражения $\Energy_{\omega}\, \NVols_{\omega}$ и $\NVols_{\omega}$
с подстановкой $\NVols_{\omega}$ и $\Energy_{\omega}$ из (\ref{558433091})
по частоте от $\omega = 0$ до бесконечности.
В результате получаем следующие выражения для полной энергии и числа солитонов в объёме $\Vols$:
}
\begin{align}
\label{571574991}
\Ugas  &= \frac{\vdlina^{3}\,\Tenerg^{4}\,6\,\pollog{4}{\e{-\underline{\Energy}/\Tenerg}}}{\pi^2\,\hbar^{3}}
+ \NVols\,\underline{\Energy}
\;,\\
\label{571574992}
\NVols &= \frac{\vdlina^{3}\,\Tenerg^{3}\,2\,\pollog{3}{\e{-\underline{\Energy}/\Tenerg}}}{\pi^2\,\hbar^{3}}
\;,
\end{align}
\engrus{0.5ex}{0.5ex}{%
\noindent
where $\pollog{s}{z}$ is polylogarithm function.
}{%
\noindent
где $\pollog{s}{z}$ -- полилогарифмическая функция.
}
\end{subequations}

\engrus{0.5ex}{0.5ex}{%
For connection between energy parameter $\Tenerg$ of distribution $\{\NVols_{i}\}$ (\ref{361607831})
and absolute temperature $\Tabs$ we take
}{%
Для связи между энергетическим параметром $\Tenerg$ распределения $\{\NVols_{i}\}$ (\ref{361607831})
и абсолютной температурой $\Tabs$ принимаем
}
\begin{equation}
\label{770372491}
\Tenerg = \kBol\, \Tabs
\;,
\end{equation}
\engrus{0.5ex}{0.5ex}{%
\noindent
where $\kBol$ is Boltzmann constant.
}{%
\noindent
где $\kBol$ -- постоянная Больцмана.
}

\begin{subequations}\label{389469291}
\engrus{0.5ex}{0.5ex}{%
The relation (\ref{770372491}) can be validated
by means of comparison between statistical determination for entropy $\Sentr$ and its thermodynamic
one for the case of constant volume
($\Vols = \const$):
}{%
Соотношение (\ref{770372491}) может быть подтверждено
посредством сравнения между статистическим определением энтропии $\Sentr$ и её термодинамическим определением для случая
постоянного объёма ($\Vols = \const$):
}
\begin{align}
\label{776201011}
  \Sentr &= \kBol\, \ln\Wspos
\;,\\
\label{776201012}
\df\Sentr  &= \frac{\df \Ugas}{\Tabs}
\;.
\end{align}
\engrus{0.5ex}{0.5ex}{%
But because the equivalence of these determinations must be postulated,
it is reasonable here to postulate the relation (\ref{770372491}).
}{%
Однако, поскольку эквивалентность этих определений должна постулироваться, резонно здесь постулировать соотношение
(\ref{770372491}).
}
\end{subequations}

\engrus{0.5ex}{0.5ex}{%
Let us write the equilibrium energy spectral density for the twisted
lightlike solitons in the volume $\Vols$.
According to
(\ref{558433091}) and (\ref{770372491}) we have
}{%
Выпишем равновесную спектральную плотность энергии для закрученных светоподобных солитонов
в объёме $\Vols$.
Согласно
(\ref{558433091}) и (\ref{770372491}) имеем
}
\begin{equation}
\label{405144411}
 \uVols (\omega,\Tabs) \eqdef \frac{\Energy_{\omega}\,\NVols_{\omega}}{\Vols}
 = \frac{\omega^2}{\pi^2}\,\frac{\hbar\,\omega + \underline{\Energy}}{\exp{\frac{\hbar\,\omega + \underline{\Energy}}{\kBol \Tabs}} - 1}
\;.
\end{equation}

\engrus{0.5ex}{0.5ex}{%
For the case of negligible static soliton energy $\underline{\Energy}\to 0$,
we have from (\ref{405144411}) the following known Planck formula for photons:
}{%
Для случая пренебрежимо малой статической энергии солитона $\underline{\Energy}\to 0$
имеем из (\ref{405144411}) следующую известную формулу Планка для фотонов:
}
\begin{equation}
\label{401602181}
\uVols (\omega,\Tabs) =  \frac{\omega^2}{\pi^2}\,\frac{\hbar\,\omega}{\exp{\frac{\hbar\,\omega}{\kBol \Tabs}} - 1}
\;.
\end{equation}

\engrus{0.5ex}{0.5ex}{%
Thus we can consider the relation between \textaddb{the} twisted lightlike solitons and photons.
}{%
Таким образом мы можем рассматривать соответствие между закрученными светоподобными солитонами и фотонами.
}

\engrus{0.5ex}{0.5ex}{%
Now let us estimate the possible values of soliton parameters in the volume $\Vols$ using
certain suppositions.
}{%
Теперь оценим возможные значения солитонных параметров в объёме $\Vols$, используя некоторые предположения.
}

\engrus{0.5ex}{0.5ex}{%
Taking into account (\ref{371726911}), we put \mbox{$\btam^{\prime} = 0$} and without \textadd1{the} loss of generality
\mbox{$\btam = 1$}.
}{%
Учитывая (\ref{371726911}), полагаем \mbox{$\btam^{\prime} = 0$} и без ограничения общности \mbox{$\btam = 1$}.
}

 \engrus{0.5ex}{0.5ex}{%
 Let the longitudinal size of the soliton
 in (\ref{374782791}) and the external diameter of \textadd1{the} cylindrical integration domain
 in (\ref{354384451})
 be equal to the side of the considered cubic volume:
 }{%
  Пусть продольный размер солитона в
 (\ref{374782791}) и внешний диаметр цилиндрической области интегрирования в  (\ref{354384451})
равны стороне рассматриваемого кубического объёма:
 }
 \begin{equation}
 \label{685819681}
  \zdlina = 2\,\trhoinf = \vdlina
 \;.
 \end{equation}

\begin{subequations}\label{681494511}
\engrus{0.5ex}{0.5ex}{%
Then, taking into account (\ref{374045971}), we have the following values contained in
(\ref{374782791}) -- (\ref{817526421}) for the metric signature (\ref{43842964a}):
}{%
Затем, учитывая (\ref{374045971}), имеем следующие значения, содержащиеся в
(\ref{374782791}) -- (\ref{817526421}) для сигнатуры метрики (\ref{43842964a}):
}
\begin{align}
\label{636939441}
 &\pint_{0}=\tilde{\pint}_{1}=\omega\,\vdlina\;,\quad\pint_{2}=0
\;,\\
\nonumber
 &\pcoef_{0} =\frac{1}{3}  - \frac{2\,\cbrho^{2}}{\vdlina^{2}} + \frac{32\,\cbrho^{6}}{3\,\vdlina^{6}}
 \;,\quad \pcoef_{2} = 1
 \;,\\
 \label{43305056}
 &\pcoef_{1} =\ln\biggl(\frac{\vdlina}{2\,\cbrho}\biggr)   + \frac{13}{72}
- \frac{8\,\cbrho^{4}}{3\,\vdlina^{4}}
- \frac{32\,\cbrho^{8}}{9\,\vdlina^{8}}
\;.
\end{align}
\end{subequations}

\engrus{0.5ex}{0.5ex}{%
\textadd1{The} condition (\ref{76338054}) with expression (\ref{38894581}) gives relation
}{%
Условие (\ref{76338054}) с выражением (\ref{38894581}) даёт соотношение
}
\begin{equation}
\label{745250891}
\cbrho^{4}\,\omega\,\vdlina\,\pcoef_{1}  = \xxx^2\,\hbar
\;.
\end{equation}

\engrus{0.5ex}{0.5ex}{%
Thus, by virtue of \textadd1{the} fixedness of the angular momentum of the soliton (\ref{76338054}), the radius of its shell $\cbrho$ depends on frequency $\omega$. But to calculate $\cbrho$ we must have
the value of the constant $\xxx$.
}{%
Таким образом, в силу фиксированности момента импульса солитона (\ref{76338054}), радиус его оболочки $\cbrho$ зависит от частоты $\omega$. Однако для вычисления $\cbrho$ мы должны иметь значение константы $\xxx$.
}

\engrus{0.5ex}{0.5ex}{%
Nevertheless, to make a very rough estimate, we assume that for a visible light frequency
$\omega$
the shell radius $\cbrho$ has an order of values in the range from the electron classical radius
to half of soliton wave-length.
}{%
Тем не менее, чтобы осуществить очень грубую оценку, предположим, что для частоты видимого света
$\omega$ радиус оболочки $\cbrho$ имеет порядок значений в диапазоне от классического радиуса электрона до половины
длины волны солитона.
}

\begin{subequations}\label{681636311}
\engrus{0.5ex}{0.5ex}{%
Let
}{%
Пусть
}
\begin{align}
\label{506214901}
 & \omega =k \sim  10^{7}\, \text{m}^{-1}
\;,\\
\label{506214902}
 &\cbrho \sim 3\cdot \left(10^{-15} \div 10^{-7}\right)\,\text{m}
 \;,\quad
 \vdlina \sim 0.1\, \text{m}
\;.
\end{align}
\engrus{0.5ex}{0.5ex}{%
Expressions (\ref{43305056}) with (\ref{506214902}) give
}{%
Выражения (\ref{43305056}) с (\ref{506214902}) дают
}
\begin{equation}
\label{508541741}
 \pcoef_{0} \approx \frac{1}{3}
 \;,\quad
 \pcoef_{1}\sim \left(12 \div  31\right)
 \;,\quad
 \pcoef_{2} = 1
\;.
\end{equation}
\end{subequations}

\engrus{0.5ex}{0.5ex}{%
\textadd1{The} standard value of Planck constant must be multiply by the velocity of light for
used unit of frequency (\ref{506214901}):
}{%
Для используемой единицы частоты (\ref{506214901}) стандартное значение постоянной Планка должно быть умножено на скорость света:
}
\begin{equation}
\label{362825251}
 \hbar\approx 2\cdot 10^{-7}\,\text{eV}\cdot\text{m}
\;.
\end{equation}

\begin{subequations}\label{416060451}
\engrus{0.5ex}{0.5ex}{%
Then \textadd1{the} relation (\ref{745250891}) with  (\ref{681636311}) and (\ref{362825251}) gives
}{%
Тогда соотношение (\ref{745250891}) с  (\ref{681636311}) и (\ref{362825251}) дают
}
\begin{align}
\label{416164631}
 &\xxx  \sim \left(1\cdot 10^{-22} \div 7\cdot 10^{-7}\right)\,\text{m}^{3/2}\cdot\text{eV}^{-1/2}
\;,\\
\label{416164632}
 &\xxx^{-2}  \sim \left(2\cdot 10^{12} \div 1\cdot 10^{44}\right)\,\frac{\text{eV}}{\text{m}^{3}}
 \sim \left(3\cdot 10^{-7} \div 2\cdot 10^{25}\right)\,\frac{\text{J}}{\text{m}^{3}}
\;,
\end{align}
\engrus{0.5ex}{0.5ex}{%
\noindent
where the minimal value of $\xxx$
in (\ref{416164631}) and the maximal value of $\xxx^{-2}$ in (\ref{416164632})
correspond to the minimal value of  $\cbrho$ in (\ref{506214902}).
}{%
\noindent
где минимальное значение $\xxx$ в (\ref{416164631}) и максимальное значение $\xxx^{-2}$ в (\ref{416164632})
соответствуют минимальному значению $\cbrho$ в (\ref{506214902}).
}
\end{subequations}

\engrus{0.5ex}{0.5ex}{%
According to \textadd1{the} formula (\ref{338678749}) and taking into account
(\ref{681494511}), (\ref{681636311}), and (\ref{416060451}),
we have the following values for the static part of soliton energy:
}{%
В соответствии с формулой (\ref{338678749}) и учитывая
(\ref{681494511}), (\ref{681636311}) и (\ref{416060451}),
имеем следующие значения для статической части солитонной энергии:
}
\begin{equation}
\label{442495801}
\underline{\Energy}  \approx  \frac{\cbrho^2\,\vdlina}{3\,\xxx^2}
\sim \left(6\cdot 10^{-3} \div 2\cdot 10^{13}\right)\,\text{eV}
\;,
\end{equation}
\engrus{0.5ex}{0.5ex}{%
\noindent
where the minimal value of $\underline{\Energy}$ corresponds to the minimal value of $\xxx^{-2}$ in (\ref{416164632})
and the maximal value of $\cbrho$ in (\ref{506214902}).
}{%
\noindent
где минимальное значение $\underline{\Energy}$ соответствует минимальному значению $\xxx^{-2}$ в (\ref{416164632})
и максимальному значению $\cbrho$ в (\ref{506214902}).
}

\engrus{0.5ex}{0.5ex}{%
Thus to provide the condition $\underline{\Energy} \ll \omega\,\hbar$, the diameter
of soliton shell $2\,\cbrho$ must be closer to the soliton wavelength
than to the electron classical diameter.
}{%
Таким образом, чтобы обеспечить условие $\underline{\Energy} \ll \omega\,\hbar$, диаметр солитонной оболочки
$2\,\cbrho$ должен быть ближе к длине волны солитона, чем к классическому диаметру электрона.
}

\engrus{0.5ex}{0.5ex}{%
Expressing $\cbrho$ from (\ref{745250891}) and substituting it to formula for
$\underline{\Energy}$ in (\ref{442495801}), we obtain
from  (\ref{490939841}) the following formula for soliton energy:
}{%
Выражая $\cbrho$ из (\ref{745250891}) и подставляя в формулу для
$\underline{\Energy}$ в (\ref{442495801}), получаем
из  (\ref{490939841}) следующую формулу для солитонной энергии:
}
\begin{equation}
\label{512588021}
\Energy  \approx \hbar\,\omega + \frac{1}{3\,\xxx}\,\sqrt{\frac{\hbar\,\vdlina}{\omega\,\pcoef_{1}}}
\;,
\end{equation}
\engrus{0.5ex}{0.5ex}{%
\noindent
where $\pcoef_{1}$ is considered to be constant. Here we disregard the dependence $\pcoef_{1} (\cbrho)$ (\ref{43305056})
what is justified for \textadd1{the} used approximation.
}{%
\noindent
где $\pcoef_{1}$ считается константой. Здесь мы пренебрегаем зависимостью $\pcoef_{1} (\cbrho)$ (\ref{43305056}),
что является обоснованным для используемого приближения.
}

\engrus{0.5ex}{0.5ex}{%
The dependence (\ref{512588021}) is shown on Fig. \ref{66179834} for the explicit values of parameters.
Of course, it \textremb{can}\textaddb{must} be considered \textremb{only}\textaddb{mainly} for a qualitative analysis.
}{%
Зависимость (\ref{512588021}) показана на Рис. \ref{66179834r} для определённых значений параметров.
Конечно, она должна рассматриваться в основном для качественного анализа.
}

\begin{figure}[h]
\begin{center}
\ifpdf
  {
  \unitlength 1mm
  \begin{picture}(120,75)
   \put(3,0){\includegraphics[width=110mm]{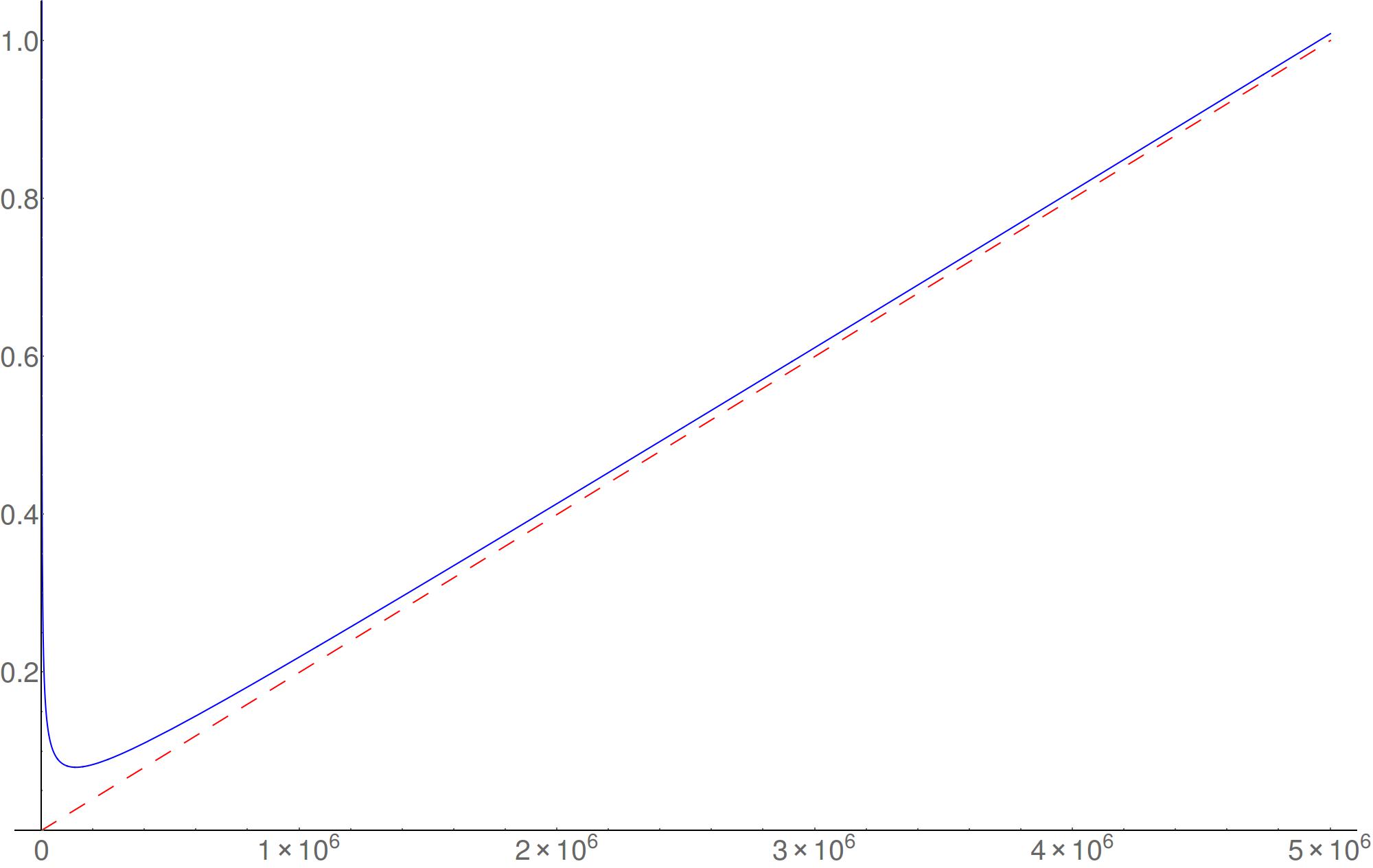}}
    \put(30,65){$\xxx = 7\cdot 10^{-7}\,\text{m}^{3/2}\cdot\text{eV}^{-1/2}$}
   \put(30,60){$\hbar = 2\cdot 10^{-7}\,\text{eV}\cdot\text{m}$}
   \put(30,55){$\vdlina = 0.1\,\text{m}$}
   \put(30,50){$\pcoef_{1} = 12$}
   \put(8,69){$\Energy,\,\text{eV}$}
   \put(103,5){$\omega,\,\text{m}^{-1}$}
  \end{picture}
  }
 \else
  \begin{pspicture}(0,0)(12,7.5)
   \put(0.3,0){\includegraphics[width=110mm]{LSOESTF_Fig61.eps}}
   \put(3,6.5){$\xxx = 7\cdot 10^{-7}\,\text{m}^{3/2}\cdot\text{eV}^{-1/2}$}
   \put(3,6.0){$\hbar = 2\cdot 10^{-7}\,\text{eV}\cdot\text{m}$}
   \put(3,5.5){$\vdlina = 0.1\,\text{m}$}
   \put(3,5.0){$\pcoef_{1} = 12$}
   \put(0.8,6.9){$\Energy,\,\text{eV}$}
   \put(10.3,0.5){$\omega,\,\text{m}^{-1}$}
  \end{pspicture}
\fi
\end{center}
\engrus{0.5ex}{0.5ex}{%
\caption{\label{66179834}Dependence of soliton energy from
%\textremc{soliton}
\textaddc{its} frequency.}
}{%
 \mifeng{\addtocounter{figure}{-1}}{}
  \caption{\label{66179834r}Зависимость энергии солитона от его частоты.}
}
\end{figure}

\engrus{0.5ex}{0.5ex}{%
As we see on Fig. \ref{66179834}, the distinction of soliton energy function from
the linear one $\hbar\,\omega$
(dashed line) can be noticeable in \textadd1{a} low-frequency region.
\textremb{But in a bit core of the plot}We see also a confirmation of the approximate condition (\ref{460697971})
at the center of the plot.
}{%
Как мы видим на Рис. \ref{66179834r}, отличие функции солитонной энергии от линейной $\hbar\,\omega$
(пунктирная линия) может быть существенно в области низких частот.
Мы видим также подтверждение приближённого условия (\ref{460697971}) в центральной части графика.
}

\engrus{0.5ex}{0.5ex}{%
The question arises as to whether there is a static part of energy for real photons.
The appropriate experimental check may be possible with the help of \textadd1{the} extrinsic photoeffect.
If the photon energy not exactly equals to $\hbar\,\omega$, then the frequency dependence of
photoelectron energy may have a weak nonlinearity near photoemission threshold.
The substances with low photoemission threshold is preferable for such experiments.
}{%
Возникает вопрос о том, существует ли статическая часть энергии у реальных фотонов. Соответствующая экспериментальная проверка может быть осуществлена
при помощи внешнего фотоэффекта. Если энергия фотона точно не равна $\hbar\,\omega$, то частотная зависимость энергии фотоэлектронов может
иметь слабую нелинейность вблизи красной границы фотоэффекта.
Вещества с низкой красной границей фотоэффекта являются предпочтительными
для таких экспериментов.
}

\engrus{0.5ex}{0.5ex}{%
Let us next consider all values of \textadd1{the} twist parameter $m$ for \textaddb{the} lightlike solitons.
For \mbox{$m=1$} we have \textaddb{obtained} the known expression for photon energy in the case
\mbox{$\hbar\,\omega\gg \underline{\Energy}$}.
}{%
Рассмотрим далее все значения параметра закрученности $m$ светоподобного солитона.
Для \mbox{$m=1$} мы получили известное выражение энергии фотона в случае
\mbox{$\hbar\,\omega\gg \underline{\Energy}$}.
}

\engrus{0.5ex}{0.5ex}{%
Thus for \mbox{$m\geqslant 2$} here we could be considered a fractional photon with
the following energy expression, according to (\ref{417457381}):
}{%
Таким образом для \mbox{$m\geqslant 2$} мы могли бы здесь рассмотреть
дробный фотон со следующим выражением для энергии, согласно (\ref{417457381}):
}
\begin{equation}
\label{672795211}
\Energy  = \frac{\hbar\,\omega}{m} + \underline{\Energy}
\;.
\end{equation}

\engrus{0.5ex}{0.5ex}{%
But we must pay attention once again to the fact that the twisted lightlike solitons
with \mbox{$m\geqslant 2$} have \textadd1{the} qualitative distinction from ones with \mbox{$m = 1$} in the part of
\textadd1{the} energy representation.
}{%
Однако мы должны ещё раз обратить внимание на тот факт, что закрученные
светоподобные солитоны с \mbox{$m\geqslant 2$} имеют качественное отличие от
солитонов с \mbox{$m = 1$} в части энергетического представления.
}

\engrus{0.5ex}{0.5ex}{%
The energy of longitudinally limited \textremb{and}twisted lightlike soliton with \mbox{$m = 1$}
logarithmically diverges in infinite space, but for \mbox{$m\geqslant 2$} its energy is finite.
In this point of view the solitons with \mbox{$m = 1$} more closely resemble
the plane waves with constant amplitude, the energy of which also diverges in infinite space.
}{%
Энергия продольно ограниченного закрученного светоподобного солитона с
\mbox{$m = 1$}
логарифмически расходится в бесконечном пространстве, тогда как для \mbox{$m\geqslant 2$} его энергия конечна.
С этой точки зрения солитон с \mbox{$m = 1$} более напоминает плоскую волну
постоянной амплитуды, энергия которой также расходится в бесконечном пространстве.
}

\engrus{0.5ex}{0.5ex}{%
Let us consider the representation of \textadd1{the} polarization property of light by twisted lightlike solitons.
}{%
Рассмотрим представление свойства поляризации света закрученными светоподобными солитонами.
}

\engrus{0.5ex}{0.5ex}{%
A beam of these solitons with right or left twist has a necessary symmetry
of  right or left circularly polarized light wave accordingly.
This beam, in particular,
\textadd1{can}
provide the Sadovskii effect \cite{SokolovIV1991UFNEng}, which is
a mechanical angular momentum transfer to absorbent
by circularly polarized electromagnetic wave.
This effect has the experimental verification \mbox{\cite{Beth1935PhysRev,Beth1936PhysRev}},
including one for electromagnetic centimeter waves \cite{Carrada1949Nature}.
}{%
Луч таких солитонов с правой или левой закрученностью имеет необходимую симметрию право или лево циркулярно-поляризованной световой волны соответственно. Этот луч, в частности, может обеспечить эффект Садовского,
заключающийся в передаче механического момента импульса поглотителю
циркулярно поляризованной электромагнитной волной. Этот эффект имеет экспериментальное подтверждение \mbox{\cite{Beth1935PhysRev,Beth1936PhysRev}},
включая эксперимент с сантиметровыми электромагнитными волнами \cite{Carrada1949Nature}.
}

\engrus{0.5ex}{0.5ex}{%
As it is known, the plane circularly polarized
electromagnetic wave with constant amplitude does not have angular momentum \cite{SokolovIV1991UFNEng}. Thus
this wave does not provide the Sadovskii effect.
But the twisted lightlike solitons as well as photons have angular momentum and \textadd1{can} provide this effect \textaddb{respectively}.
}{%
Как известно, плоская циркулярно поляризованная электромагнитная волна
 постоянной амплитуды не обладает моментом импульса \cite{SokolovIV1991UFNEng}. Таким образом эта волна не обеспечивает эффект
 Садовского.
 Однако закрученные светоподобные солитоны также как и фотоны имеют момент
 импульса и соответственно могут обеспечить этот эффект.
}

\engrus{0.5ex}{0.5ex}{%
\textadd1{The} elliptical polarization and, as limiting case, linear one of the soliton beam
could be provided by a coherent combining of solitons twisted to the right and to the left.
}{%
Эллиптическая и, как предельный случай, линейная поляризация солитонного луча могла бы быть обеспечена когерентной комбинацией солитонов, закрученных вправо и влево.}

\engrus{0.5ex}{0.5ex}{%
This representation for elliptical polarization conforms to one \textremb{for}\textaddb{in}
 the beam of photons, which have two helicity states only.
}{%
Это представление для эллиптической поляризации согласуется с представлением
поляризации в луче фотонов, которые имеют только два состояния спиральности.
}

\engrus{0.5ex}{0.5ex}{%
\textadd1{The} peculiarity of the value \mbox{$m=1$} for \textadd1{the} twist parameter becomes apparent here.
According to \textadd1{the} solution symmetry for this case (see Fig. \ref{53247958} and Fig. \ref{53247959}),
the coherent combining of equal quantities of such right and left
twisted solitons can give a beam having
a crystal like
symmetry with axes of the first order.
This case can be interpreted as a linear polarization.
}{%
Выделенность значения \mbox{$m=1$} для параметра закрученности становится здесь очевидной.
Согласно солитонной симметрии для этого случая (см. Рис. \ref{53247958r} и Рис. \ref{53247959r}),
когерентная комбинация одинаковых количеств таких право и лево закрученных
солитонов может дать подобную кристаллографической симметрию первого порядка. Этот случай может быть интерпретирован как линейная поляризация.
}

\engrus{0.5ex}{0.5ex}{%
But for the case of \textaddb{the} solitons with higher \textaddb{values for the} twist parameter, we have for the same conditions the
appropriate crystal like
symmetry with axes of $m\geqslant 2$ order (see Fig. \ref{35836648} and Fig. \ref{35836659}).
This case can not be interpreted as a linear polarization.
}{%
Однако для случая солитонов с высшими значениями параметра закрученности при тех же условиях имеем соответствующую кристаллоподобную симметрию с осью
порядка $m\geqslant 2$ (см. Рис. \ref{35836648r}  и Рис. \ref{35836659r}).
Этот случай не может интерпретироваться как линейная поляризация.
}

\engrus{0.5ex}{0.5ex}{%
Thus the lightlike solitons with \textaddb{the} twist parameter \mbox{$m = 1$} can be considered as usual photons
in some approximation.
But the solitons of \textadd1{the} higher twist \mbox{$m\geqslant 2$} have qualitative differences from the
solitons of the lowest twist \mbox{$m = 1$}.
}{%
Таким образом светоподобные солитоны с параметром закрученности \mbox{$m = 1$} могут рассматриваться как обычные фотоны в некотором приближении.
Однако солитоны с высшей закрученностью \mbox{$m\geqslant 2$} имеют качественные отличия от солитонов низшей закрученности \mbox{$m = 1$}.
}

\mifengrus{\newpage}{}
\engrus{3ex}{2ex}{%
\section{Conclusions}
\label{concl}
}{%
\mifeng{\addtocounter{section}{-1}}{}
\section{Заключение}
\mifeng{}{\label{concl}}
}
\engrus{0.5ex}{0.5ex}{%
Thus we have considered the field model \textremb{for}\textaddb{of} extremal space-time film, which is sometimes
called Born -- Infeld type scalar field model.
}{%
Таким образом мы рассмотрели полевую модель экстремальной пространст\-венно-временной плёнки, которую иногда называют
скалярной моделью типа Борна -- Инфельда.
}

\engrus{0.5ex}{0.5ex}{%
We have obtained \textadd1{the} new class of exact solutions for this model that
is \textaddb{the class of} lightlike solitons.
We have considered the significant subclass of these solutions that \textremb{is}\textaddb{are} twisted lightlike solitons.
It is notable that the energy of these solitons is proportional to its angular momentum in high-frequency approximation.
}{%
Мы получили новый класс точных решений этой модели в виде светоподобных солитонов.
Мы рассмотрели важный подкласс этих решений, а именно, решения, представляющие собой закрученные светоподобные солитоны.
Примечательно, что в высокочастотном приближении энергия этих солитонов пропорциональна их моменту импульса.
}

\engrus{0.5ex}{0.5ex}{%
The soliton under consideration has a singularity which is a moving two-dimensional \textaddb{tubelike} surface
or shell.
The lightlike soliton can have a set of \textaddb{such} tubelike shells with the appropriate cavities.
}{%
Рассматриваемый солитон имеет сингулярность, представляющую собой движущуюся двумерную трубчатую поверхность или
оболочку. Светоподобный солитон может иметь множество таких трубчатых оболочек с соответствующими полостями.
}

\engrus{0.5ex}{0.5ex}{%
A relatively simple twisted lightlike soliton with one cavity \textrem1{has}\textadd1{was} considered in details.
This soliton is characterized, in particular, by a twist parameter $m$ which is a natural number.
The energy of longitudinally limited this soliton in infinite space is finite for $m\geqslant 2$, but for $m = 1$ its energy is logarithmically divergent.
For the case $m=1$ we have the asymptotic relation between soliton energy, momentum, and angular momentum, which is characteristic for photon.
}{%
Детально рассмотрен относительно простой закрученный светоподобный солитон с одной полостью. Этот солитон характеризуется,
в частности, параметром закрученности $m$, представляющим собой натуральное число.
Энергия продольно ограниченного такого солитона в бесконечном пространстве является конечной для $m\geqslant 2$,
однако для  \mbox{$m = 1$} его энергия логарифмически расходится.
Для случая $m=1$ мы имеем асимптотическое соотношение между энергией, импульсом и моментом импульса солитона, которое является характерным для фотона.
}

\engrus{0.5ex}{0.5ex}{%
Then we have investigated relations of the twisted lightlike solitons with $m=1$ to photons.
The model of ideal gas of the twisted lightlike solitons in a bounded volume has considered for this purpose.
Planck formula for the soliton energy spectral density in the volume
has obtained with explicit assumptions in some approximation.
}{%
Затем мы исследовали отношение закрученных светоподобных солитонов с \mbox{$m = 1$} к фотонам.
Для этой цели рассмотрена модель идеального газа закрученных светоподобных солитонов в ограниченном объёме.
Получена формула Планка для спектральной плотности энергии солитонов в объёме при определённых предположениях в некотором
приближении.
}

\engrus{0.5ex}{0.5ex}{%
An experimental check for a validity of the obtained soliton energy
\textremd{exact}\textaddd{true} formula for real photon is proposed.
}{%
Предложена экспериментальная проверка справедливости полученной
истинной
формулы энергии солитона для реальных фотонов.
}

\engrus{0.5ex}{0.5ex}{%
A beam of twisted lightlike solitons \textadd1{was} considered.
\textremb{We have shown}\textaddb{It was noted} that this beam \textadd1{can} provide
the effect of mechanical angular momentum transfer to \textadd1{an} absorbent
by \textadd1{the} circularly polarized beam. This effect well known for photon beam.
}{%
Рассмотрен луч закрученных светоподобных солитонов. Отмечено, что этот луч может обеспечить эффект передачи
механического момента импульса поглотителю циркулярно поляризованным лучём. Этот эффект хорошо известен для фотонного луча.
}

\engrus{0.5ex}{0.5ex}{%
It has been found that a beam of the twisted lightlike solitons with \mbox{$m=1$} can provide \textaddb{the} polarization
property of light as well as photon beam.
}{%
Обнаружено, что луч закрученных светоподобных солитонов с  \mbox{$m=1$} может обеспечить свойство поляризации
света, также как фотонный луч.
}

\engrus{0.5ex}{0.5ex}{%
Thus we have a correspondence between photon and \textaddb{the} lightlike twisted soliton with the minimal value of
\textadd1{the} twist parameter.
}{%
Таким образом мы имеем соответствие между фотоном и светоподобным закрученным солитоном с минимальным значением
параметра закрученности.
}

\engrus{3ex}{2ex}{%
\section*{Acknowledgements}
}{%
\section*{Благодарности}
}
\label{acknowl}
\engrus{0.5ex}{3.5ex}{%
The author is grateful to participants of ``XVI Workshop on High Energy Spin Physics (D-SPIN2015)''
(JINR, Dubna, Russia, September 8--12, 2015)
for helpful discussions (see also the conference paper \cite{Chernitskii2016a}).
}{%
Автор признателен участникам ``XVI Симпозиума по высокоэнергетической спиновой физике (D-SPIN2015)''
(ОИЯИ, Дубна, Россия, 8--12 сентября, 2015)
за полезные обсуждения (см. также статью в трудах конференции \cite{Chernitskii2016a}).
}

\end{document}